\documentclass[useAMS,usenatbib]{mn2e}
\usepackage{rotating}
\usepackage{subfigure}
\RequirePackage{setspace}
\usepackage{graphicx}
\usepackage{fancyhdr}
\usepackage{natbib}
\usepackage{amsmath}
\usepackage{amssymb}
\usepackage{aas_macros}
\usepackage{epstopdf}
\usepackage{lscape}
\usepackage{rotating}

\title[High redshift clustering in star-forming galaxies]{Identifying clustering at high redshift through actively star-forming galaxies}
\author[L. J. M. Davies et. al.]{L. J. M. Davies$^{1,4}$\thanks{E-mail:
luke.j.davies@uwa.edu.au}, M. N. Bremer$^{1}$, E. R. Stanway$^{2}$, K. Husband$^{1}$, M. D. Lehnert$^{3}$, 
\newauthor E. J. A. Mannering$^{1}$ \\
$^{1}$Department of Physics, University of Bristol, H.H. Wills Physics Laboratory, Tyndall Avenue, Bristol, BS8 1TL, UK\\
$^{2}$Department of Physics, University of Warwick, Gibbet Hill Road, Coventry, CV4 7AL, UK\\
$^{3}$Institut dÕAstrophysique de Paris, UniversitŽ Pierre et Marie Curie/CNRS,98 bis Bd Arago, 75014 Paris, France \\
$^{4}$ICRAR, The University of Western Australia, 35 Stirling Highway, Crawley, WA 6009, Australia  }

\begin{document}

\date{Re-draft: Nov 2013}

\pagerange{\pageref{firstpage}--\pageref{lastpage}} \pubyear{2013}

\maketitle

\begin{abstract}

Identifying galaxy clustering at high redshift (i.e. $z>1$) is essential to our understanding of the current cosmological model. However, at increasing redshift, clusters evolve considerably in star-formation activity and so are less likely to be identified using the widely-used red sequence method. Here we assess the viability of instead identifying high redshift clustering using actively star-forming galaxies (SMGs associated with over-densities of BzKs/LBGs). We perform both a 2- and 3-D clustering analysis to determine whether or not true (3D) clustering can be identified where only 2D data are available. As expected, we find that 2D clustering signals are weak at best and inferred results are method dependant. In our 3D analysis, we identify 12 SMGs associated with an over-density of galaxies coincident both spatially and in redshift - just 8\% of SMGs with known redshifts in our sample. Where an SMG in our target fields lacks a known redshift, their sightline is no more likely to display clustering than blank sky fields; prior redshift information for the SMG is required to identify a true clustering signal. We find that the strength of clustering in the volume around typical SMGs, while identifiable, is not exceptional. However, we identify a small number of highly clustered regions, all associated with an SMG. The most notable of these, surrounding LESSJ033336.8-274401, potentially contains an SMG, a QSO and 36 star-forming galaxies (a $>20\sigma$ over-density) all at $z \sim 1.8$. This region is highly likely to represent an actively star-forming cluster and illustrates the success of using star-forming galaxies to select sites of early clustering. Given the increasing number of deep fields with large volumes of spectroscopy, or high quality and reliable photometric redshifts, this opens a new avenue for cluster identification in the young Universe.

\end{abstract}


\section{Introduction}
\label{Into}

High redshift ($z>1$) galaxy clusters and groups are sensitive probes of cosmology and the dependance of galaxy evolution on environment.  According to current paradigms, today's galaxy clusters form in the early Universe ($z>3$) and grow hierarchically over cosmological timescales.  

Both the exceptionally tight colour-magnitude relation in local clusters \citep[$e.g.$][]{Bower98} and stellar population models \citep{Thomas10}, suggest that today's massive cluster galaxies formed in a brief (perhaps as little as 1\,Gyr) period at $z > 2$. Assuming a typical cluster core of a few $\times10^{12}$\,M$_\odot$ at $z=0$ forms in this interval, it would host a few $\times1000$\,M$_\odot$ yr$^{-1}$ of star-formation in a comparatively small volume at $z\gtrsim2$. Recent observations have supported the interpretation that at increasing redshifts ($z>2$), galaxy clusters become increasingly populated with star-forming galaxies \citep{Tran10, Tanaka13}, but whether these systems formed in a monolithic collapse scenario \citep[see,][for discussion]{Davies13b}, through mergers of many small starburst galaxies \citep[$e.g.$ Lyman Break galaxies, LBGs, with SFRs $<200$\,M$_{\odot}$\,yr$^{-1}$,][]{Davies13} or as a region of small star-forming systems \citep[such as LBGs or $sBzK$s,][]{Daddi04} surrounding a more massive obscured star-forming galaxy \citep[a submillimeter galaxy, or SMG, with SFR $\gtrsim$1000\,M$_{\odot}$\,yr$^{-1}$,][]{Casey13} is unclear. While \cite{VanDokkum08} and \cite{Cimatti08} find evidence in favour of the merger scenarios, others such as \cite{Nipoti09}, \cite{Nipoti12} and \cite{Newman12} predict that mergers alone cannot account for the rapid structure evolution observed at $z > 1.3$. \cite{Huang13} take a middle ground, finding that local elliptical galaxies have a three-component structure that implies the central core of such systems formed from major (wet) mergers at high redshift, while the outer regions formed through minor (dry) mergers at a later epoch. 

Evidence for formation processes from galaxy cluster morphology is similarly ambiguous. \cite{Bassett13}, studying a $z\sim1.6$ cluster, and \cite{Bauer11}, \cite{Grutzbauch12} and \cite{Santos13}, all targeting the same $z\sim1.4$ cluster, find that star-forming galaxies do not reside in the cluster core but are confined to large radii ($>$200 kpc), consistent with the SFR-radius relation found in low redshift clusters \citep{Treu03, Haines07}. Conversely, other studies \citep[$e.g.$][]{Hilton10, Tran10, Strazzullo13} suggest that high redshift clusters display an inverse SFR-radius relation, with star-forming galaxies residing in the cluster core, or even no SFR-radius relation at all \citep{Ziparo13}! Clearly a fuller understanding and consensus on the structure and hence formation mechanism of distant galaxy clusters requires further study.

Of course, investigation of these early stages of structure growth requires the identification of clusters in their infancy at high redshift. One successful method for doing this is the galaxy red-sequence technique \citep[first discussed in][ and used in many studies since]{Gladders00}. This searches for a population of mature, passively evolving galaxies occupying a well-defined region in colour-magnitude space, inferring their near-coeval formation and hence likely association as cluster members. While this technique has proved successful in identifying some clusters at $z>1$ \citep[$e.g.$ using the Spitzer Adaptation of the Red-Sequence Cluster Survey, ][]{Wilson06}, it is biased towards systems which formed at a much earlier epoch, and hence have a significant number of old, passively evolving members. It is insensitive to less evolved structures dominated by active star-forming galaxies. 

An alternative is to consider other tracers. It has long been thought that massive high-z galaxies (such as SMGs) trace the most significant over-densities of mass at any epoch \cite[$e.g.$][]{Mo02,Springel06}, and the recent simulations of \cite{Gonzalez11} suggest that the majority of Abell-richness clusters contain at least one bright and distant ($z > 1.5$, S$_{\mathrm{850\mu m}}$ $>$ 5 mJy) SMG in their merger tree.  If so, the volumes surrounding such systems should also contain an excess of other, more numerous, galaxy populations relative to the mean for the given epoch: high redshift SMGs should act as signposts for sites of cluster formation at early times. This premise has been explored at very high redshift with the identification of proto-cluster regions surrounding both SMGs and QSO/SMGs at $z\sim5$ \citep{Capak11,Husband13}, and would seem to be supported by \cite{Huang13}'s suggestion that the environs of forming, massive ellipticals must host a supply of relatively small, gas poor, post-starburst systems for later accretion.

So, if the red sequence technique misses forming clusters, but the most massive galaxies signpost the most massive regions, do we observe clustering around those we know about?  And are our methods sufficient to be sure of our answer to that question?

Some studies have suggested that these questions merit further study.  Using deep multi-wavelength coverage of the Cosmic Evolution Survey (COSMOS) field, \citet[][hereafter A10]{Aravena10} identified candidate star-forming galaxies at $1.4 < z < 2.5$ ($sBzK$s). Using a two dimensional analysis (i.e. projected distribution of sources in the plane of the sky), they compared regions of high $sBzK$ number density with the positions of SMGs across a small $\sim0.135$ deg$^2$ region with millimeter data. They identified three over-densities of systems surrounding 1.1 mm (MAMBO-2) detected sources at $z = 1.4 − 2.5$, suggesting that $\sim$30\% of bright SMGs at this redshift are embedded in a galaxy over density. However their analysis was limited by the small field area and their use of two-dimensional clustering data.

\begin{figure*}
\begin{center}

\includegraphics[scale=0.42]{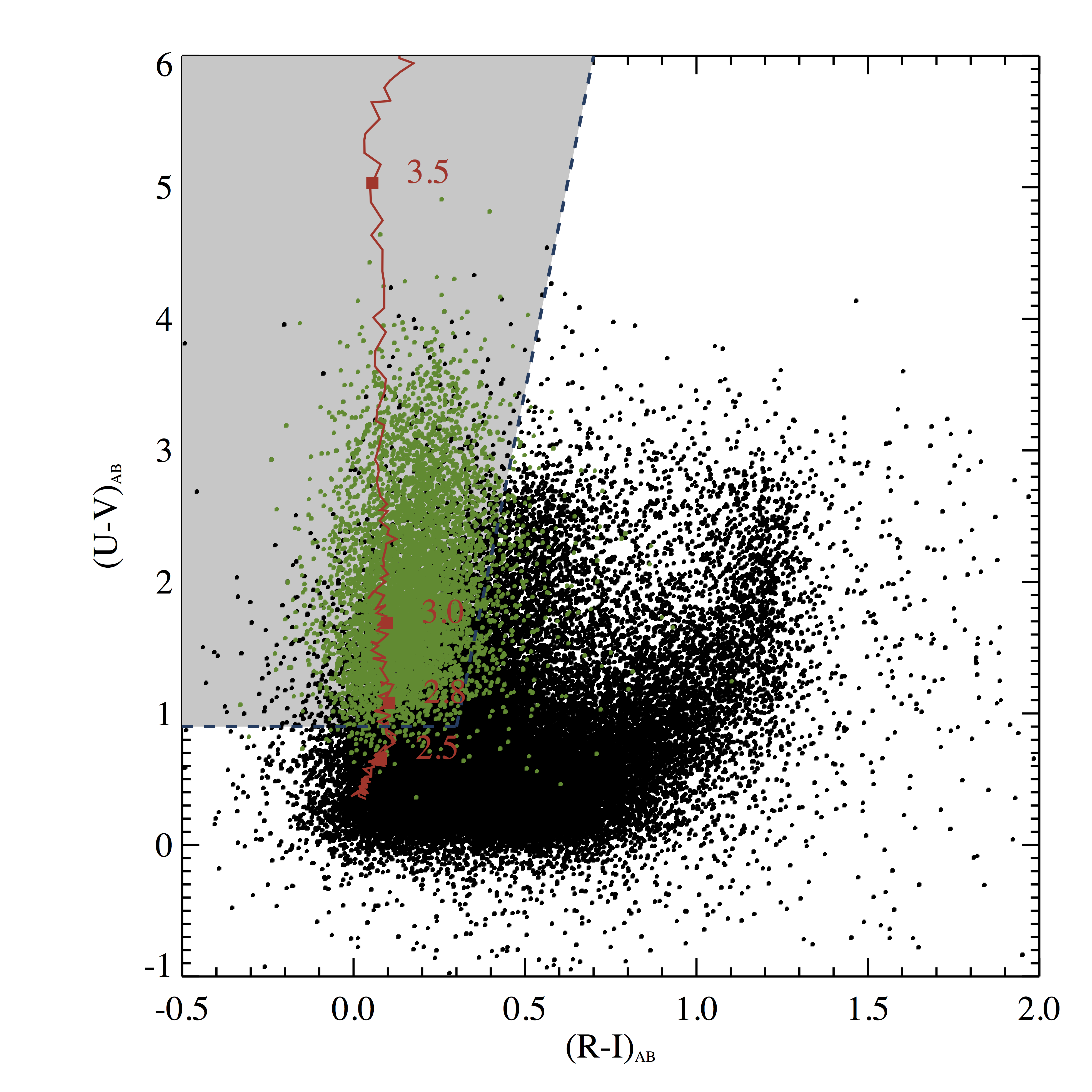}
\includegraphics[scale=0.42]{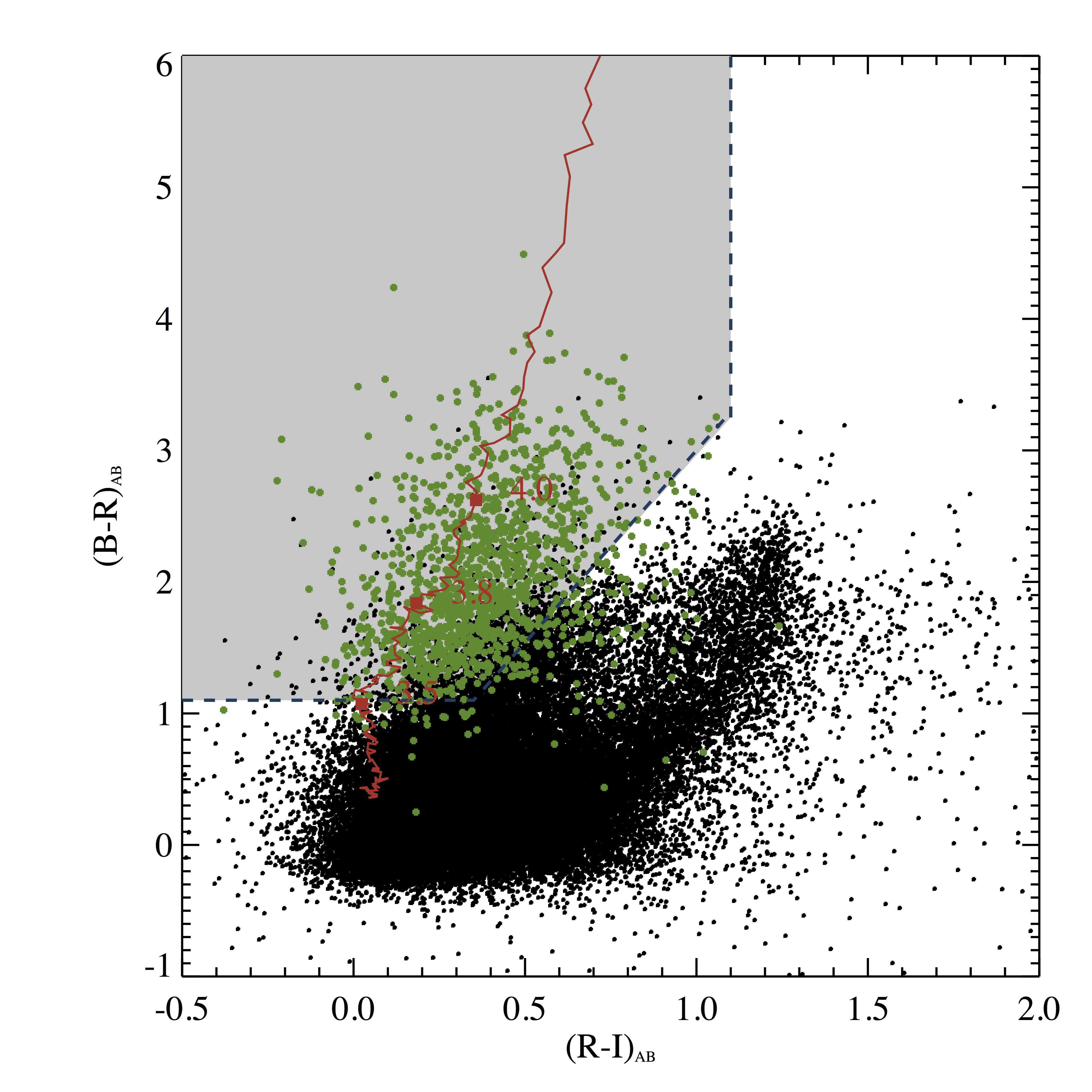}

\caption[]{The colour selection of LBGs at $z\sim3$ (left) and $z\sim4$ (right) in the COSMOS field. Black points show a representative sample of all sources in the field. The red line indicates the colours of a young, metal poor galaxy at various redshifts, taken from the stellar population models of \citet{Maraston05}. Green points show sources with photometric redshifts within our desired colour selection range ($2.8<z<3.5$ in the left panel and $3.5<z<4.5$ in the right panel). Our colour selection regions are highlighted by the grey filled box.}    

\label{fig:cols_z34}
\end{center} 
\end{figure*}

In this work, we extend the analysis of Aravena et al to the bulk ($\sim$0.75 deg$^2$) of the COSMOS region, using SMG catalogues taken from the AzTEC observations of the field \citep{Aretxaga11}. This area encompasses the region discussed in A10 - allowing a comparison between our methods. We also select star-forming galaxies over a much larger range of epochs than that covered by A10 \citep[$1.4<z<4.5$, targeting the bulk of the SMG redshift distribution,][]{Chapman05}, ensuring that we do not miss clustered regions around SMGs at an earlier time. We also perform a similar analysis over the 0.25 deg$^2$ of the Extended Chandra Deep Field South (ECDF-S) using the deep LABOCA observations of the field \citep{Weiss09}. In these fields, we have the benefit of deep many-band photometric and spectroscopic data which we can use to verify or refute any clustering observed in 2D through a 3D analysis (i.e. incorporating redshift information).

We consider an analysis of spatial clustering in these two fields and address the following primary questions:

 \vspace{1mm}
 
\noindent $\bullet$ Do SMGs truly reside in the most clustered regions?

 \vspace{1mm}
 
\noindent $\bullet$ What is the typical 3D environment of an SMG, as traced by star-forming galaxies?

 \vspace{1mm}
 
\noindent $\bullet$ Can we identify clustered regions of typical star-forming galaxies around SMG using a purely 2D analysis (i.e. without prior redshift information)?

 \vspace{1mm}
 
\noindent $\bullet$ Do clustered regions of star forming galaxies exist, within the targetted redshift range) that are not associated with a known SMG?

 \vspace{1mm}
 
\noindent $\bullet$ Are regions clustered in star forming galaxies (LBGs or sBzKs) also clustered in more passive systems at the same redshift (pBzKs, Daddi et al. 2004), and which is the better tracer? 

In sections 2 through 6 we discuss our methods and observations, while in sections 7 to 10 we discuss the implications of our work and also present two strong candidates for $z>1.4$ galaxy clusters, marked out through star formation rather than passively evolving cluster members. We propose these (and other similar regions) as key sites in the formation of today's massive structures. Finally, in section 11 we present our conclusions.

Throughout this paper all optical magnitudes are quoted in the AB
system \citep{Oke83}, and we use a standard $\Lambda$CDM cosmology with
{H}$_{0}$\,=\,70\,kms$^{-1}$\,Mpc$^{-1}$,
$\Omega_{\Lambda}$\,=\,0.7 and $\Omega_{M}$\,=\,0.3.

\section{Data Sets and Source selection}
\label{sec:cands}

\subsection{Star-forming galaxies and passive galaxies}

In order to evaluate the clustering in our two target fields, we must first select samples of star forming galaxies for analysis. Both the COSMOS field \citep[$e.g.$][]{Guillaume06, Capak08} and the ECDF-S \citep[e.g.]{Gawiser06} are abundantly supplied with publicly-available, multi-wavelength archival data across a broad wavelength range although the use of different filter sets in the two fields requires some adjustment of galaxy selection criteria to ensure that like-for-like comparisons are undertaken. Bearing this in mind, we use public catalogue data and well-established colour selection methods to identify star-forming systems at  $1.4\lesssim z \lesssim4.5$  - using three separate colour/magnitude cuts. 

Initially, we select sources at $2.8\lesssim z \lesssim4.5$ using the Lyman break colour selection criteria \citep[hereafter LBGs, $e.g.$][]{Steidel95}. This method essentially performs a crude photometric redshift estimate, selecting sources with a large spectral break and relatively flat continuum long-ward of the break. This selection is aimed at identifying the Lyman-$\alpha$ continuum break produced via intervening neutral hydrogen in relatively unobscured star-forming galaxies at high redshift \citep[for further details of the well documented Lyman Break Technique, see][etc and references therein]{Steidel95, Lehnert03,Stanway03,Verma07, Douglas09}. We apply two sets of colour selection criteria to identify sources at $2.8<z<4.5$, outlined in Figure \ref{fig:cols_z34} and Table \ref{tab:col_sel}. These colour cuts are designed to select potential high redshift sources while limiting the contamination fraction in our sample. 

Colour selection of similar LBG sources at $z\lesssim2.8$ proves much more problematic as the Lyman break moves from optical wavelengths to the UV - and is therefore only observable by space based instruments. However, many ground based studies have been successful in identifying large numbers of $z\lesssim2.8$ star-forming galaxies through other colour selection techniques. Although these sources vary in intrinsic properties to LBGs at higher redshift, they are typical of the general population at $z<2.8$. The most notable of these is the $BzK$ colour selection outlined in \citet{Daddi04}. This method essentially selects sources with greater z-K than B-z colour and targets the 4000\AA/Balmer breaks in systems at $z\sim2$. While these systems appear more massive and reddened than true LBG selected sources at this epoch \citep[see][]{Haberzettl12}, and comprise both passive and starforming subclasses, they represent typical star-forming galaxies at $z\sim2$ and are easily selected in the publicly available data covering the COSMOS field. We therefore use the $BzK$ method as a third selection criterion in this study.         

It is well documented that simple colour selections such as these still contain significant numbers of contaminating sources such as lower-redshift passively evolving galaxies and (at higher redshifts) cool stars \citep[$e.g.$][]{Stanway08}. In order to mitigate the effects of such contaminants, we also make use of the many-band photometric redshift catalogues that have been published for sources in both fields.  

In the ECDF-S we use the deep multi-wavelength data taken from the MUSYC survey \citep{Gawiser06}, which provides deep coverage in 32 (broad and intermediate) bands in the optical and near-IR. Such detailed photometric coverage allows the determination of accurate photometric redshifts over a large redshift range \citep[see][for further details]{Cardamone10}. Our ECDF-S LBG sample is identical to that described in \cite{Davies13}. Hence we refer the reader to that work for the details of our selection. 

In the COSMOS field, we use the photometric redshift catalogues of \cite{Ilbert08}. We only select sources with a best fit photometric redshift within our selection redshift range \citep[for photometric redshift calculations, see][]{Ilbert08}. We also remove objects whose $68\%$ confidence (1$\sigma$) error range extends below the lower boundary of our selection redshifts to eliminate sources with poorly-fit photometric redshifts. Therefore, we essentially treat the well-constrained and reliable photometric redshifts of  \cite{Ilbert08} as a redshift confirmation. To avoid any additional stellar contaminants we remove all ACS point sources with a stellar or AGN type spectral energy distribution \cite[see][]{Robin07} and sources which display $BzK$ colours which are consistent with stars at K$_{AB}\,<\,22$ \citep{Daddi04}. In Section \ref{sec:results} we shall briefly discuss the affects of using a simple colour cut with no photometric classification applied.  This removes any bias applied by selecting only those sources with good photometric redshifts (effectively a magnitude cut) but uses a more highly contaminated sample.

\begin{figure}
\begin{center}

\includegraphics[scale=0.28]{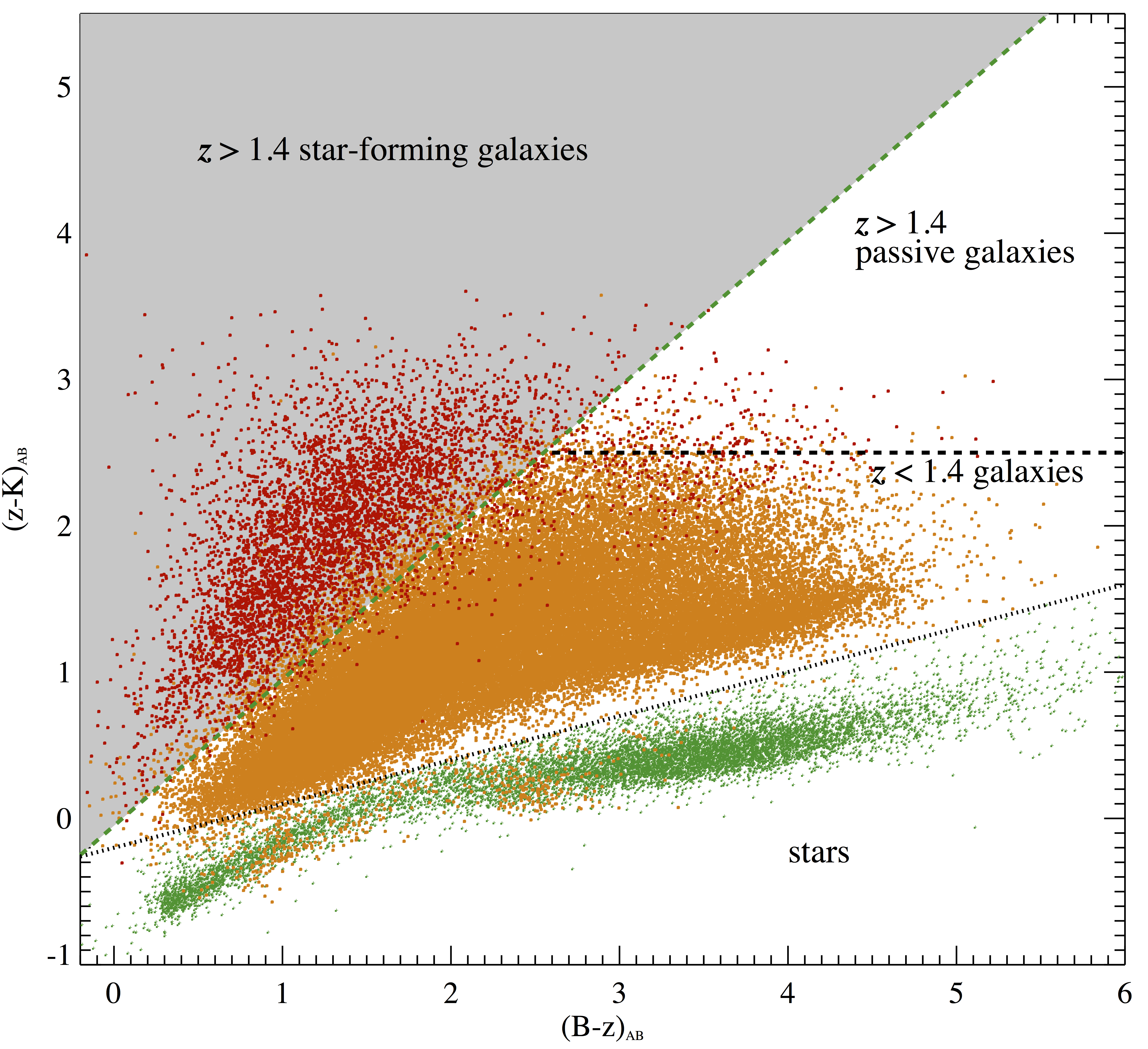}

\caption[]{The $BzK$ colour selection of galaxies at $z>1.4$. Objects with a clear z- and K-band detection, and a reliable photometric redshift are plotted as points.  The colour of the points represent a photometric redshift estimation of $z>1.4$ (red), $z<1.4$ (orange) and star (green). We select $z>1.4$ star-forming galaxies as sources with (z-K)$\,>\,$(B-z)-0.2 (grey shaded region) and passive galaxies a sources with (z-K)$\,<\,$(B-z)-0.2 and (z-k)$\,>\,$ 2.5.}    

\label{fig:bzk}
\end{center} 
\end{figure}

For $BzK$ galaxies, we apply the same selection as that outlined in \cite{Greve10} in the ECDF-S, with colour selections for these samples given in Table \ref{tab:col_sel}. In COSMOS  we produce a sample of $BzK$ selected galaxies (both star-forming and passive) using similar colour selections to those outlined in \citet{Daddi04}, supplementing the published catalogues by the K-band magnitudes from the year 1 UltraVISTA data set \citep[K$_{\mathrm{AB}}\,<\,23.4$,][] {McCracken2012}. To be consistent with our higher redshift samples, we also apply a photometric redshift cut restricting the sample to sources with $1.4<z_{\mathrm{photo}}<2.5$. Figure \ref{fig:bzk} and Table \ref{tab:col_sel} illustrate the $BzK$ colour selection of sources at $1.4<z<2.5$ in the COSMOS field.             

\begin{table*}
\centering

\begin{scriptsize}

\begin{tabular}{c c c c c c c c }
\hline
\hline
Field & Sources & Redshift & Colour & Colour & Colour & Magnitude & IR colour\\
\hline

COSMOS & sBzK &1.4-2.5 & z - K $ > $ (B - z) - 0.2 & - & - & $17.0 < $ K $< 23.4$ & - \\
COSMOS & LBGs &2.8-3.6 & U - V $ > $ 0.9 & U - V $ > $ 12.75(V - R) + 2.925& - & 21.1$<$ V $ < $ 27.2 & - \\
COSMOS & LBGs & 3.6-4.5 & B - R $ > $ 1.1 & B - R $ > $ 2.92(R - I) -0.08& - &  22.0$<$ R $ < $ 27.0& I - J $ < $ 1.0 \\
COSMOS & pBzK &1.4-2.5 & z - K $ < $ (B - z) - 0.2 & z - K $ > $ 2.5 & - & $17.0 < $ K $< 23.4$ & - \\
 & & & & & & \\
 
ECDF-S & sBzK &1.4-2.5 & z - K $ > $ (B - z) - 0.2 & - & - & $17.0 < $ K $< 23.4$ & - \\
ECDF-S & LBGs & 2.8-3.6 & U - V $ > $ 1.2 & V - R $ < $ 1.1 & U - V $ > $ 3.63(V - R) + 0.58&V $ < $ 26.2 & - \\
ECDF-S & LBGs & 3.6-4.5 & B - R $ > $ 1.6 & R - I $ < $ 1.4 & B - R $ > $ 1.27(R - I) + 1.1&R $ < $ 27.0& I - J $ < $ 1.0\\
ECDF-S & pBzK &1.4-2.5 & z - K $ < $ (B - z) - 0.2 & z - K $ > $ 2.5 & - & $17.0 < $ K $< 23.4$ & - \\

\hline
\end{tabular}
\end{scriptsize}
\caption{The colour selection criteria used to identify star-forming galaxies at $z\sim2$, 3 and 4. An additional I-J colour selection is applied at $z\sim4$  in order to reduce contamination from low redshift sources \citep[see][]{Stanway08}. }

\label{tab:col_sel}        
\end{table*}

\subsection{Sub-mm bright sources}

As discussed in the introduction, submillimeter bright galaxies are believed to be more massive than the systems identified above, and to be forming stars at a significantly higher rate. In order to evaluate the presence or absence of these sources in early structures, we use available surveys at (sub)-millimeter wavelengths in each field.

The COSMOS field has been surveyed at 1.1mm using the 144-pixel AzTEC camera. The AzTEC observations cover a 0.72\,deg$^{2}$ region in the centre of the COSMOS field  obtaining a uniform noise of $\sim1.26$\,mJy/beam \citep[see][for details]{Aretxaga11}. We use the catalogue produced in \cite{Aretxaga11}  which contains 189 sources detected at S/N$\,>\,$3.5. As the region with uniform noise properties is smaller than the area covered by the UV and optically selected sources, we limit all our samples to this region (which incorporates that considered in earlier, similar studies, $e.g.$ A10). In order to identify 3D clustering of sources surrounding the SMGs, we use both the spectroscopically-confirmed \citep[43 sources at $z>1.4$,][]{Smolcic12, Casey12a, Casey12b} and photometrically-derived \citep[21 sources with best-fit $z_{\mathrm{photo}}>1.4$,][]{Smolcic12} redshifts for the sample. We exclude any SMG from our analysis which has a redshift $z<1.4$. Combined with the AzTEC sources without redshifts this gives a total of 234 SMGs in the field (hereafter the COSMOS-SMG sample). A brief summary of the surveys used in this work can be found in Table \ref{tab:submm_obs}.

The ECDF-S has been surveyed at 870$\mu$m using the 295-element Bolometer camera LABOCA \citep[the Large APEX Bolometer Camera,][]{Siringo09}. We used data from the Large APEX Bolometer Camera Survey of the Extended Chandra Deep Field South (LESS) \cite[for full details see ][]{Weiss09}. The LESS map comprises of 200\,h of integration time, covering the 0.25\,deg$^2$ of the ECDF-S with a uniform coverage of rms=1.2\,mJy/beam and a beam FWHM of 19 arcsec. The LESS survey identifies 126 submm bright sources at $>3.7\sigma$ in the ECDF-S. We glean redshift data and a small number of additional sources from the data, of which 17 sources have spectroscopic redshifts (Casey et al. 2011, 2012b) and 67 sources have best-fit photometrically derived redshifts of $z_{\mathrm{photo}}>1.4$ \citep{Biggs10, Wardlow11}, forming a total sample of 136 unique submillimeter sources. We note that this `ECDFS-SMG' sample contains $\sim60\%$ of the number of SMGs in the COSMOS field while covering only $\sim35\%$ of the area. The number density of SMGs in the ECDF-S remains much higher than that in COSMOS even when considering sources to comparable 1.1mm depth (see Table \ref{tab:submm_obs}) and will be discussed further in Section \ref{sec:comparisons}.

\begin{table*}
\begin{scriptsize}
\begin{center}
\begin{tabular}{c c c c c c c c c c c }
\hline
\hline
Field & Survey & $\lambda$ & Area & Depth & N$_{\mathrm{gals}}$ & N$_{\mathrm{photo-}z}$  &   N$_{\mathrm{spec-}z}$ & Depth$_{1.1\mathrm{mm}}$ & N$_{\mathrm{gals}}>3.5\sigma$ \\
(1)   & (2)    & (3)       & (4)  & (5)   & (6)               & (7)                  &  (8)                  & (9)                   & (10) \\

\hline
COSMOS & AzTEC + & 1.1mm     & 0.72 & 1.26\,mJy & 234 &  21  & 43 & 1.26       & 189 \\   
ECDF-S & LESS +  & 870$\mu$m & 0.25 & 1.20\,mJy & 136 &  67  & 17 & $\sim0.75$ & 96 \\ 

COSMOS & COSMOS & Optical-NIR & 1.0 & $I^{\mathrm{+}}_{\mathrm{AB}}<25.5$ & 332000 & $\sim135000$ & - & - &-\\    
ECDF-S & MUSYC & Optical-NIR & 0.25 & $R_{\mathrm{AB}}<25.3$ & 44000 & $\sim30000$ & - & - &-\\

\hline

\end{tabular}

\caption{Summary of survey data used in this work. (1) Field covered, (2) survey name, + denotes that the survey in supplemented with additional sources, (3) wavelength of survey, (4) area of survey region in deg$^{2}$, (5) survey depth - for optical-NIR surveys this is the limit for reliable photometric redshift estimation, (6) number of source identified in the region - for optical-NIR surveys this is for sources with R$<$25.5, (7) number of sources with photometric redshifts at $z>1.4$,  (8) number of spectroscopically confirmed SMGs at $z>1.4$, (9) depth of the survey at 1.1mm, assuming source flux scaling of $f\propto\lambda^{-2}$ for LESS - in order to compare depths of submm surveys directly, (10) number of galaxies which would be detected as $3.5\sigma$ sources to the 1.1mm flux limit of the AzTEC/COSMOS survey. Note the high number density of SMGs in the ECDF-S relative to COSMOS, even after adjusting for observed wavelength. }
\label{tab:submm_obs}
\end{center}
\end{scriptsize}
\end{table*}

\section{Analysis}
\label{sec:analysis}

\subsection{2D Correlations}

We first consider the projected number density of star-forming galaxies in the plane of the sky, using only the crude colour selections and no additional redshift information. We produce number density maps using the method outlined in \cite{Dressler80} and also in A10. For each of our high redshift samples, and each field, we grid the region into equally spaced positions separated by 7$^{\prime\prime}$. Each grid position is then assigned a value $\rho_{n}$, defined as:

\begin{equation*}
\rho_n=n/\pi(d_n)^2,
\end{equation*}

 where $d_n$ is the distance to the $n^{\mathrm{th}}$ nearest object. 

We scale the derived number density values for a given value of $n$ with respect to the mean and standard deviation of the density distribution over the entire field, in order to assess the significance of any over-dense regions\footnote{Note that the distribution of projected number densities is not Gaussian.}. For consistency with earlier work by A10, we consider the case $n=7$ for the $sBzK$ samples. A $\sim3\times$ standard deviation galaxy over-density (relative to the mean of our $sBzK$ sample, $d_7$=0.43\,Mpc) occurs when 7 source occupy a region defined by $d$=0.375\,Mpc or less. For our other photometric samples, at $z\sim3$ and 4 and in our $pBzK$ sample, we use $n=3$ to produce our density maps - selected such that an over-density of the same significance occurs on the same scale as that defined above for $sBzKs$. In selecting these values of $n$, we are sensitive to structures on small projected angular scales, and also study the density with a reasonably large sample of data points. In the Section \ref{sec:vary} we shall also discuss how varying the choice of $n$ affects the over-densities which are identified.   

\subsection{3D Correlations}

We also undertake a clustering analysis in three dimensions, using the accurate photometric redshifts of \cite{Ilbert08} in COSMOS, and \cite{Cardamone10} in the ECDF-S. Where either an SMG or a peak in the projected number density has been identified, we generate redshift distributions surrounding that location in $\Delta z=0.1$ bins (similar to the typical photometric redshift error on the star-forming galaxies). For SMGs with known redshifts we centre the redshift bins on the precise SMG redshift. 
We investigate these redshift distributions on 0.375, 0.75, 1.5, and 6\,Mpc scales (co-moving). In this manner we aim to identify clustered sources on a broad spectrum of scales, ranging from nascent galaxy clusters right down to the formation locations of individual galaxies. We also consider both $\Delta z=0.2$ and $\Delta z=0.3$ binning. We find that - as expected - any clustering signal is reduced by averaging over a larger redshift interval, and so do not present detailed results for these cases. In the following, we discuss results derived from the $\Delta z=0.1$ binning. 

We consider four different types of regions: 

\vspace{1mm}
\noindent $\bullet$ Volumes centred on SMGs and also associated with a region of high-projected number density in the 2D analysis, are examined for peaks in the galaxy photometric redshift distribution coincident with the SMG redshift, assessing the validity of the 2D analysis (Section \ref{sec:2d_in_3d}).

\vspace{1mm}
\noindent $\bullet$ Volumes surrounding all SMGs with known redshift are examined for three dimensional clustering at that redshift, irrespective of 2D results (Section \ref{sec:2d_in_3d_2}). 

\vspace{1mm}
\noindent $\bullet$ Volumes surrounding all SMGs \textit{without} known redshifts are also examined for three dimensional galaxy clustering, irrespective of redshift (Section \ref{sec:no_red}). 

\vspace{1mm}
\noindent $\bullet$ Volumes containing a high projected number density of star-forming galaxies in the 2D analysis, are also searched for three dimensional clustering signals, irrespective of the presence of an SMG (Section \ref{sec:no_SMG}).

We define the significance of any redshift spike in terms of the standard deviation, $\sigma$, of number counts in regions on the same angular scale and redshift bin size, measured across the full field under examination. We define $>3\sigma$ peaks as significant if these occur either at the SMG redshift (if known) or at $z>1.4$ (if unknown). We consider a number of angular scale griddings and so are sensitive to both over-densities with a small number of galaxies on small scales, and more populated structures on larger scales. This method also selects systems of varying number density and size as a function of redshift  - an identical number of clustered galaxies on a fixed scale becomes more significant as redshift increases. As such, while the structures identified in this work vary dramatically in absolute numbers of galaxies and scale, they are all significant over-densities for their redshift and physical size. 

We summarise the structures identified using this method in Table \ref{tab:summary} which compiles the number of clustered regions identified at each stage of our analysis. A more detailed discussion of individual sources, including estimated stellar mass of individual systems and comments on their counterparts in simulations, is given in Appendix A, while our main results consider the sample as a statistical whole.

\section{Two Dimensional Over-density Analysis}
\label{sec:results}

\subsection{Over-densities in the Projected Number Density}
\label{sec:COS_2d}\label{sec:ECDFS_2d}

\begin{figure*}
\begin{center}

\includegraphics[scale=0.85]{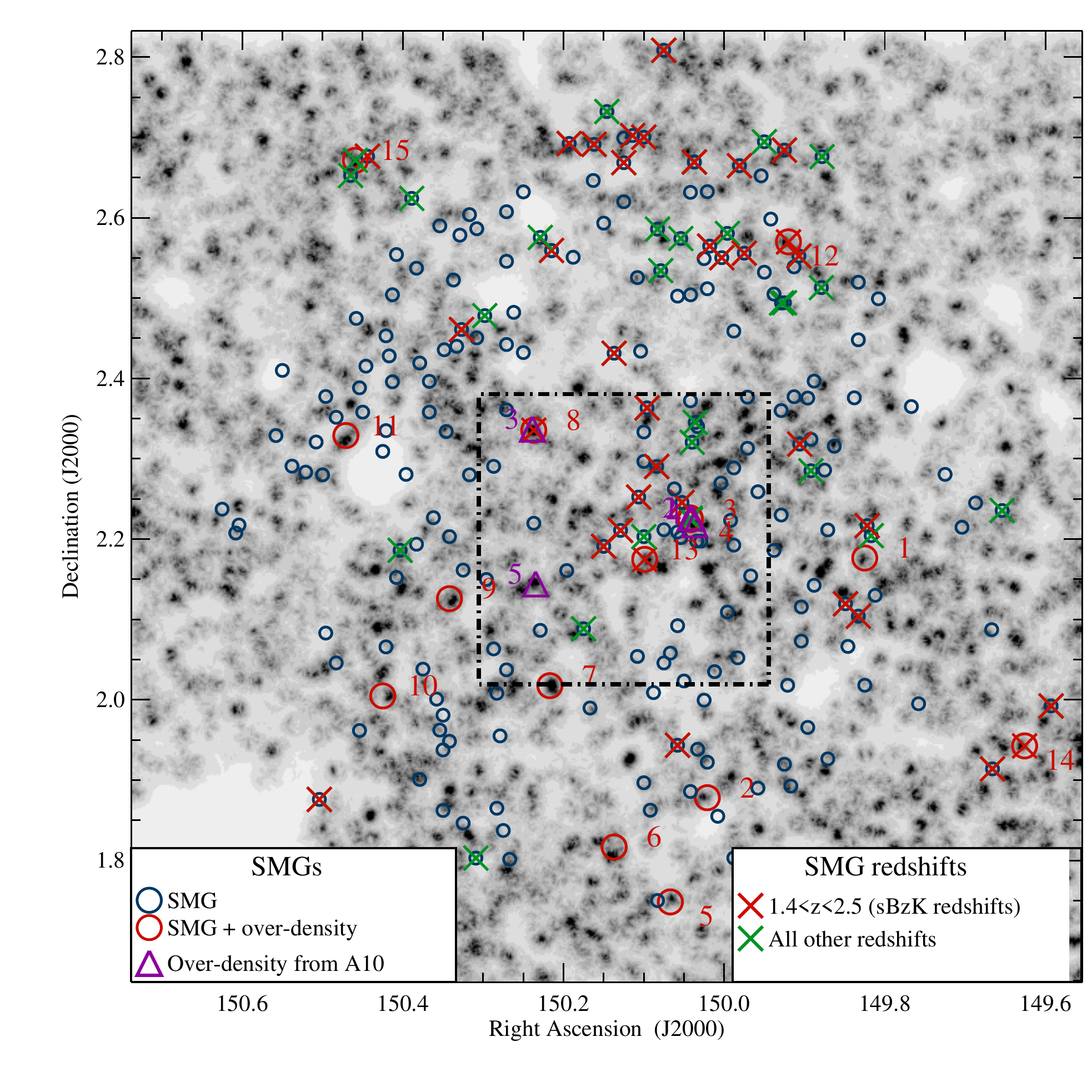}\\
 \hspace{13mm} \includegraphics[scale=0.75]{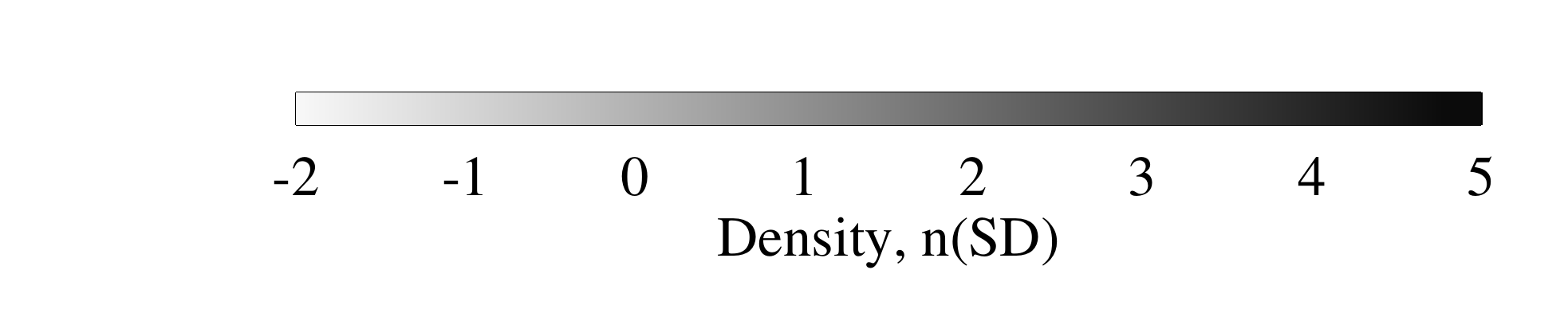}\\

\caption[]{The projected number density map of sources in the $sBzK$ in the COSMOS field. Dark colours represent regions of high projected number density. The circles represent the COSMOS-SMG sources in the field. The region covered by the MAMBO-2 map used by A10 is bounded by the dashed line. The red circles represent COSMOS-SMG sources which are within 30$^{\prime\prime}$ of a $>6 \times$ standard deviation over-density of star-forming galaxies (these are labeled with red numbers and referenced in Section \ref{sec:COS_2d}). The purple triangles are MAMBO-2 submm sources which are within 30$^{\prime\prime}$ of a $>6 \times$ standard deviation over-density of $BzK$ galaxies and are discussed in A10 (labeled with purple numbers and also referenced in Section \ref{sec:COS_2d}). Crosses denote SMGs with either a spectroscopic redshift or best fit photometric redshift taken from \citet{Smolcic12, Casey12a, Casey12b} and references therein. Red crosses highlight sources where the SMG falls within the redshift range of the colour selection for star-forming galaxies ($1.4<z<2.5$), while green crosses mark all other sources which have a known redshift.}   

\label{fig:num_dense}
\end{center} 
\end{figure*}

An example of the projected number density maps produced by the method described in section \ref{sec:analysis} is shown in Figure \ref{fig:num_dense}. Regions over-dense relative to the field are shown in dark greyscale while under-dense regions are relatively pale. The locations of SMGs are circled, and comparisons of the two distributions allow us to evaluate the correlation between these.

\subsubsection{COSMOS}

\textit{sBzKs:}\ As Figure \ref{fig:num_dense} shows, we identify a total of 15 regions of high projected number density ($>$6 times the mean) in  $sBzK$ galaxies in the COSMOS field that are within 30$^{\prime\prime}$ of the position of a sub-mm bright source. We recover three of the four over-dense regions discussed in the A10 work (which considered a similar redshift range in a subset of this field). These are shown on the figure as purple triangles 2, 3 and 4. A10's Source 5 is not identified in our COSMOS-SMG sample but is associated with an over-density of $sBzK$s and we add it to our list.  Outside of the subfield considered by A10, we locate 11 additional regions where an over-density of $sBzK$ selected galaxies are found within 30$^{\prime\prime}$ of the position of a sub-mm bright source. However, of these 15 regions (across the full field), just 4 are coincident with an SMG with a known redshift at $1.4<z<2.5$ (red circles with red crosses). 

\textit{pBzKs:}\ Similar maps (not shown here) are produced for each galaxy selection, in each field. Only one SMG is associated with an over-density of more passive galaxies $pBzK$ at $1.4<z<2.5$. This region is not associated with any of the over-densities identified in the $sBzK$ sample. Hence for $pBzK$ sources with photometric redshifts, we find no strong evidence for clustering associated with SMGs. However, only a small fraction of the sources colour-selected as $pBzKs$ have photometric redshifts and make our cuts. We shall investigate clustering of $pBzKs$ without photometric redshifts later.   

\textit{LBGs:}\ Moving to higher redshifts, we find find 13 (8) coincident positions between a submillimeter source and LBGs selected to lie at $z\sim3$ (4). None of these are coincident with an overdensity in the $sBzK$ ($z<2.5$) sample. However none of the SMGs in these regions have a known redshift, and - as in $sBzK$ cases without SMG redshifts - we must consider the possibility that the small projected separations represents chance alignments. We do so in section \ref{sec:chance}

\subsubsection{ECDF-S}

We repeat the above analysis for systems in the ECDF-S region. Due to the smaller field coverage (and therefore number of significant over-densities), we relax our criteria for correlations between submm bright sources and regions of high-projected number density, to include all positions with a $>4.5 \times$ the mean over-density of sources (previously $>6 \times$ the mean). 

We find just 6, 1, and 0 such correlations in our $sBzK$, $z\sim3$ LBGs and $z\sim4$ LBGs sample respectively.  As the ECDF-S covers 1/3 of the area of the COSMOS field discussed in this work, the number density of 2D correlations is roughly consistent between both fields. Of the 6 regions identified in the $sBzK$ sample, 4 have SMGs with a redshift consistent with the star-forming galaxies. This matches the number in the COSMOS field. However, as noted previously, the ECDF-S contains much higher number density of submm bright sources than COSMOS (136 compared to 234 while only covering $\sim1/3$ of the area, to the same depth). This is even more apparent when only considering SMGs with either spectroscopic or photometric redshifts at $1.4<z<2.5$ (the range covered by the $sBzK$ selection), where the ECDF-S contains 53 SMGs in comparison to just 37 in COSMOS. This suggests that the number of 2D correlations in the ECDF-S is less than might be expected, if the results in the COSMOS field are representative (see Section  \ref{sec:comparisons}), even before the probability of chance alignments (which is obviously higher for ECDF-S) is taken into account.

\subsection{Likelihood of Chance Alignment}
\label{sec:chance}

\begin{figure*}
\begin{center}

\includegraphics[scale=0.27]{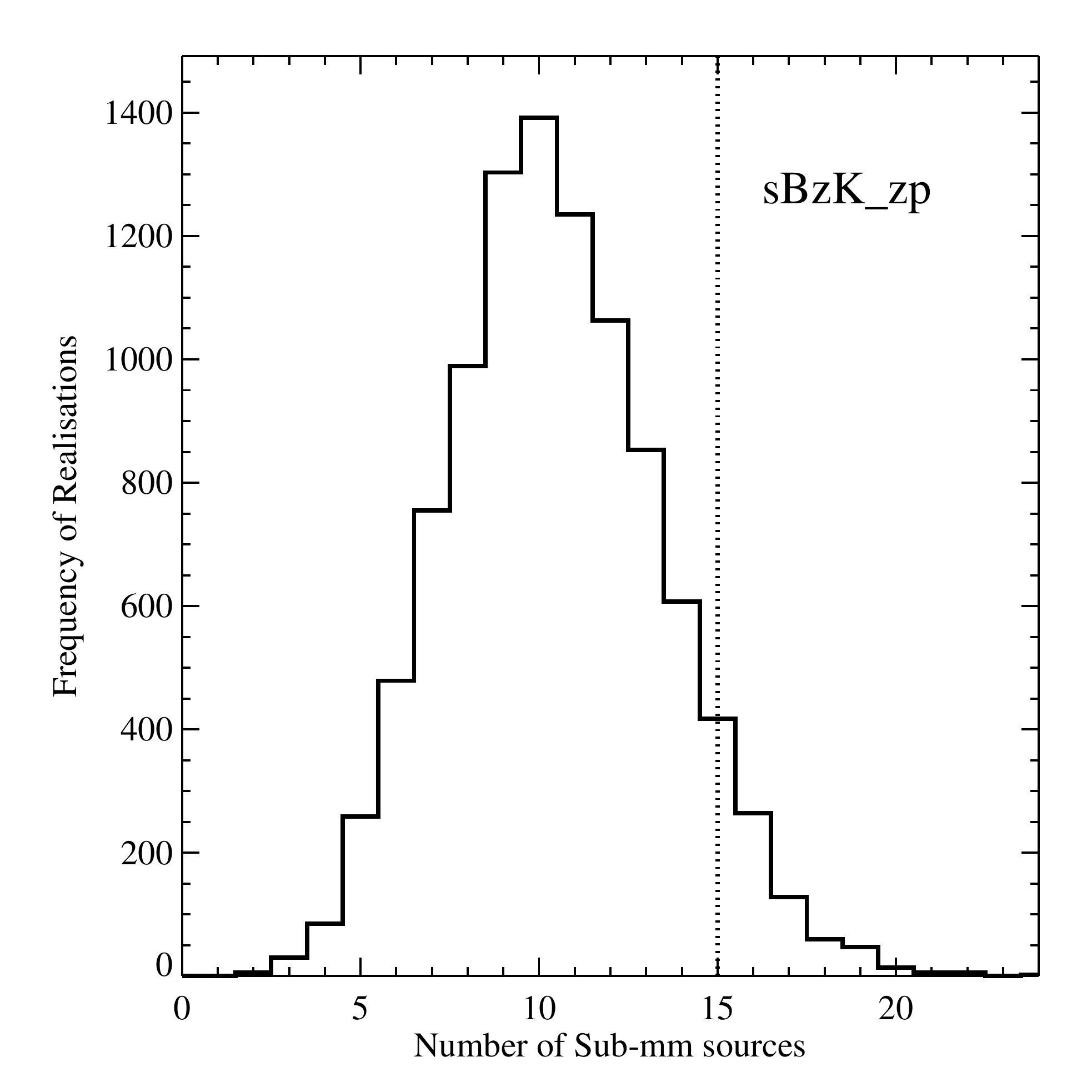}
\includegraphics[scale=0.27]{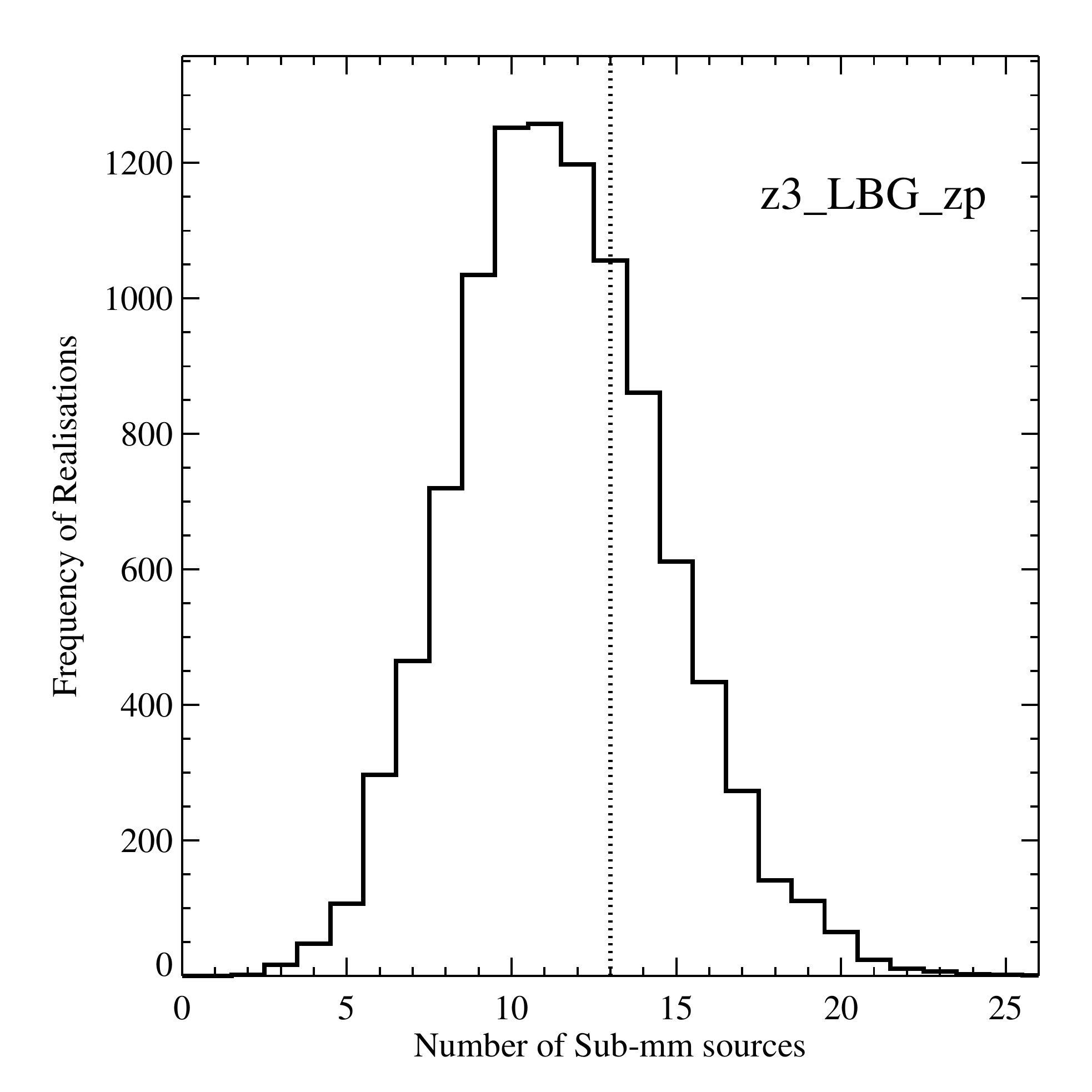}
\includegraphics[scale=0.27]{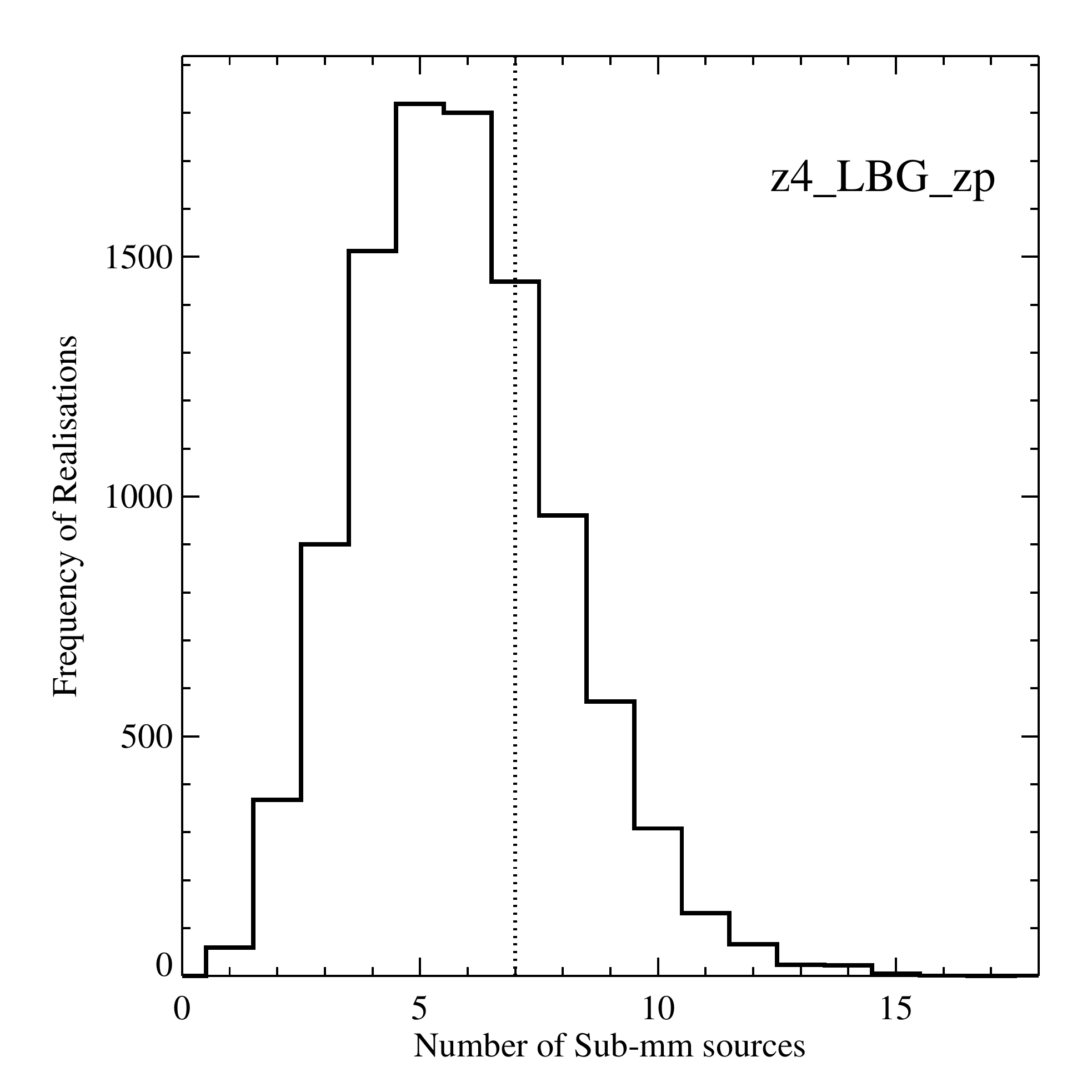}

\caption[]{The distribution of the number of sub-mm sources with a chance association with a random distribution of star-forming galaxies for the $sBzK$ (left), $z\sim3$ (middle) and $z\sim4$ (right) samples. We produce 10,000 realisations of a random distribution of points of the same number as the sources in each of our samples. We then identify the number of sub-mm sources that have a $>6 \times$ the standard deviation over-density within 30$^{\prime\prime}$ in each sample. The distribution of the number of sub-mm sources with a chance $>6 \times$ the standard deviation association is plotted as the solid histogram. The dashed line represents the number of sub-mm sources with a $>6 \times$ the standard deviation association in our data sample. Clearly, in all samples the number of sub-mm sources with an association is consistent with a random distribution of points. }    

\label{fig:chance}
\end{center} 
\end{figure*}

In order to estimate the incidence of chance alignments between SMGs and star forming galaxy over-densities, we randomly reposition all SMGs in each field and repeat our analysis. We note that produces a crude estimate since the real galaxy population is not drawn from a random distribution, but is spatially correlated, even when projected on the sky. Furthermore, this method does not account for the effect of over-densities of galaxies on background SMGs (which are at a much higher redshift). Over-densities in the mass distribution of the cosmic web may cause lensing of background SMGs, increasing their observed flux and allowing them to be identified in submm surveys. As such, any lensing will cause background SMGs in the line of sight of galaxy over-densities to be preferentially identified and enhance any 2D clustering signal. While this effect is likely to be small, it can be removed by only considering SMGs with known redshifts. In addition, the above analysis does not take into account that only a fraction of the SMGs are likely to fall at the same redshift as the star-forming galaxies in each of our redshift samples - the SMGs themselves will have a broad range of redshifts. Including sources which are not at the same redshift as the star-forming galaxies will wash out any true 2D clustering signal.  

Nonetheless, producing 10,000 Monte Carlo realisations of this randomisation process allows some indication of the probability of chance alignment to be assessed. We undertake the analysis in the COSMOS data (the larger of our two fields). Figure \ref{fig:chance} presents the incidence of chance alignments of an SMG within 30$^{\prime\prime}$ of an over-density in the COSMOS field. We find that in each sample the number of high density regions which are coincident with an SMG in our COSMOS sample is consistent with a random distribution. However, the number of correlations in our $sBzK$ sample is somewhat above average for the random distribution ($\sim2\sigma$ from the mean predicted number, assuming a Gaussian distribution). This suggests that there is a weak correlation between SMGs and $sBzK$ star-forming galaxies when considering a 2D analysis alone (with no redshift information on the SMG), although the caveats mentioned above hold for any 2D analysis. As already noted, the ECDF-S region is both sparser in observed SMG-overdensity coincidences, and has a higher probability of chance alignments - thus it seems unlikely that SMGs are significantly correlated with $sBzK$ density in this field.

\subsection{Affects of changes to our method: choice of $n$ and colour selections without $z_{\mathrm{photo}}$}
\label{sec:vary}

The above results suggest that at best there is a very weak correlation between star-forming galaxy density and SMGs when only the crudest redshift information (i.e. $z_{\mathrm{photo}}>1.4$) is taken into account. However, it is interesting to consider how this is affected by our method. How does varying our choice of the density scale parameter $n$ affect the number and distribution of correlations we find? And does removing $z_{\mathrm{photo}}$ cuts in our selection improve our clustering signal? If our 2D approach, and that used by A10, is strongly sensitive to our method, then we can not draw reliable conclusions from the analysis above. 

To investigate this, we vary our choice of $n$ by $\pm5$. We find that varying $n$ has a reasonably significant affect on the number of potential clustered regions we find. For example, in our COSMOS $sBzK$ sample, fluctuating our choice of $n$ by just $\pm1$ varies the number of regions identified by $^{+2}_{-3}$, with the lower $n$ value producing a higher number of correlations. This is unsurprising, as at lower values of $n$ fewer individual galaxies are required for a region to be classified as over-dense. This result is true for our samples at all redshifts and in both COSMOS and the ECDF-S. Hence, any statistical correlations derived from our 2D analysis are largely dependant on our choice of $n$ and as such, can not be deemed reliable. However, while this affects our number statistics, it does not rule out diminish the importance of $individual$ clustered regions identified using this method, which may still be significantly over-dense structures by any definition. 

We find a different but equally significant uncertainty is generated when considering the effects of omitting a photometric redshift cut. In both ECDFS and COSMOS we find that the same selection method identifies different regions than in samples with photometric redshift cuts applied. This is compounded by changes in the number of regions selected due to varying our choice of $n$ in this sample. 

Hence both the number and distribution of clustered regions is dependant on details of our method. As a result we conclude that a 2D analysis is a valid method primarily for identifying regions with which to follow up with a 3D investigation, rather than to generate statstical samples.

\begin{figure}
 \begin{center}

  \includegraphics[scale=0.34]{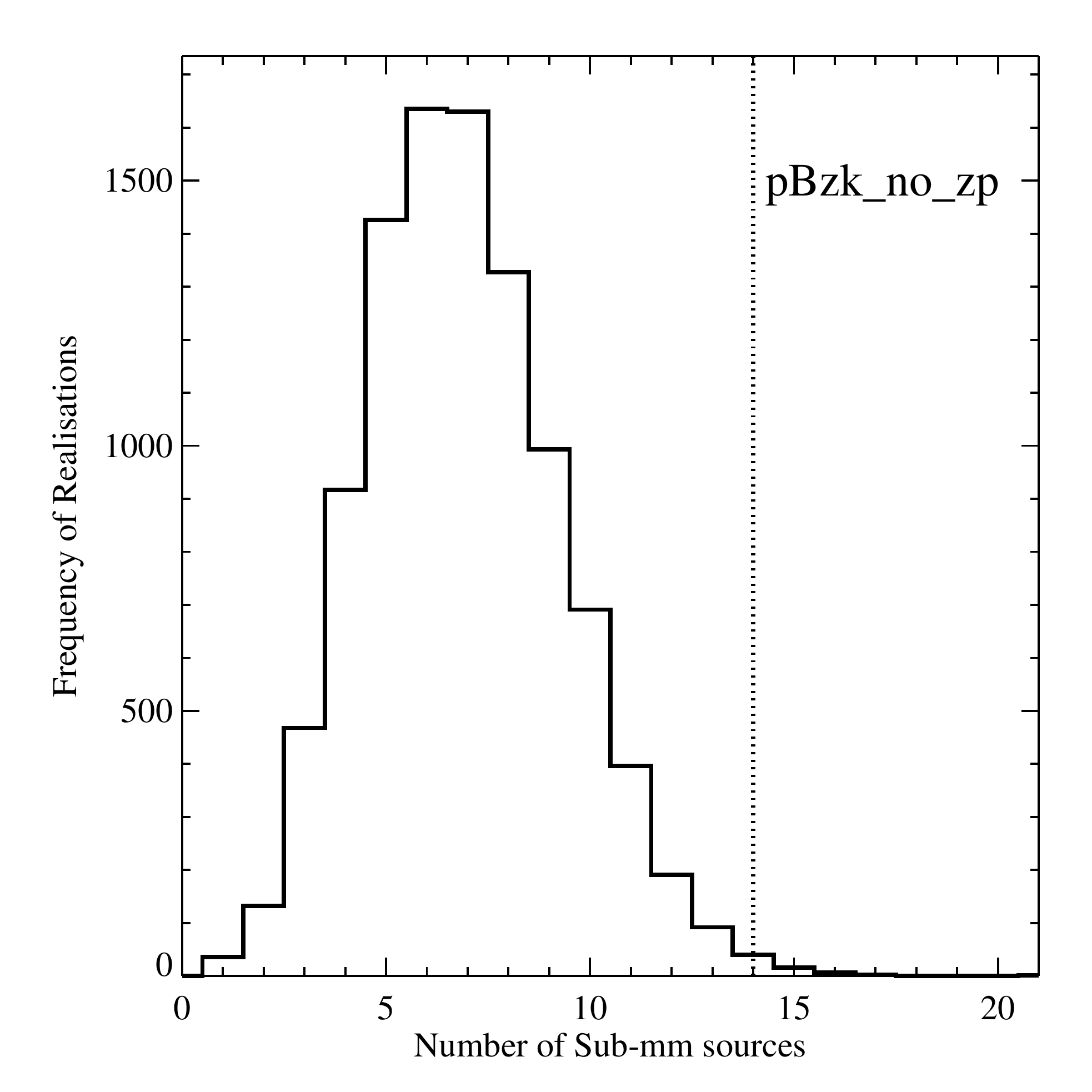}

  \caption[]{Same as Figure \ref{fig:chance} but for the sample of $pBzk$ galaxies in COSMOS, with no photometric redshift cut applied to the sample.}    

  \label{fig:pbzk_asso}
 \end{center} 
\end{figure}

However, we do note one interesting corrollary that arises from this analysis. The relatively smooth spectral energy distributions of passive $pBzK$ selected sources mean that only a small fraction of those selected based on colour have accurate photometric redshifts - only $\sim10\%$ of $pBzK$ sources in the COSMOS field have a known photometric redshift, compared to $\sim35\%$ of $sBzK$ sources in the MUSYC sample. When the cut on photometric redshift quality is removed, our estimate of the clustering strength of passive $BzK$ sources dramatically changes. Using $n$=3, as in the analysis discussed above, we would now find 14 spatial coincidences between over-densities of $pBzK$ sources and an SMG in COSMOS. This is significantly higher than the mean expected from a random distribution ($\sim6$, Figure \ref{fig:pbzk_asso}), and suggests that $pBzK$ systems are potentially the most effect method for identifying clustering at these redshifts in the absence of detailed 3D redshift information. However, it is difficult confirm any of these clustered structures without detailed spectroscopic follow up.

Comparing the potential $pBzK$ over-densities to those identified using actively star-forming galaxies at the same redshift (our $sBzK$ sample), we find that only one structure is identified in both samples. As such, using both passive and actively star-forming galaxies identifies different systems at $z\sim2$. This strengthens the motivation for searching for actively star-forming clusters at high redshift, as we are likely to identify different structures to those found using passive galaxies alone.   

\vspace{2mm}

\subsection{Summary}

In summary, while a 2D analysis identifies potential weak clustering of high redshift sources, the individual regions which are identified and their number density is extremely sensitive to the method used (both in choice of $n$ and in the use of photometric redshifts). As a result, we can not perform a statistical analysis on the number and distribution of sources identified in 2D, and a full 3D approach (utilising the maximum possible redshift information in addition to projected alignment) is required. Despite this we do see a potential weak 2D correlation between SMGs and more typical galaxies at $z\sim2$ in the COSMOS field. This is most evident in the passive $BzK$ sample when a photometric redshift cut is omitted from the procedure. However, we can not currently confirm any of the structures identified through $pBzK$s in 3D, as only a small fraction of $pBzK$ sources have a photometric redshift. The regions identified using the $pBzK$ sample are different to those found using $sBzK$s and as such, this validates an approach that uses star-forming galaxies for cluster searches - as by using passive galaxies we may be missing a large fraction of high redshift galaxy clusters.

\section{Three Dimensional Over-density Analysis}
\label{sec:3d_COSMOS}

\subsection{SMGs with known redshifts}
            
\subsubsection{SMGs identified with two-dimensional over-densities}
\label{sec:2d_in_3d}\label{sec:ECDFS_3d}

Four SMGs with known redshifts were found to be spatially coincident with a 2D over-density of galaxies in the COSMOS field, and a further four in the ECDF-S (see Section \ref{sec:COS_2d}).

Now exploiting the full redshift information available for each source, we probe clustering on a number of scales by identifying $>3\sigma$ excesses in the redshift distribution, relative to that over the entire field in the relevant survey. We also apply the constraint that there must be at least three sources in a $\Delta z=0.1$ peak for it to be selected (removing shot noise of single sources at high redshifts and on small scales). For spectroscopically confirmed sources we only identify peaks which are within $\Delta z = \pm0.1$ of the SMG redshift, while for SMGs with photometric redshifts we select over-densities within a window defined by combining the photometric uncertainty on the SMG with the typical, $\Delta z_{\mathrm{photo}}=0.1$, uncertainty on the star-forming galaxies. 

We find two of the four SMGs with known redshifts in COSMOS, and just one of the four in the ECDF-S (that of LESSJ033336.8-274401), show an over-density of galaxies in redshift as well as projected space in their surrounding volume (see the appendix and Figure \ref{fig:a_10_2d} for more details). 

As such, only $\sim40\%$ of regions identified in our 2D analysis (with the caveat of effects due to our choice of $n$ and galaxy selection method) are confirmed in a fuller analysis. Once a three dimensional over-density has been identified we refine the size of the surrounding region for which we measure the photometric redshift distribution, rather than allowing this to vary, but restrict this region to a square to avoid complex clustering selection (as shown by the dashed region in the bottom panels of Figure \ref{fig:a_10_2d}).

\subsubsection{SMGs with over-densities identified only in three dimensions}
\label{sec:2d_in_3d_2}

In addition to the structures identified above, we also consider the volume surrounding all SMGs in the COSMOS field with known redshifts (irrespective of the previous determination of a 2D galaxy over-density). We find a further five regions in COSMOS where an SMG is coincident with a redshift peak (Figure \ref{fig:spec_COS}), and three such regions in the ECDF-S. 

The over-densities we identify around SMGs with known redshifts in COSMOS (both in this and the previous subsection) constitute either a small number of sources over a small volume producing a highly significant over-density (as in the case of AzTEC/C8 and  1HERMESX24J100023.75+021029) or less significant over-densities on larger scales (as in 1HERMESX24J100024.00+024201 and 1HERMESX24J100038.88+024129). The former of these are likely to represent individual massive galaxies in formation, while the latter are potential forming clusters and groups. The most intriguing of these regions is 1HERMESX24J100046.31+024132 which contains a significant over-density of 10 sources spanning $\Delta z=0.2$ at $z\sim1.9$ over a $0.5\times0.5$\,Mpc region. We note that the region surrounding AzTEC/C6 is the same as that identified around COSBO-3 in the previous work of A10. COSBO-3 is the only SMG in an over-density from A10 which has a known redshift at $z>1.4$ and is also detected in the AzTEC maps (and so is used in our analysis). Hence, this detection is in agreement with the A10 results and displays consistency between our methods. Properties of all of the over-dense regions selected in the paper are given in Table A1 of the Appendix.

This number density of SMG + over-densities in ECDF-S is consistent with that which is found in the COSMOS field. The ECDF-S contains around half the number of clustered regions showing a three dimensional excess around SMGs with known redshifts (4 compared to 7), while also containing roughly half the number of SMGs. We find that two of the regions in the ECDF-S are exceptionally interesting, those surrounding LESSJ033136.9-275456 and LESSJ033336.8-274401, the former of the two is discussed further in Section \ref{sec:indu} of the Appendix, while the latter is discussed in Section \ref{sec:LESS24}. The other two regions contain SMGs with poorly constrained redshifts, and only moderate over-densities of sources. Hence, we consider these less secure identifications of early clusters.

In total, only $\sim30\%$ (3/11) of over-densities identified around SMGs with known redshifts, using full three dimensional information, were selected in our 2D analysis. As such, in addition to being dependant on method, the 2D analysis misses a significant fraction of clustered regions which can be selected using detailed redshift data.

\subsubsection{Population-averaged 3D results}

In Section \ref{sec:COS_2d} we identified a potential weak 2D spatial correlation between SMGs and star-forming galaxies, however it would appear that only a small fraction of SMGs show clear ($>3\sigma$) 3D clustering in their surrounding volume. This suggests that any clustering associated with SMGs may be weak, and may be missed when investigating individual sources. To investigate this, we consider all 63 SMGs in COSMOS (the larger of our target) with known redshifts and assess the statistical significance (which varies with redshift) of the number of lower mass sources in their surrounding volume.     

\begin{figure}
\begin{center}

\includegraphics[scale=0.4]{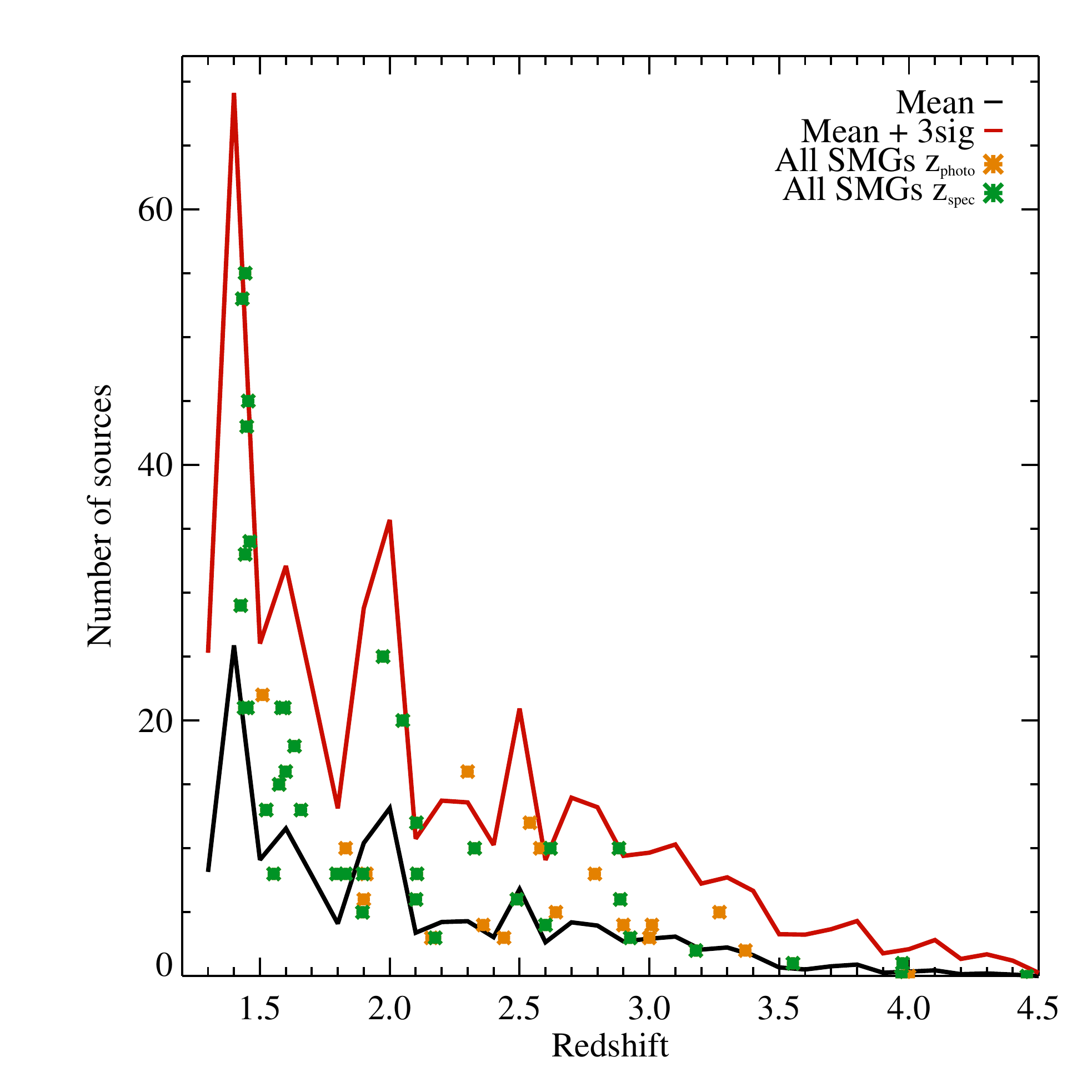}

\caption{The predicted number of COSMOS galaxies in a $4^{\prime} \times 4^{\prime} \times \Delta z=0.1$  volume, as a function of redshift, in comparison to a similarly sized volume in regions surrounding SMGs with known redshifts. The black line displays the mean predicted number of sources, while the red line shows the $3\sigma$ excess beyond which an overdensity might be considered significant as a function of redshift. Stars show the number of sources surrounding individual SMGs, similarly divided by redshift. }

\label{fig:detect_num}
\end{center} 
\end{figure}

Figure \ref{fig:detect_num} shows the number of sources predicted from random regions of 4 arcmin diameter, as a function of redshift (black and red lines, $\Delta z=0.1$ binning). We over-plot the number of galaxies in the volume surrounding COSMOS SMGs with known redshifts (on the same scale). For SMGs with only photometric redshifts (10 of the 63) we take the volume surrounding the best fit photometric redshift. A submillimeter galaxy with a significant excess for its redshift would be expected to lie more than three standard deviations above the mean of the random fields. We find that only one SMG is significantly over-dense for its redshift on 4 arcmin scales - the SMG at $z\sim2.3$ (this is in fact AZTEC 7 identified in Section \ref{sec:2d_in_3d}). This suggests that there is no strong clustering around the bulk of SMGs on these scales (as we showed found in Sections \ref{sec:2d_in_3d} and \ref{sec:2d_in_3d_2}). 

However, Figure \ref{fig:detect_num} also indicates that the majority ($81\%$) of the SMGs have a galaxy density in their surrounding volumes that exceeds the mean of the random distribution.
This echoes the results found in our two-dimensional analysis - that SMGs and typical galaxy over-densities correlate, but only weakly. As such, it is likely that SMGs reside in over-dense regions relative to the field, but generally not sufficiently so to be individually detected as clustered regions. This result is consistent with that found for the surrounding environments of other massive galaxies (QSO hosts) at much higher redshifts \citep{Husband13}.  

We repeat this analysis on 0.25, 0.5, 1\,arcmin scales but do not see a clear signal - potentially due to low number statistics - in small projected regions, very few star-forming galaxies occupy each redshift bin.

\subsection{SMGs with unknown redshifts}
\label{sec:no_red}

In addition to the SMGs with known redshifts, there are also 171 SMGs in the COSMOS region, and 53 in the ECDF-S, which have neither a spectroscopic nor photometric redshift. While we cannot confirm that these SMGs and any redshift spike in their vicinity are coincident in three dimensions, it is interesting to consider the fraction of SMG fields which contain a spike in their line of sight, relative to blank sky fields. Are redshift over-densities of star-forming galaxies more frequently identified in sight-lines which contain an SMG than those without? In addition, could a line of sight correlation between an SMG and over-density be used to constrain the SMG redshift where no other diagnostic is available? 

We find 20 regions in COSMOS with a $>3\sigma$ galaxy excess in a $\Delta z=0.1$ bin at $1.4<z<4.5$, coincident with an SMG,  and a single additional region in the ECDF-S. (examples are shown in Figure \ref{fig:pos_COS} - for full list of regions identified see Table A1). 

\begin{figure}
\begin{center}

\includegraphics[scale=0.38]{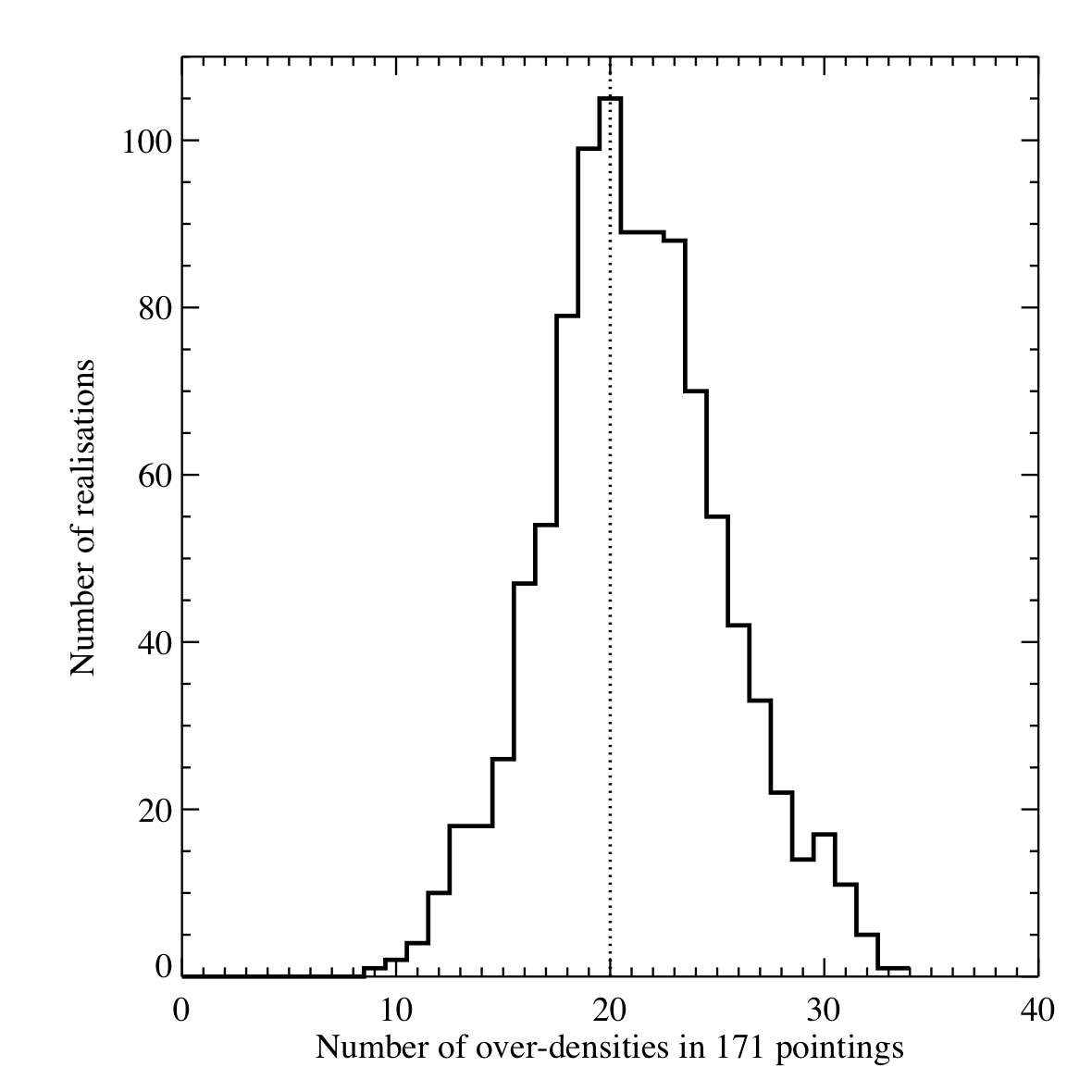}

\caption{The distribution of the number of pointings that contain a $>3\sigma$ over-density of galaxies at $1.4<z<4.5$ when taking 1000 realisations of 171 pointings (see text for details). The number of SMG fields which contains an over-density is plotted as the dashed vertical line. Clearly this is consistent with a random distribution. }

\label{fig:rand_red}
\end{center} 
\end{figure}

As we have no redshift information about the SMGs in these regions, it is possible that any number of these structures may be chance superpositions of sources. In order to test this further, we take 1000 realisations of 171 random pointing in the COSMOS region and identify the fraction of these pointings which contain a $>3\sigma$ over-density in a $\Delta z=0.1$ bin at $1.4<z<4.5$ (Figure \ref{fig:rand_red}). We only select random pointings which are centred at least $2^{\prime}$ away from an SMG, to exclude reselecting over-densities associated with the SMGs.  Figure \ref{fig:rand_red} clearly shows that the number of correlations around SMGs with unknown redshifts is completely consistent with a random distribution. Similarly, the equivalent analysis on the ECDF-S predicts two such chance alignments (where we found one).

We conclude that all of these regions are potentially chance projections of a galaxy over-density with an SMG. Without further spectroscopic information or SMG redshifts, we can not confirm any of the potential cluster candidates identified here, and also cannot use line of sight correlations between SMG and galaxy over-densities to constrain the SMG redshifts, as they are highly likely to be chance superpositions.

The potential lack of line of sight correlations between SMGs and galaxy over-densities is surprising. Given that 7 of 63 SMGs with redshifts in COSMOS display an over-density, we may expect $\sim20$ COSMOS SMGs without redshifts to be associated with a true over-density. While this is identical to the number of line of sight matches we find, the bulk of these are likely to be chance superpositions and not truly clustered regions. As such, the number of true line of sight associations is lower than that predicted from the SMGs with known redshifts. 

Of course, sources with either spectroscopic or photometric redshifts are likely to have either strong emission line features \citep[$e.g.$][]{Casey12a, Casey12b} or bright rest-frame UV-optical fluxes \citep[$e.g.$][]{Smolcic12}. Therefore, they are likely to form a subsample of the most UV-optically bright SMGs and as such, may display different clustering statistics to the UV-optically faint SMGs without redshifts. Our results may suggest that the SMGs with redshifts and those without may not form a single population. It may also suggest that background SMGs are not significantly lensed by clustered regions at $z\sim2$. If background SMGs were lensed, we may expect sight-lines containing an SMG to preferentially have an over-density of star-forming galaxies (in fact, it would be the over-density that preferentially causes the detection of an SMG through lensing, but the effect should be apparent when targeting either SMGs or over-densities). As we find no excess of over-densities in SMG sight-lines in comparison to random fields, it is likely that any lensing effect from $z>1.4$ is minimal.

\subsection{Regions of high projected number density and no SMG} 
\label{sec:no_SMG} 

We have identified a certain degree of clustering around SMGs, but it is also interesting to investigate whether similar structures exist which are not associated with an SMG but can be targeted simply by galaxy number density. These regions will potentially contain less stellar material (the SMGs are likely to dominate the stellar budget in their surrounding volume) but may be equally likely to form a massive cluster at low redshift. The assumption of lower mass is not clear cut, however: the relatively short SMG duty cycle \citep[$<500$\,Myr,][]{Hayward11} means that such structures may potentially be associated with an SMG-mass galaxy, detectable either in the past or future.

To investigate this further we repeat the process outlined in Section \ref{sec:2d_in_3d}, but target the regions surrounding peaks in the projected number density which are not associated with an SMG (from Figure \ref{fig:num_dense}). We once again identify $>3\sigma$ peaks in the photometric redshift distribution between $1.4<z<4.5$ on a number of angular scales. We find 32 regions in COSMOS and 3 in the ECDF-S which meet our selection criteria. Figure \ref{fig:pos_COS2} displays a selection of these regions (further discussion of individual regions can be found in Section \ref{sec:indu} of the Appendix, while properties of the full sample can be found in Table A1). The ECDFS BzK\_1 region is particularly interesting as it contains a Lyman-alpha blob \cite[LAB, $e.g.$][]{Fynbo99} falling at the exact same redshift \citep{Yang10} as the over-density of galaxies (for further discussion see Section \ref{sec:ly_blob}). However the structures identified using this method show a wide variety of properties from large over-densities at moderate redshift containing a relatively large number of sources (such as COSMOS BzK$\_$6) to highly significant over-densities of a small number of sources in a small volume at high redshift (such as COSMOS z3$\_$13/14). Once again, the former of these are likely to form a massive structure at low redshift, while the latter, given the timescales involved, are likely to represent massive galaxies in formation.

\subsection{Summary of three dimensional over-densities}
\label{sec:sum}

In total, we identify 59 regions clustered both spatially and in redshift in the COSMOS field at $1.4<z<4.5$. Of these, 27 are spatially coincident with an SMG (7 with redshifts), while the other 32 are simply identified as 3D over-densities in the star-forming galaxy population. We identify 8 such regions in the ECDF-S, 5 spatially coincident with SMGs and 3 not associated with SMGs. For a breakdown of the numbers of over-densities identified using each method see summary Table \ref{tab:summary}.

The number of lines of sight which contain both a redshift spike and an SMG is slightly lower than those which just display a redshift spike (even assuming that the SMG and galaxy over-density are associated where there is no redshift information about the SMG - hence an upper limit, see Section \ref{sec:no_red}). This suggests that SMGs are not necessarily the most reliable tracers of distant structure. Potentially we are equally likely to identify clustering by targeting regions of high projected number density of star-forming galaxies selected using colour methods (without any detailed redshift information). However, these numbers place no weight on the strength of the over-densities found using each tracer. However,  we note that the two most interesting over-dense regions (those around LESSJ033136.9-275456 and LESSJ033336.8-274401 in the ECDFS), potentially the most extreme clusters/proto-clusters in our sample, are both associated with an SMG in 3D. This implies that while the majority of SMGs are found in typical to moderately over-dense environments, a small number of SMGs are found in the most over-dense structures. The region surrounding LESSJ033336.8-274401 is the most extreme example in our study and this is discussed further in Section \ref{sec:LESS24}. However a second LESS region, LESSJ033136.9-275456, is the second strongest potential protocluster and thus itself of significant interest. It is discussed in detail in the Appendix. 

If it is assumed that the SMGs and over-densities discussed are correlated in all cases with spatial coincidence, we find that at best only $\sim12\%$ of SMGs in the field show $any$ evidence, even loosely, of potential 3D clustering. As a significant fraction of these are likely to be chance superpositions of sources, this represents an upper limit to the true number of clustered regions around SMGs. 

Considering the entire volume probed by this work (at $1.4<z<4.5$) we find that the number density of volumes with evidence for 3D clustering is a few $\times 10^{-7}$\,Mpc$^{-3}$. This is comparable to the number density of the most massive ($>10^{14.5}$\,M$_{\odot}$) galaxy clusters at $z=0$ \citep[$e.g.$][]{Wen10}. While this does not necessarily mean that we have identified all regions which will eventually form a massive cluster at $z=0$ or that all of the structures identified in this work are massive clusters in formation (this has been shown to be extremely unlikely as potentially all of the regions identified without SMG redshifts are likely to be chance superpositions). It is interesting to note that we do not find significantly higher number densities than those of $>10^{14.5}$\,M$_{\odot}$ clusters at low redshift - if we had, it would be unlikely that these regions represent the progenitors of massive clusters. Simply considering the volume over which the majority of clustered regions surrounding SMGs are found ($1.4<z<2.6$ - assuming we are less sensitive to over-dense regions at $z>2.6$) we find a number density of  $\sim10^{-6}$\,Mpc$^{-3}$ - comparable to that of $\sim10^{14.3}$\,M$_{\odot}$ clusters at $z=0$ and once again consistent with the most massive structures.

\subsection{Comparisons between COSMOS and ECDF-S}
\label{sec:comparisons}

In our 2D analysis, we find 5 times as many projected galaxy over-densities around SMGs in COSMOS than in the ECDF-S, while the ECDF-S covers a third of the COSMOS area (see Table \ref{tab:summary} for details). As shown previously, the ECDF-S region contains a higher number density of SMGs, even when considering sources to a similar flux limit (Table \ref{tab:submm_obs}). Hence, the discrepancy between the number of 2D clustered regions is even more problematic (with a higher number density of SMGs, we would expect more correlations). Both fields have a similar number density of colour-selected star-forming galaxies satisfying our selection at $z_{\mathrm{photo}}>1.4$ and hence it is unlikely that the fidelity of the photometric redshifts of star-forming galaxies is a significant factor in this discrepancy. In combination with the significant effects of varying our method, this discrepency in the number of 2D over-densities identified continues to suggest that a 2D analysis alone is not a reliable method for identifying high redshift clustering.       

Considering the number of 3D over-densities around SMGs with known redshifts (7 in COSMOS and 4 in ECDF-S), we find these to be much more consistent with the number density of SMGs. COSMOS contains roughly twice as many SMGs, when considering the surveys to the same depth, and roughly twice as many spectroscopically confirmed SMGs. Therefore, the number of 3D over-densities identified around spectroscopically confirmed SMGs is consistent with the total number density of SMGs.

\section{Identification with previously known clusters} 

\subsection{Regions in COSMOS}

Three regions in our COSMOS sample have previously been identified in the H-band selected HIROCS survey \citep{Zatloukal07}. The region surrounding 1HERMESX24J100023.75+021029 is at an identical position and estimated redshift to HIROCS 100023.9+024158,  COSMOS BzK$\_$6 is identical to HIROCS 095954.2+022924 and COSMOS BzK$\_$9 to HIROCS 095957.9+024125. As mentioned previously, the region surrounding AzTEC/6 (COSBO-3) was also identified in A10. The re-identification of the HIROCS clusters using a completely different cluster finding technique once again validates our method.    
  
Despite these reconfirmations it is nonetheless interesting to investigate why other known high redshift clusters in the field were not selected by our cluster finding method. We select all $z>1.3$ (clusters at $z>1.4$ should be identified in our sample, however the typical photometric redshift error is $\sim0.1$) known clusters in the field from the literature, all taken from the HIROCS survey, and all at $z<2.0$. Of the ten $z>1.3$ clusters, only six are coincident with high projected number density region of $sBzK$s in our maps. We investigate the photometric redshift distribution around these regions on a number of angular scales and find that only three sources (those already identified in our sample and discussed above) display a significant ($>3\sigma$) redshift over-density on any scale. Therefore, the other systems are unlikely to be identified as actively star-forming clusters. The HIROCS clusters are primarily selected as over-densities of NIR-bright sources and as such, they target older, more passively evolving systems than the method described in this work. However, it is interesting to note that the HIROCS clusters not identified through star-forming galaxies are also not coincident with with a 2D over-density of $pBzK$ sources in our analysis. This suggests that these HIROCS clusters would not be identified by a purely 2D analysis alone, even when considering passive galaxies.         

We note that HIROCS 095954.2+022924 and HIROCS 095957.9+024125 (those identified here as COSMOS BzK$\_$6 and COSMOS BzK$\_$9) are actually among the weakest over-densities identified in the HIROCS survey. However, in our current analysis we identify almost twice as many potential cluster sources as those discussed in \cite{Zatloukal07}. Hence, previous work may be missing a significant fraction of the cluster population. We also note that the region surrounding 1HERMESX24J100023.75+021029 (HIROCS 100023.9+024158) was rejected from the final catalogues of \cite{Zatloukal07} as they found no clear redshift spike in their data, and their colour-magnitude plots suggested it was a chance projection of two structures. The re-confirmation of this region using our analysis, and the spectroscopic confirmation of an SMG at a similar redshift to the over-density spike, suggest that this is in fact a true forming cluster at high redshift and should not have been dismissed.

\subsection{Regions in the ECDF-S}

We compare the positions of our clustered regions in the ECDF-S with those of known clusters in the region. There are just two known high redshift clusters in the field;  [TCF2007]\,3, an over-density of sources at $z\sim1.55$ identified in the photometric redshift distribution of the field \citep{Trevese07} and CL\,J0332-2742, a spectroscopically confirmed forming cluster at $z\sim1.6$ \citep{Casellano07,Kirk09}. Both of these clusters correspond to regions of high projected number densities of $sBzK$s in our 2D maps. While [TCF2007]\,3 displays some evidence of a peak in its photometric redshift distribution, it is not significant enough ($ >3\sigma$) to be selected using the analysis outlined in this paper. However, the region around CL\,J0332-2742 is re-identified it as a $>4\sigma$ over-density in our analysis.  [TCF2007]\,3 was selected through its passively evolving galaxy population and, if it contains relatively few star-forming galaxies, it is unsurprising that it remains undetected in our sample.

\begin{table*}
\begin{center}
\begin{tabular}{c c c c c c c c }
\hline
\hline
Field & Total 2D OD & 2D OD$_{\mathrm{SMG-}z}$&  2D OD$_{\mathrm{Conf}}$ &Total 3D OD & SMG$_{z}$ OD & SMG$_{\mathrm{no-}z}$ OD & No SMG OD \\
(1) & (2) & (3) & (4) & (5) & (6) & (7) & (8)  \\
\hline

COSMOS & 35 (\S \ref{sec:COS_2d})  & 4 (\S \ref{sec:COS_2d})    & 2 (\S \ref{sec:2d_in_3d})  &  59 (\S \ref{sec:sum})    & 7 (\S \ref{sec:2d_in_3d} - \ref{sec:2d_in_3d_2}) & 20 (\S \ref{sec:no_red})  & 32 (\S \ref{sec:no_SMG}) \\
ECDFS  & 7 (\S \ref{sec:ECDFS_2d}) & 4 (\S \ref{sec:ECDFS_2d})  & 1 (\S \ref{sec:ECDFS_3d})  & 8 (\S \ref{sec:ECDFS_3d}) & 4 (\S \ref{sec:ECDFS_3d})                        & 1 (\S \ref{sec:no_red}) & 3 (\S \ref{sec:no_SMG}) \\

\hline
 Total & 42 & 8 & 3 & 67 & 11 & 21 & 35 \\
\hline

\end{tabular}

\caption{Summary of over-densities identified in both the COSMOS field and ECDF-S. For reference, section numbers of where the number is first derived are given in brackets next to the value. (1) Field name, (2) The total number of 2D over-densities where an SMG is spatially coincident with a region of high projected number density, (3) 2D over-densities which are associated with an SMG at a redshift consistent with the star-forming galaxies, (4) 2D over-densities which are confirmed in 3D, (5) total number of 3D over-densities found, (6) 3D over-densities coincident with an SMG both spatially and in redshift, (7) 3D over-densities coincident with an SMG in the line of sight, (8) 3D over-densities with no SMG in the field.}

\label{tab:summary}

\end{center}
\end{table*}


\section{The LESSJ033336.8-274401 region}
\label{sec:LESS24}

The region surrounding LESSJ033336.8-274401 is by far the strongest outlier in galaxy density identified in this study and is likely to be an actively star-forming cluster at $z\sim1.8$. The structure contains a highly significant over-density of 23 sources ($\sim27\sigma$) spanning $\Delta z=0.2$ and is consistent both spatially and in redshift with both an SMG and a QSO at $z=1.764$ \citep[][]{Treister09}. We note that the LESSJ033336.8-274401 region does not fall in the extensively well studied Great Observatories Origins Deep Survey South (GOODS-S) field region of the ECDF-S. While it is covered by the XMM observations of the field, it is situated in an off-axis position and is not obviously associated with any x-ray emission. In addition, given the depths of the submillimetre and optical-NIR data in the COSMOS field, if a system such as LESSJ033336.8-274401 were in the COSMOS region, it would have been identified in our analysis as a similar significance over-density. 

If we search for similarly over-dense regions in the Millennium Run (MR) Simulations (see Appendix A for details) we fail to  identify any volumes which are comparable in number density to the LESSJ033336.8-274401 region. This may suggest that the Millennium Run semi-analytic models do not accurately reproduce the most over-dense structures. A potential explanation for this is that the MR does not model a sufficient volume to produce the most extreme structures at high redshift. Our cluster finding search covers $\sim1$\,deg$^2$ in total and probes $1.4<z<4.5$, which equates to a volume of $\sim0.377$\,Gpc$^{-3}$. In comparison, the MR simulates a volume of $\sim0.125$\,Gpc$^{-3}$, a third of the volume probed in this work. Therefore, it is unsurprising that we do not find a LESSJ033336.8-274401 type region in the simulation volume. 

Regions such as LESSJ033336.8-274401 are likely to be key sites in testing the validity of our current paradigm and therefore are highly attractive candidates for spectroscopic follow-up observations which would confirm their nature. Comparisons of such structures with the next generation of large scale cosmological simulations \citep[which should cover sufficient volume to detect these low number density outliers, $e.g.$ the Millennium-XXL,][]{Angulo12}, will be paramount to demonstrating the validity of structure formation models.            

\begin{figure*}
\begin{center}

\includegraphics[scale=0.4]{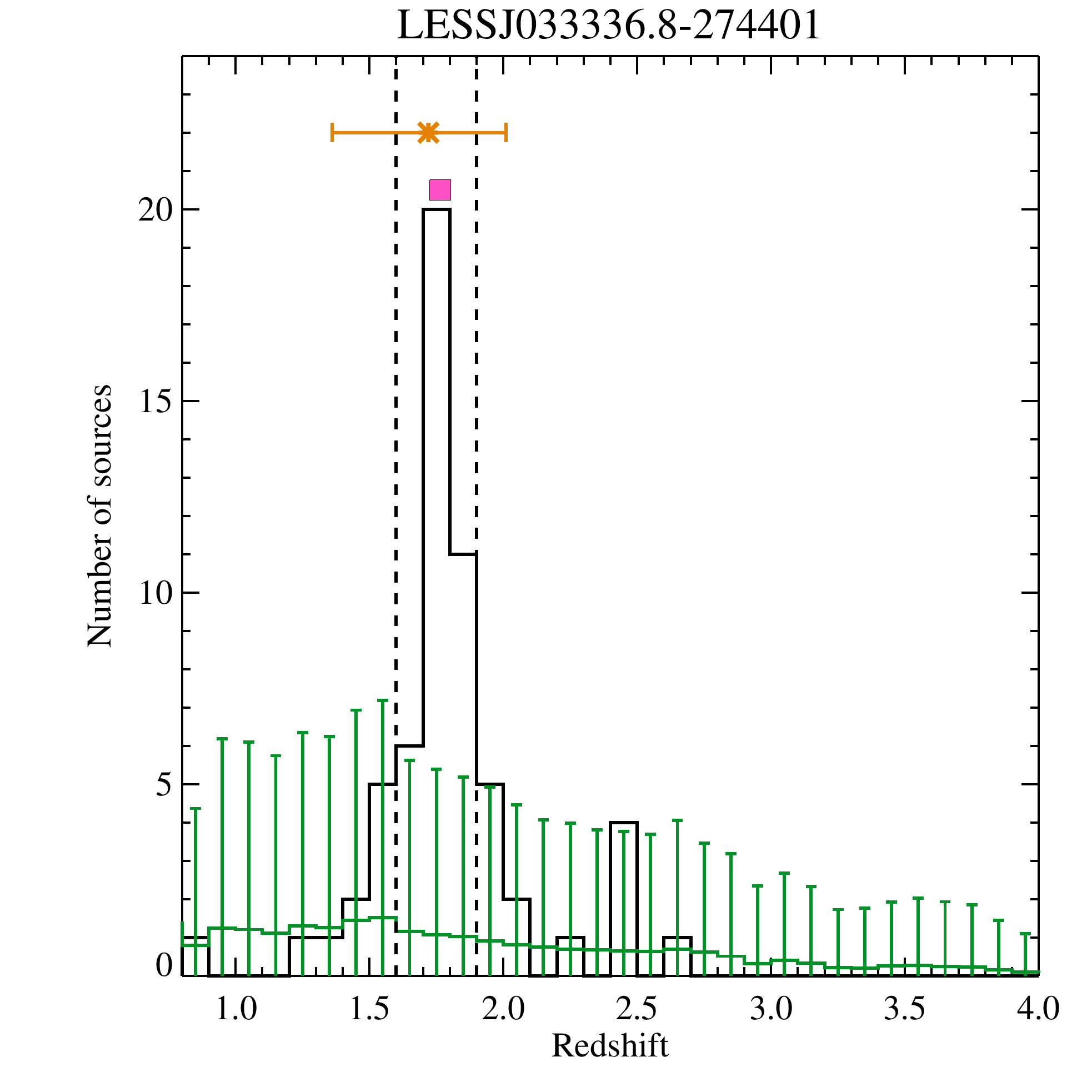}
\includegraphics[scale=0.438]{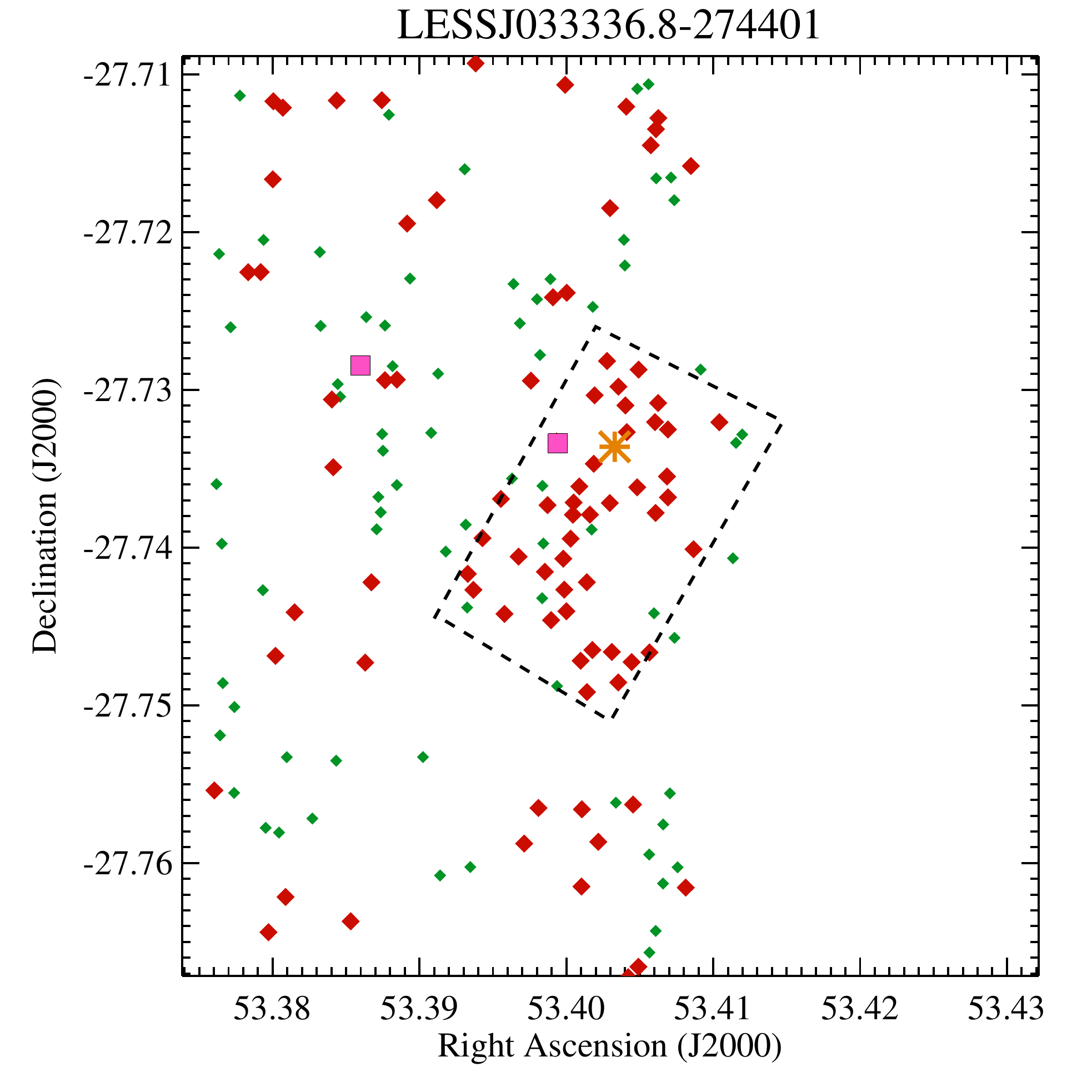}

\caption[]{The clustered region surrounding LESSJ033336.8-274401. The selected region is expanded to a rectangular shape to encompass all of the cluster members. Left: The photometric redshift distribution of sources surrounding LESSJ033336.8-274401. Right: The spatial position of sources surrounding the SMG. In the left panel, the redshift distribution (black line) is taken from the region bounded by the dashed box in the right panel. The green line shows the mean number of sources (with 3$\sigma$ error) in a region of this size over the entire ECDF-S. Orange stars mark LESSJ033336.8-27440, while pink squares indicate QSOs. In the right panel, red diamonds indicate galaxies with a photometric redshift which is consistent with the peak in the redshift distribution (bounded by dashed vertical lines in the top panel). Green diamonds are sources within $z_{SMG}\pm0.3$.  In total we find 38 potential cluster members spanning $\Delta z=0.3$ on a $\sim1.5^{\prime}$ scale ($\sim770$\,kpc at $z=1.8$). The region contains 36 high redshift galaxies, an SMG and a QSO. As such, this region is highly likely to be an actively star-forming cluster at $z\sim1.8$.}
\label{fig:LESS24}
\end{center} 
\end{figure*}

While this region is highly significant as it stands, the restriction of out analysis to square selection regions may be limiting the number of cluster members we have identified. Figure \ref{fig:pos_ECDFS} clearly shows an excess of sources at the cluster redshift falling South of our box-selected region. In addition, the LESSJ033336.8-274401 region lies at the edge of the ECDF-S and we may be missing further clustering which falls out of the MUSYC field\footnote{We note that we have checked the original MUSYC data and these sources are not artefacts of the reduction process at the CCD edge.}. To explore this further, we repeat the analysis discussed in Section \ref{sec:ECDFS_3d} but apply a rectangular selection region, covering the full extent of the over-dense structure (Figure \ref{fig:LESS24}). We find that the redshift spike extends over $\Delta z=0.3$ in redshift and spans $\sim1.5^{\prime}$ ($\sim770$\,kpc at $z=1.8$). It contains 36 galaxies (predominantly $sBzK$ sources), an SMG and a QSO essentially at the same redshift (the mean 1$\sigma$ photometric redshift error on each of the star-forming galaxies is $^{+0.04}_{-0.1}$). Follow up observations of this region are currently being undertaken (Davies et al, in prep).

\section{The ECDFS BzK\_1 region}
\label{sec:ly_blob}

As noted previously, the ECDFS BzK\_1 region is also of particular (Figure \ref{fig:pos_ECDFS_no_sub}). The structure contains a strong galaxy over-density of 7 sources at $z\sim2.3$ in a very small $\sim22^{\prime\prime}$ ($\sim170$\,kpc at $z=2.3$) region, surrounding a Lyman-alpha blob \cite[LAB, $e.g.$][]{Fynbo99} at an identical redshift \citep{Yang10}. Given the extent of this structure we are likely to be witnessing either a cluster core or massive galaxy in formation. If the former, then we may be witnessing a new mode of star formation in high redshift cluster cores - with actively star-forming systems surrounding what may be a central region of intense, unconstrained star-formation (traced by the LAB). While extended line emission is observed in radio galaxies at these redshifts, and these can can exist in forming clusters \cite[$e.g.$][]{Hatch11}, this system has no strong radio source to power the nebulosity. At low redshifts extended emission line regions exist at the centres of clusters with cooling cores. Given the similarity of the physics of cooling cores and that expected to control the formation of stars out of cooling gas at high redshift, the presence of the ionised gas here may not be surprising. If the system is instead a single massive galaxy in formation, most of the SF appears to be occurring in plain sight - with no submillimeter bright source associated with it in the LESS map. Summing the total K-band estimated stellar mass of the constituent galaxies in this system (see Appendix A for details), we already observe $\gtrsim10^{11}$\,M$_{\odot}$. While systems of this mass are found at this redshift, they are typically submm-bright ULIRGS or old passively evolving systems.

\section{Summary and conclusions}

We have identified clustering of star-forming galaxies at $z>1.4$ in both the COSMOS field and ECDF-S using a variety of different methods. Firstly, we compared the 2D distribution of the general population of high redshift galaxies with submm bright sources at the same epoch and find a weak correlation at best. This correlation is most apparent at $z\sim2$, near the peak of the SMG redshift distribution and at the latest epoch probed in this work (where clusters have had more time to coalesce). However, we find that when applying a 3D analysis to the same fields we only confirm a small fraction ($30\%$) of over-densities identified in 2D. In addition, we find that the 2D analysis misses the majority of 3D galaxy overdensities surrounding SMGs with known redshifts. Thus using a 2D analysis alone (with no redshift information beyond simple colour selection) suggests a weak general correlation between SMGs and star-forming galaxies but is an inefficient and unreliable method for selecting individual clustered regions. In addition to this, the significant affects of varying our method renders our 2D approach unsuccessfully in robustly identifying high redshift clustering.        

We also perform a full 3D analysis of clustering in the volumes surrounding SMGs (using well-calibrated photometric redshift information) and identify eleven regions where an over-density in the general galaxy population is coincident in both spatial position and redshift with the submm bright source. These regions are likely to represent a broad range of clustered regions, from actively star-forming clusters to massive galaxies in formation. Further to this, we investigate the typical environment of the volumes surrounding SMGs with known redshifts and demonstrate that, while only a small number of SMGs display a significant over-density, the majority ($\sim80\%$) are associated with an above average number of sources for their redshift. This is consistent with the premise that the progenitors of massive early type cluster galaxies at low redshift, require their seeds to form in over-dense environments where they can grow slowly via minor mergers \cite[$e.g.$][]{van de Sande13}. If SMGs are such progenitors, then the consistently above average number of sources in their surrounding volume may provide a constant supply of these minor merger systems.

We identify a further 21 regions where an SMG (without a redshift) and over-density of star-forming galaxies are coincident in the line of sight. The number of over-densities in random pointings selected in an identical manner, is $21\pm4$. As such, it is likely that all regions where an SMG and over-density are coincident in the line of sight are chance superpositions. 

In comparison, we also identify 35 regions where a 3D over-density is suggested by a region of high projected number density of star-forming galaxies, but does not have an SMG. This suggests that we are equally likely to identify 3D clustering by targeting such regions as targeting SMGs. However, the significance of the 3D over-densities identified does show a bias, as the most interesting over-densities identified in our analysis (LESSJ033136.9-275456 and  LESSJ033336.8-274401) are both associated with SMGs in three dimensions.    

\vspace{2mm}

In conclusion we find that:

\vspace{2mm}

\noindent $\bullet$ Using star-forming galaxies we can identify individual clustered regions at high redshift. This is highlighted by regions such as LESSJ033336.8-274401 and the reconfirmation of known clusters in the COSMOS field, and opens up a new channel for cluster identification at redshifts where the red sequence method fails.            

 \vspace{2mm}

\noindent $\bullet$ Using a 2D analysis we can detect a weak spatial correlation between the locations of SMGs and star-forming galaxies; it can not efficiently and reliably identify individual structures. At the very least we require SMG redshifts and accurate photometric redshifts of the star-forming galaxies in order to select clustered regions for detailed follow-up.

  \vspace{2mm}

\noindent $\bullet$ The typical 3D environments of SMGs with known redshifts generally contain moderate ($<3\sigma$) over-densities of sources. This is consistent with the weak correlation observed in 2D and suggests that, while SMGs are typically located in more over-dense regions than the field, they do not generally trace exceptionally clustered regions.        
 
   \vspace{2mm}

\noindent $\bullet$ While the typical SMG is not found in an exceptional environment, the most extreme environments do contain an SMG (LESSJ033136.9-275456 and LESSJ033336.8-274401). As such, targeting regions which contain an SMG with a known redshift and a significant photometric redshift spike in star-forming galaxies is likely to yield detections of early clustering. However, prior redshift information is required to select such systems and, as such, they can only be identified in fields with sufficiently deep submm and multi-wavelength optical data.

 \vspace{2mm}

\noindent $\bullet$ The region surrounding LESSJ033336.8-274401 is the most significant found in this study and is highly likely to represent an actively star-forming cluster at $z\sim1.8$. The structure contains 36 galaxies, an SMG and a QSO all at essentially the same redshift. This region is therefore a highly attractive candidate as, not only one of the highest redshift clusters yet identified, but also as one of the first high redshift clusters to be identified through its actively star-forming galaxy population. We do not identify similar volumes to that surrounding LESSJ033336.8-274401 in the Millennium Run simulation. This suggests that either the simulations do not probe a large enough volume to produce the most massive structures or the LESSJ033336.8-274401 region serendipitously falls in the well studied ECDF-S. Either way, if confirmed, regions such a LESSJ033336.8-274401 are likely to be key sites in testing our current cosmological model.  
     
 \vspace{2mm}

\noindent $\bullet$ The region of ECDFS BzK\_1 is also intriguing, containing a significant over-density of sources surrounding a Lyman-$\alpha$ blob at the same redshift. In this system we may be witnessing a new mode of star formation in high redshift cluster cores - with actively star-forming galaxies surrounding a central region of intense, unconstrained star-formation.

\section*{Acknowledgements}

We would like to thank the anonymous referee for their extremely helpful comments which have improved this paper. LJMD, KH and EJAM acknowledge funding from the UK science technologies and facilities council.  This paper makes use of data obtained for APEX program IDs 078.F-9028(A), 079.F-9500(A), 080.A-3023(A), and 081.F-9500(A). APEX is operated by the Max Planck-Institut fur Radioastronomie, the European Southern Observatory, and the Onsala Space Observatory.

\appendix

\section{Examples of over-densities and individual regions of interest}
\label{sec:indu}

\begin{figure*}
\begin{center}
\includegraphics[scale=0.25]{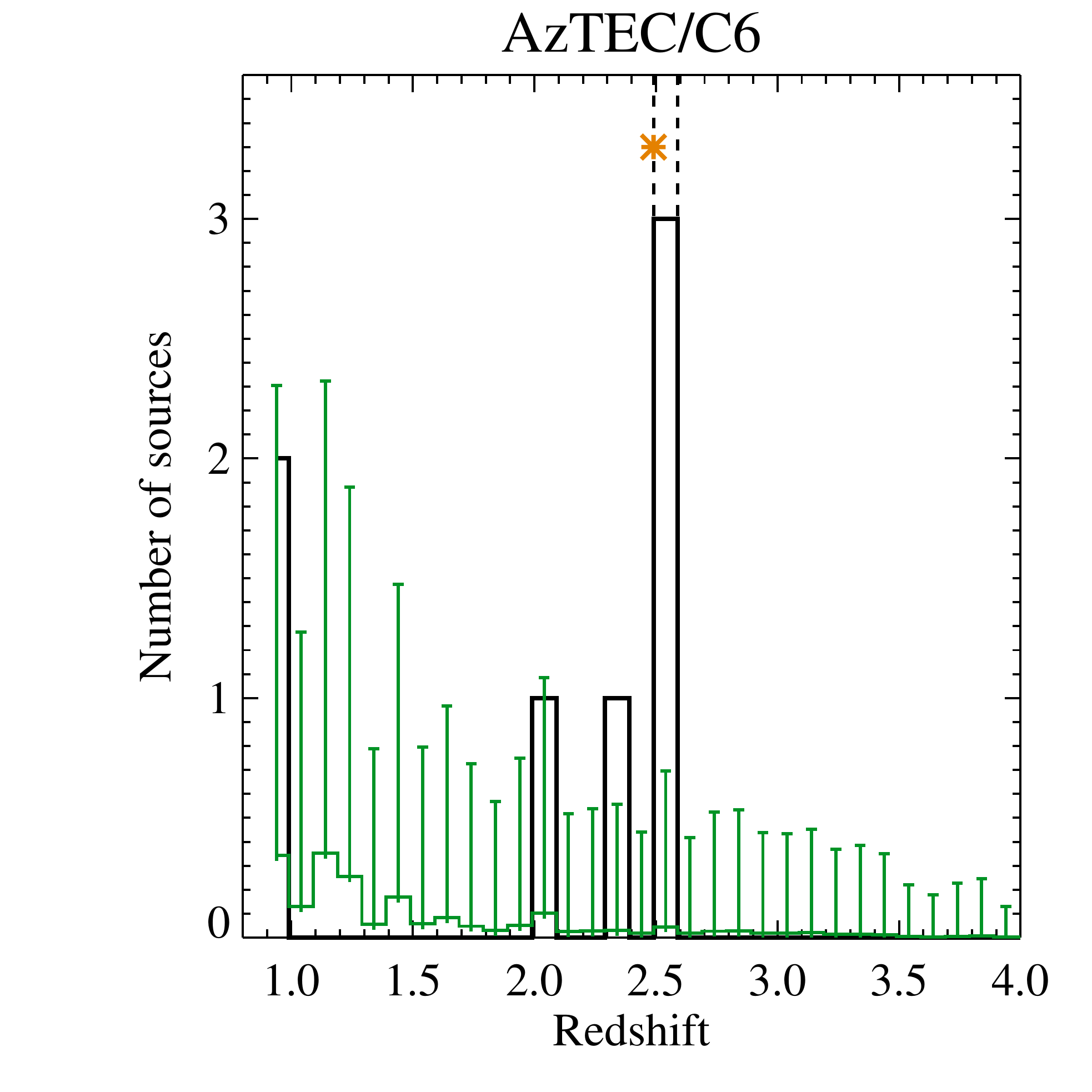}
\includegraphics[scale=0.25]{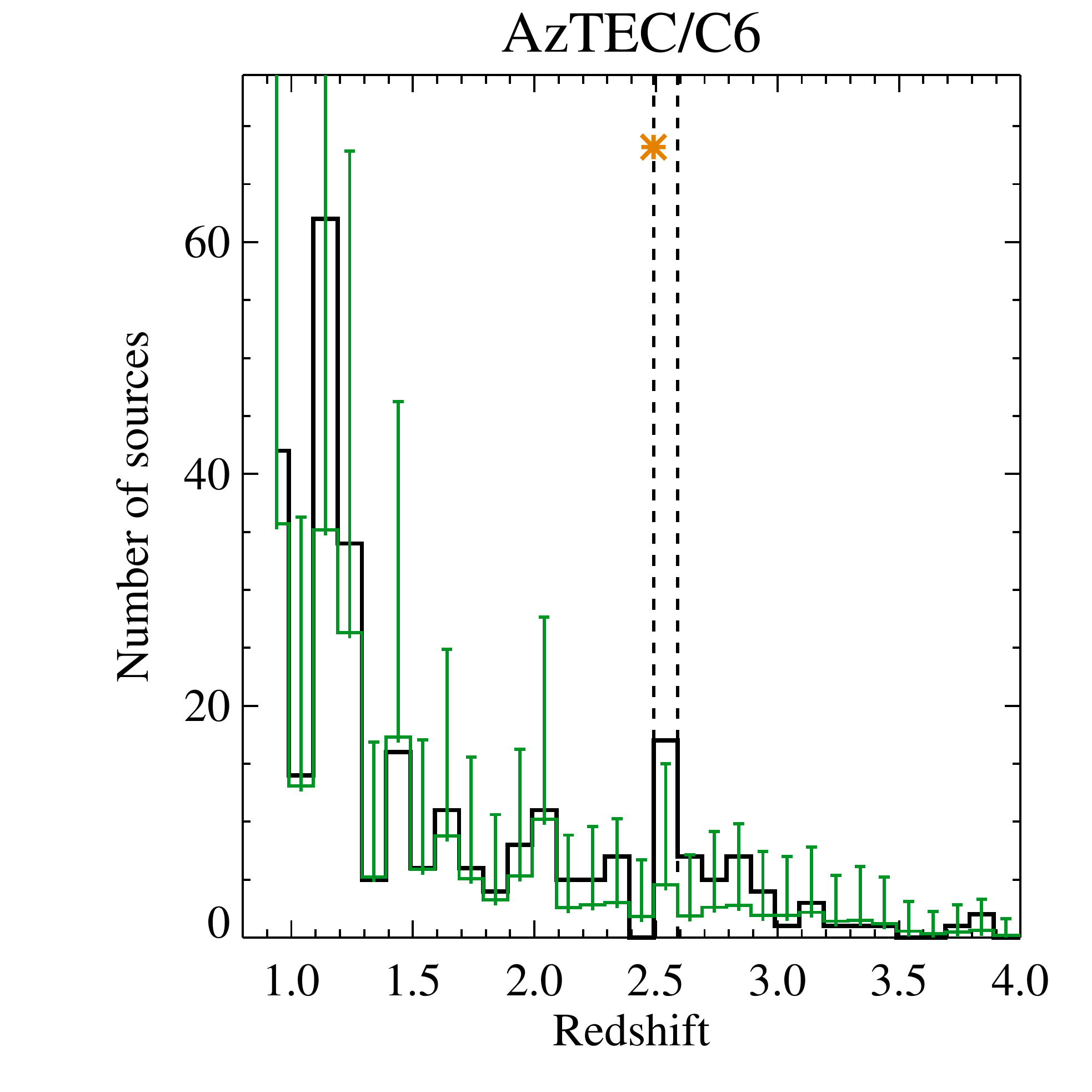}
\includegraphics[scale=0.25]{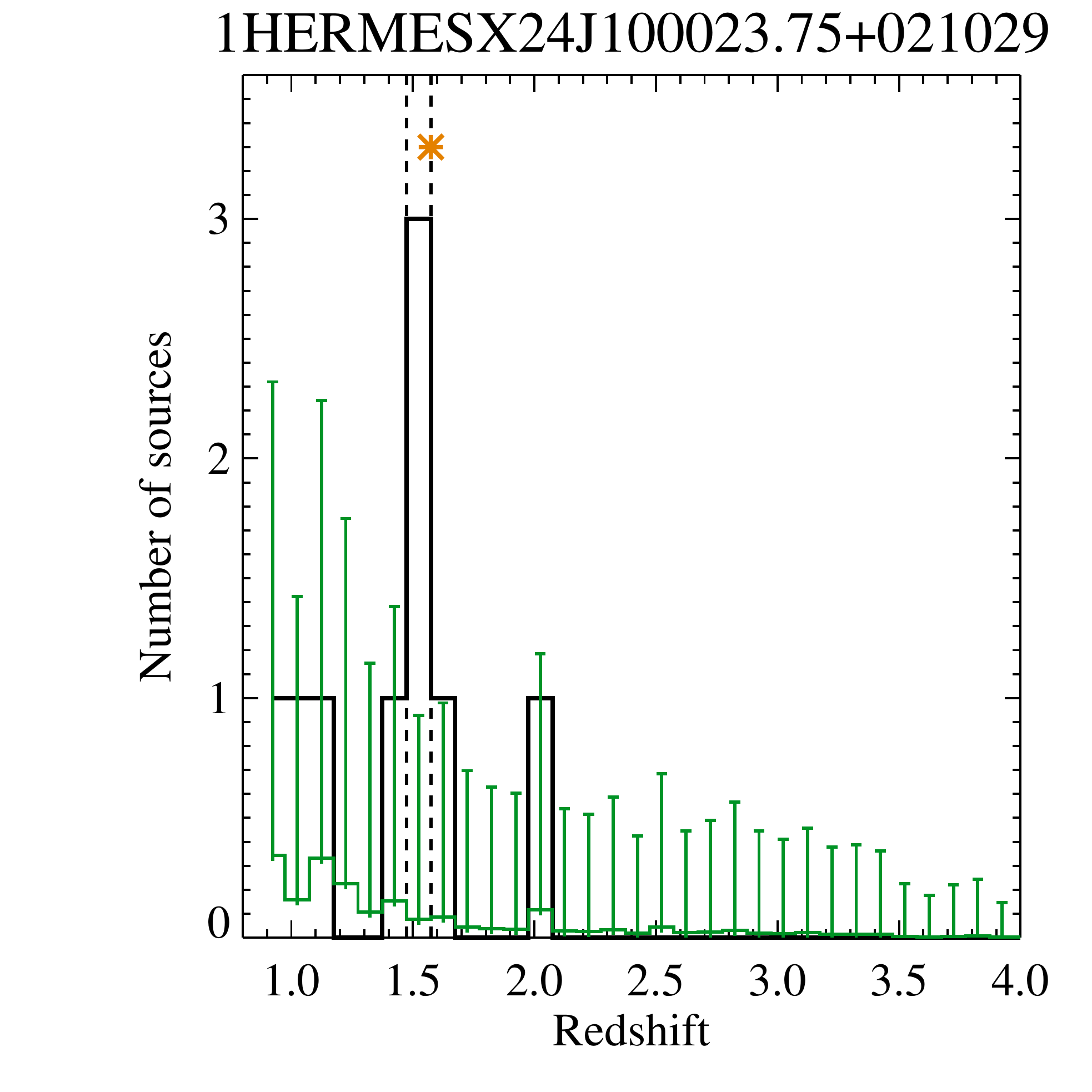}\\
\includegraphics[scale=0.25]{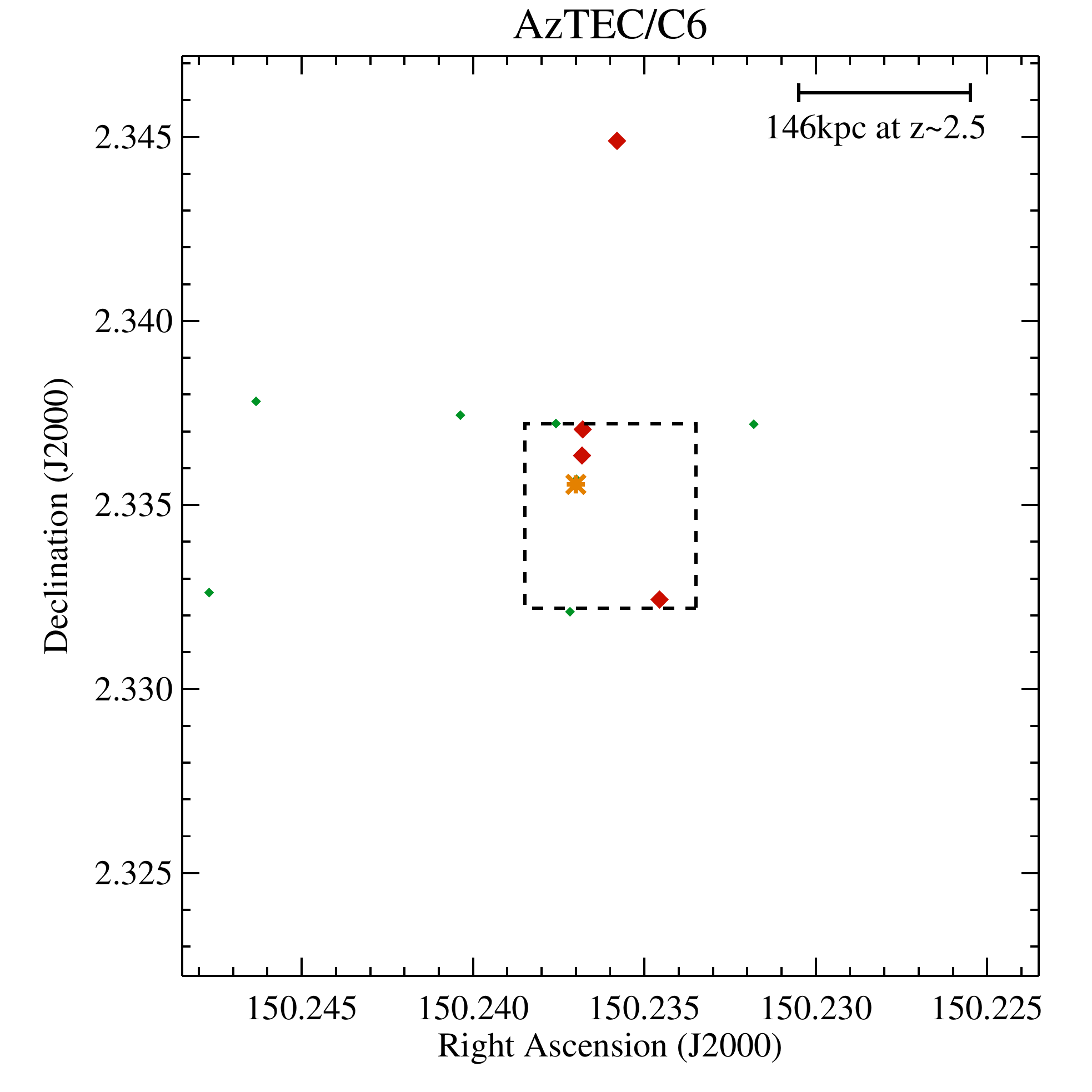}
\includegraphics[scale=0.25]{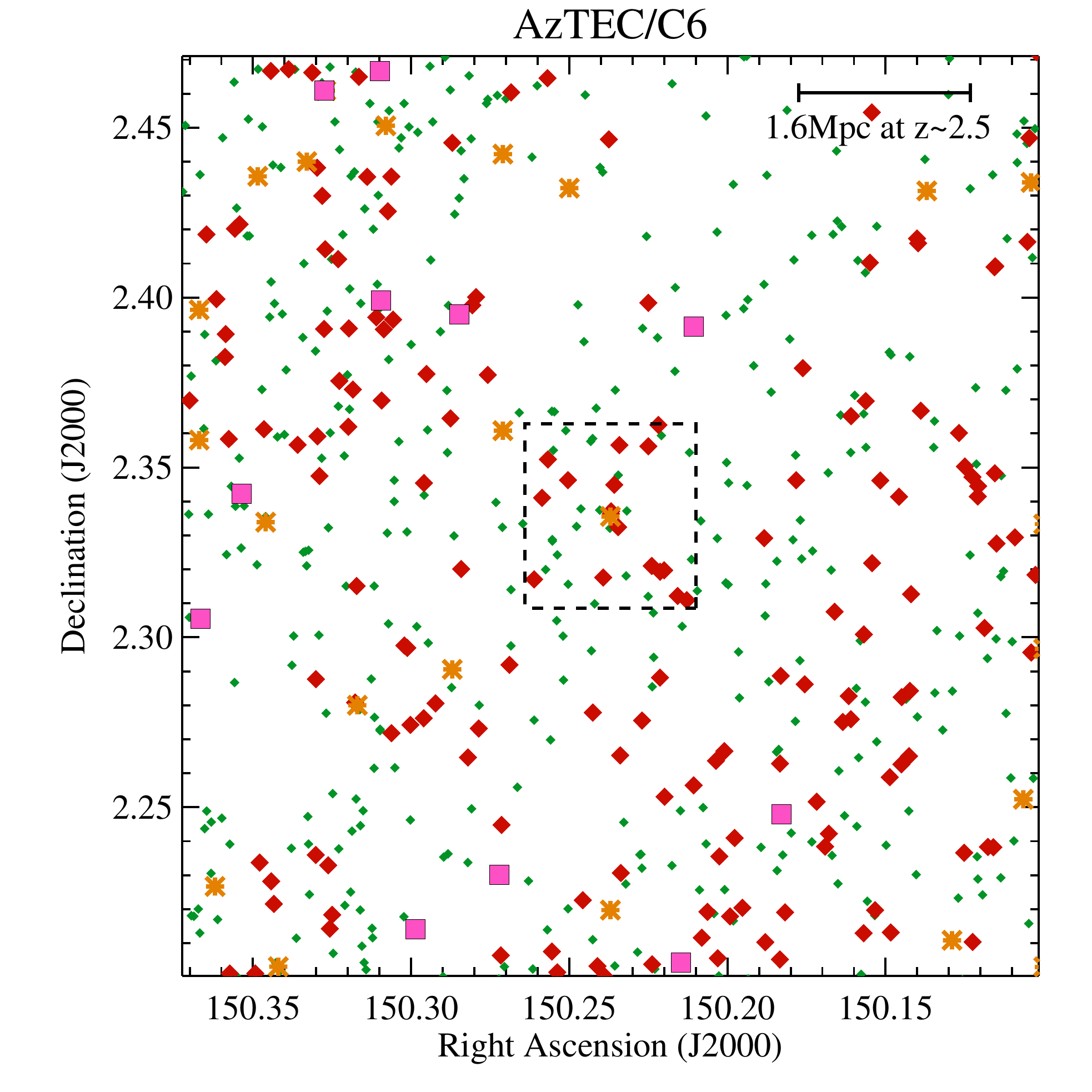}
\includegraphics[scale=0.25]{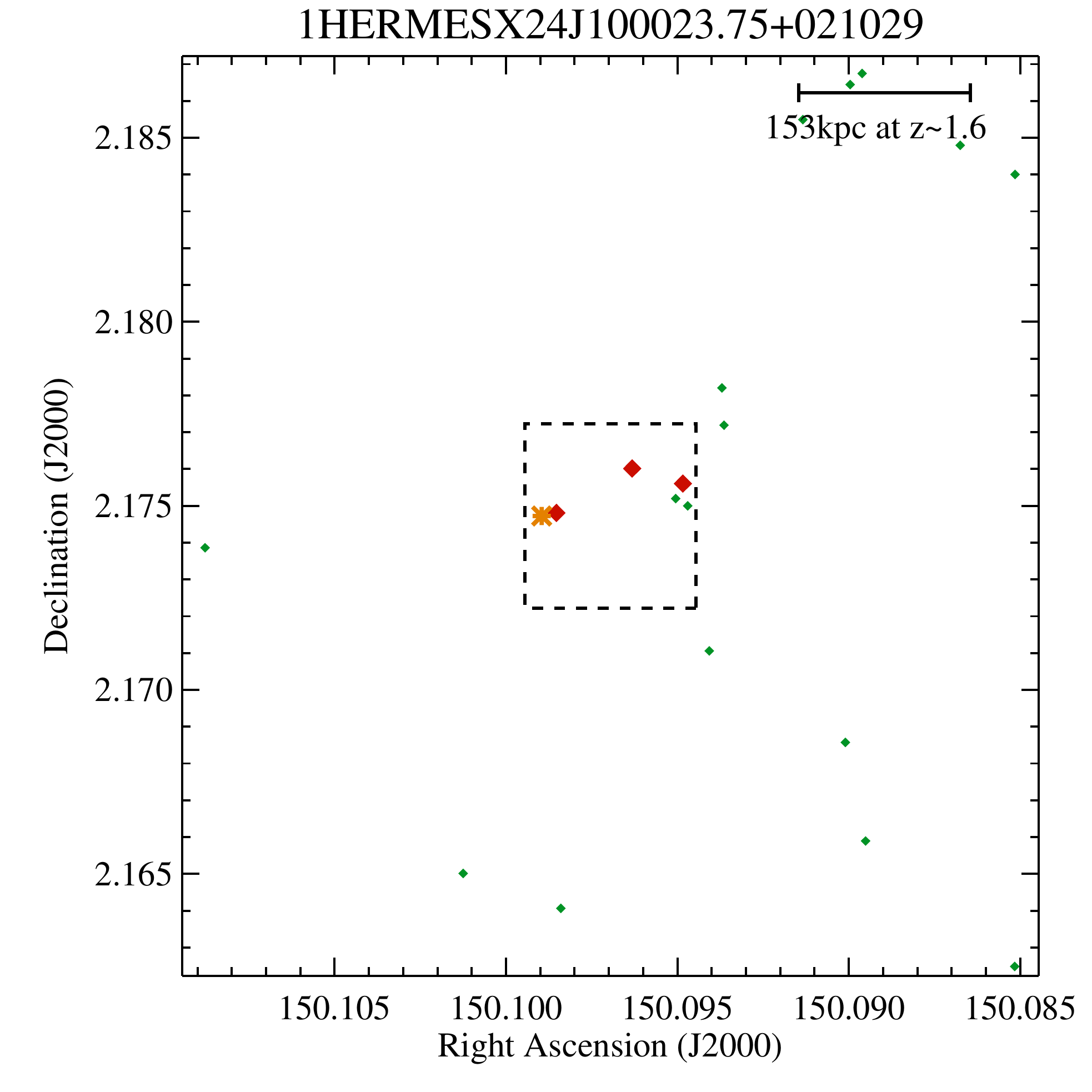}\\

\caption{Top: The photometric redshift distribution of sources surrounding COSMOS SMGs with known redshifts which are coincident with a 2D projected over-density of star-forming galaxies. Bottom: The spatial position of sources surrounding the SMG. In the top panel, the redshift distribution (black line) is taken from the region bounded by the dashed box in the bottom panel. The green line displays the mean number of sources (with 3$\sigma$ error) in a region of this size over the entire COSMOS field. Orange stars show the SMG. In the bottom panel, red diamonds indicate galaxies with a photometric redshift which is consistent with the peak in the redshift distribution (bounded by dashed vertical lines in the top panel). Green diamonds are sources within $z_{SMG}\pm0.3$ (displaying the general distribution of sources at the SMG redshift). We also display the positions of known QSOs (pink squares). Note that an over-density, as defined in the text, is identified around AzTEC/C6 on both $\sim$150\,kpc and $\sim$1.5\,Mpc scales - both regions are displayed in this figure.}

\label{fig:a_10_2d}
\end{center} 
\end{figure*}

\begin{figure*}
\begin{center}

\includegraphics[scale=0.26]{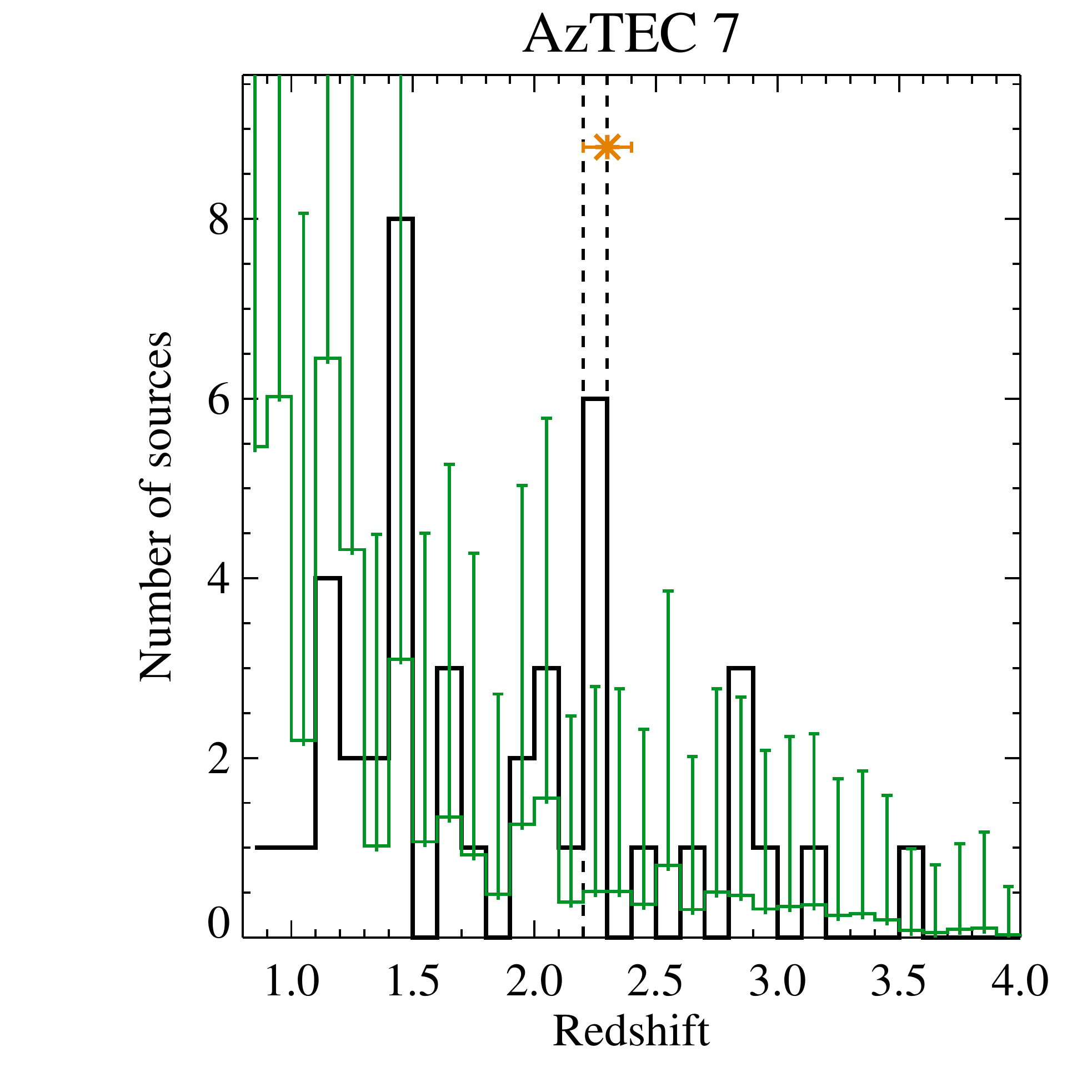}
\includegraphics[scale=0.26]{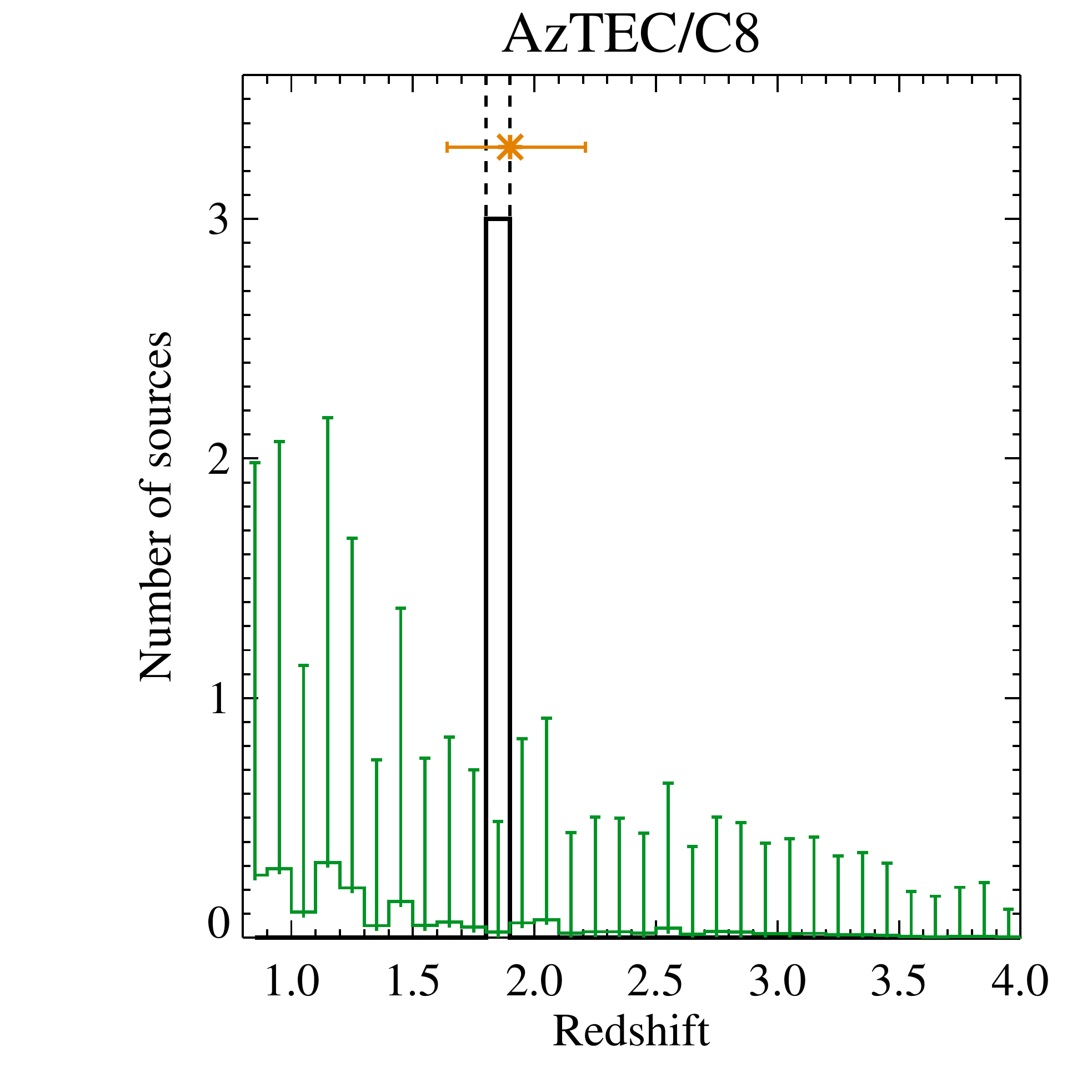}
\includegraphics[scale=0.26]{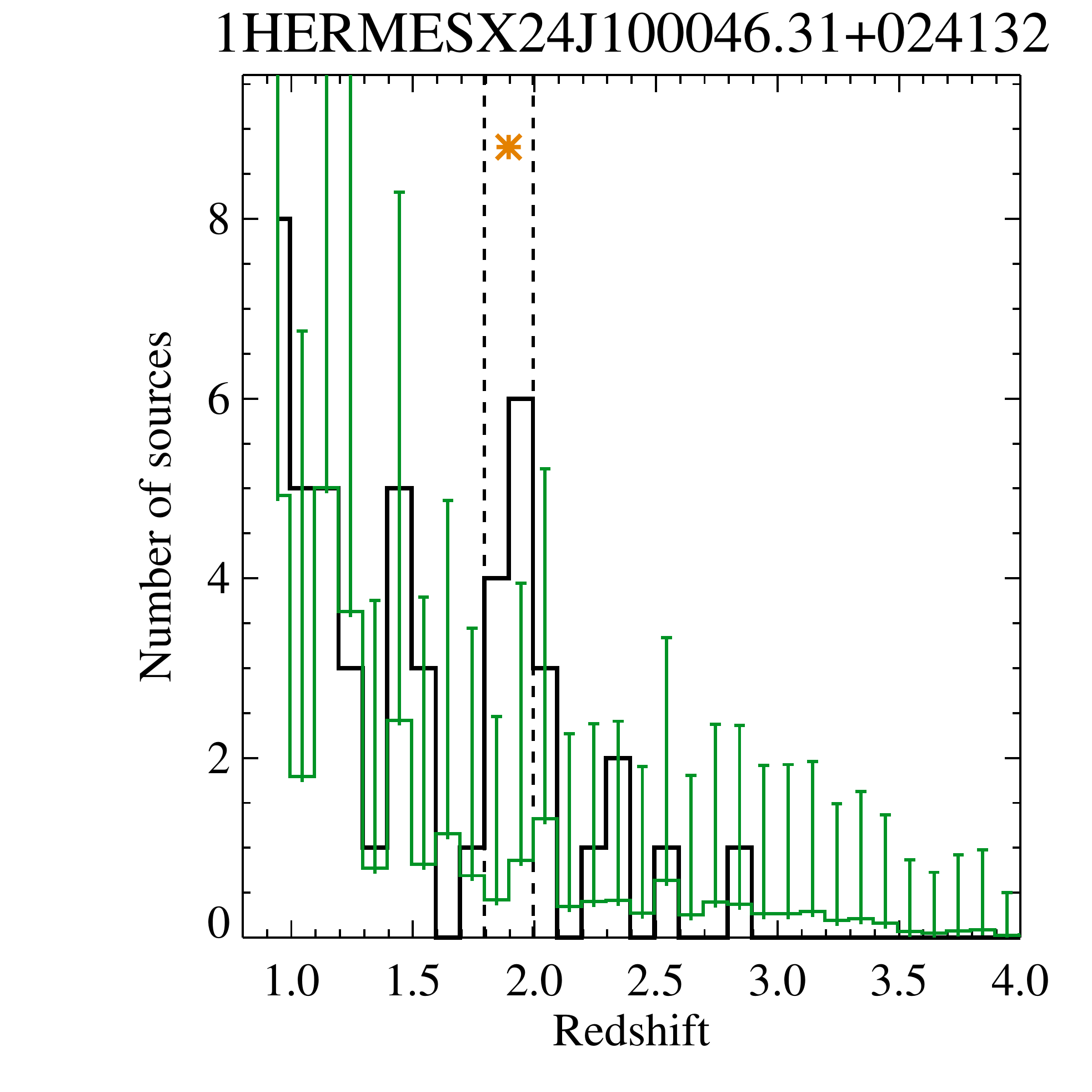}\\

\includegraphics[scale=0.26]{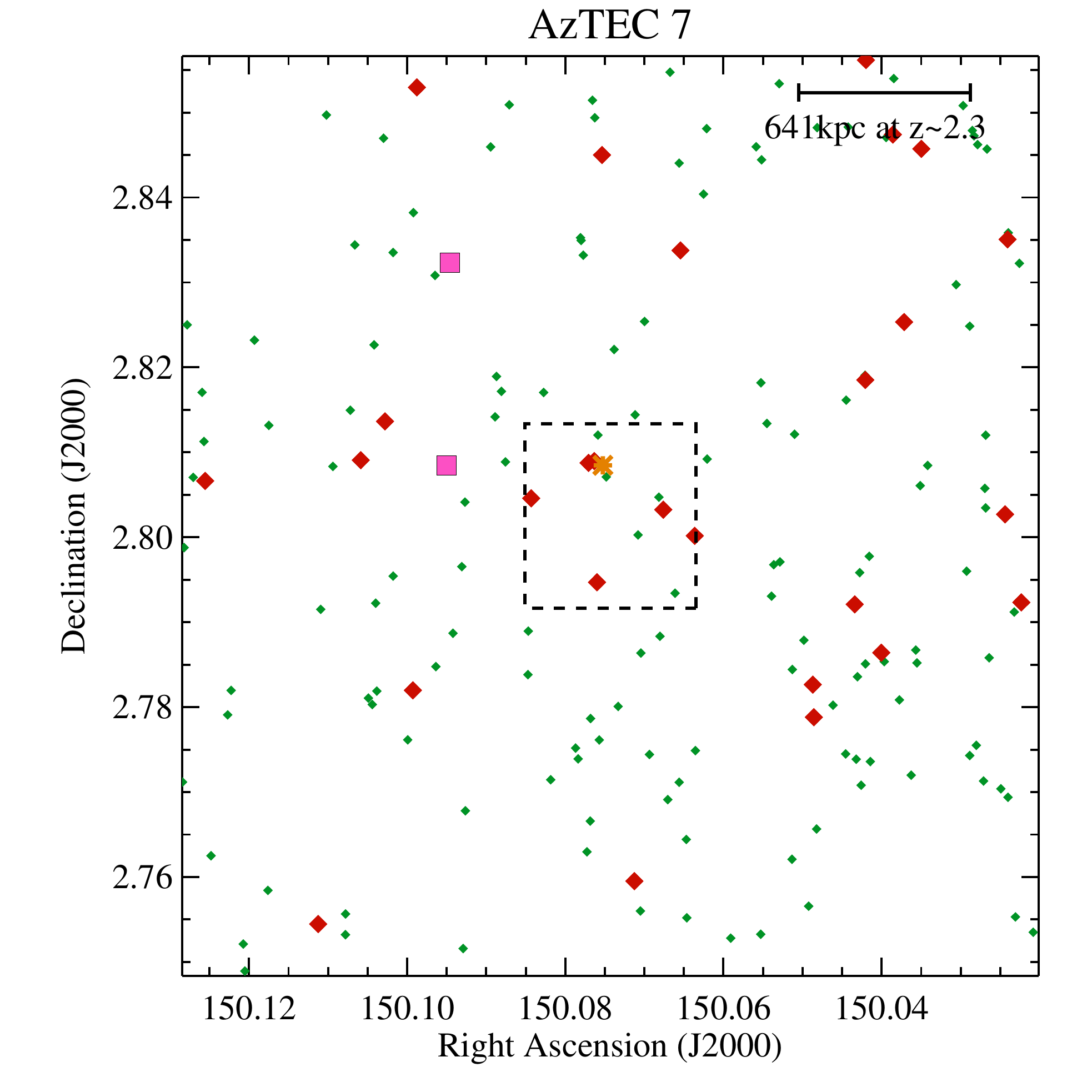}
\includegraphics[scale=0.26]{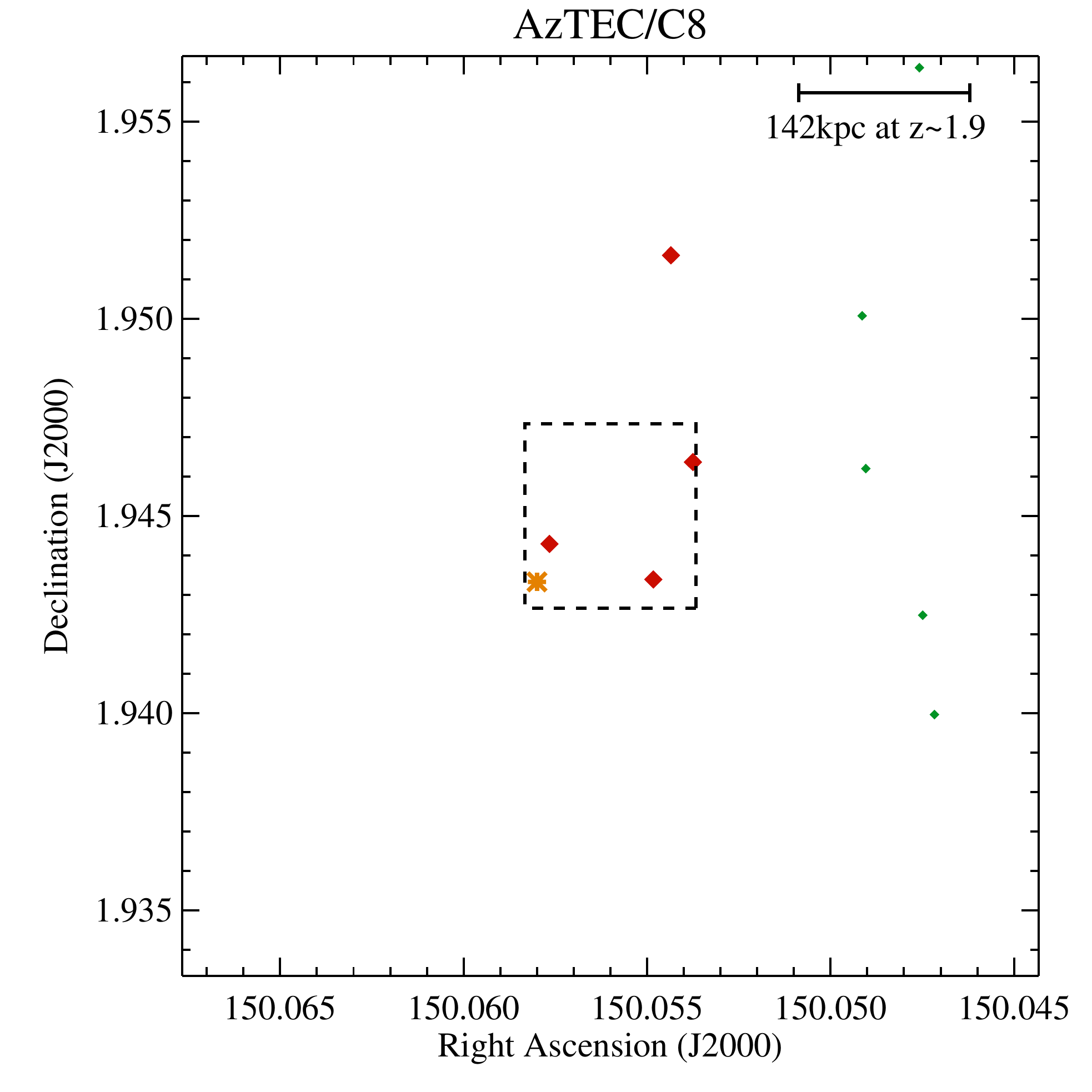}
\includegraphics[scale=0.26]{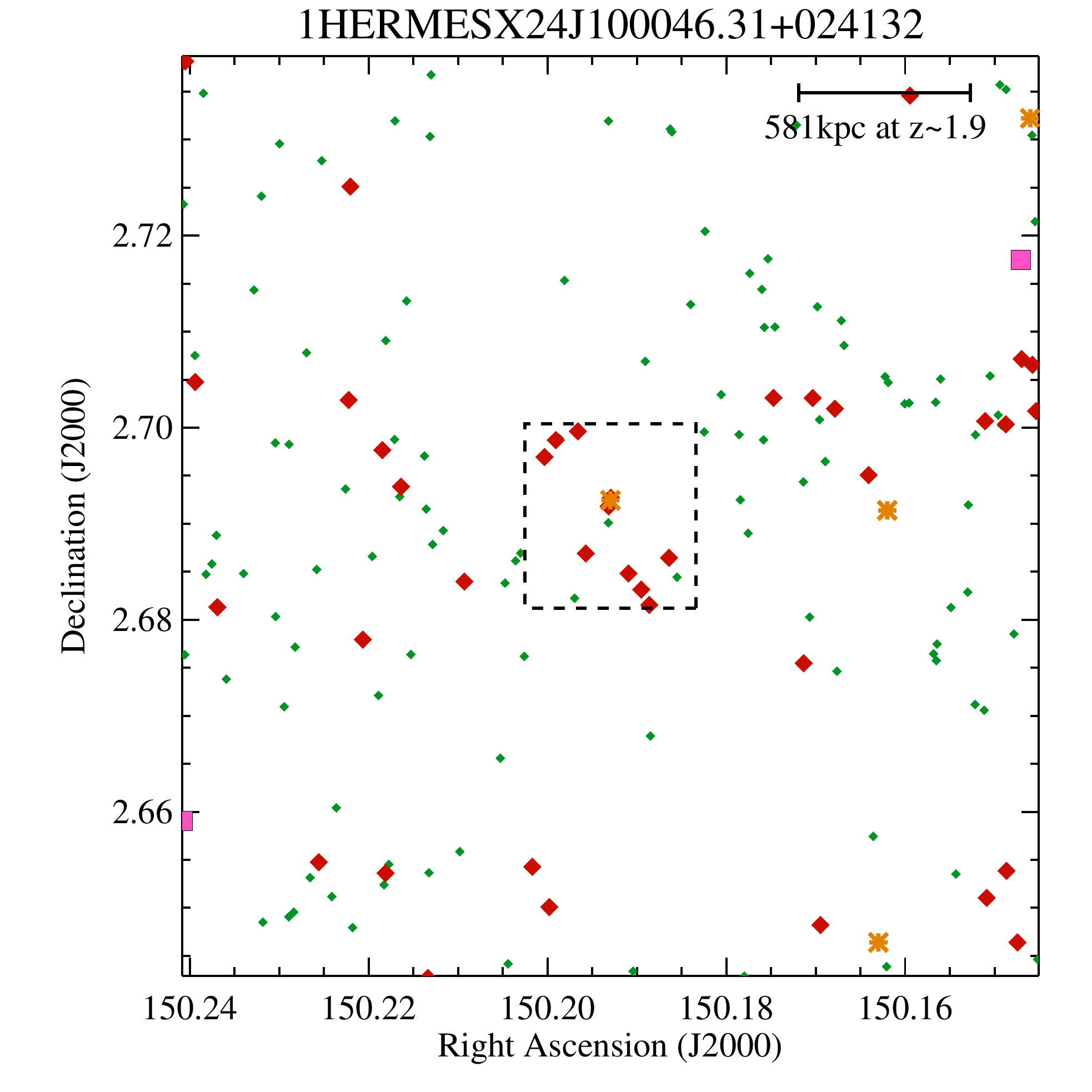}

 \rule{18cm}{0.4pt}

\includegraphics[scale=0.25]{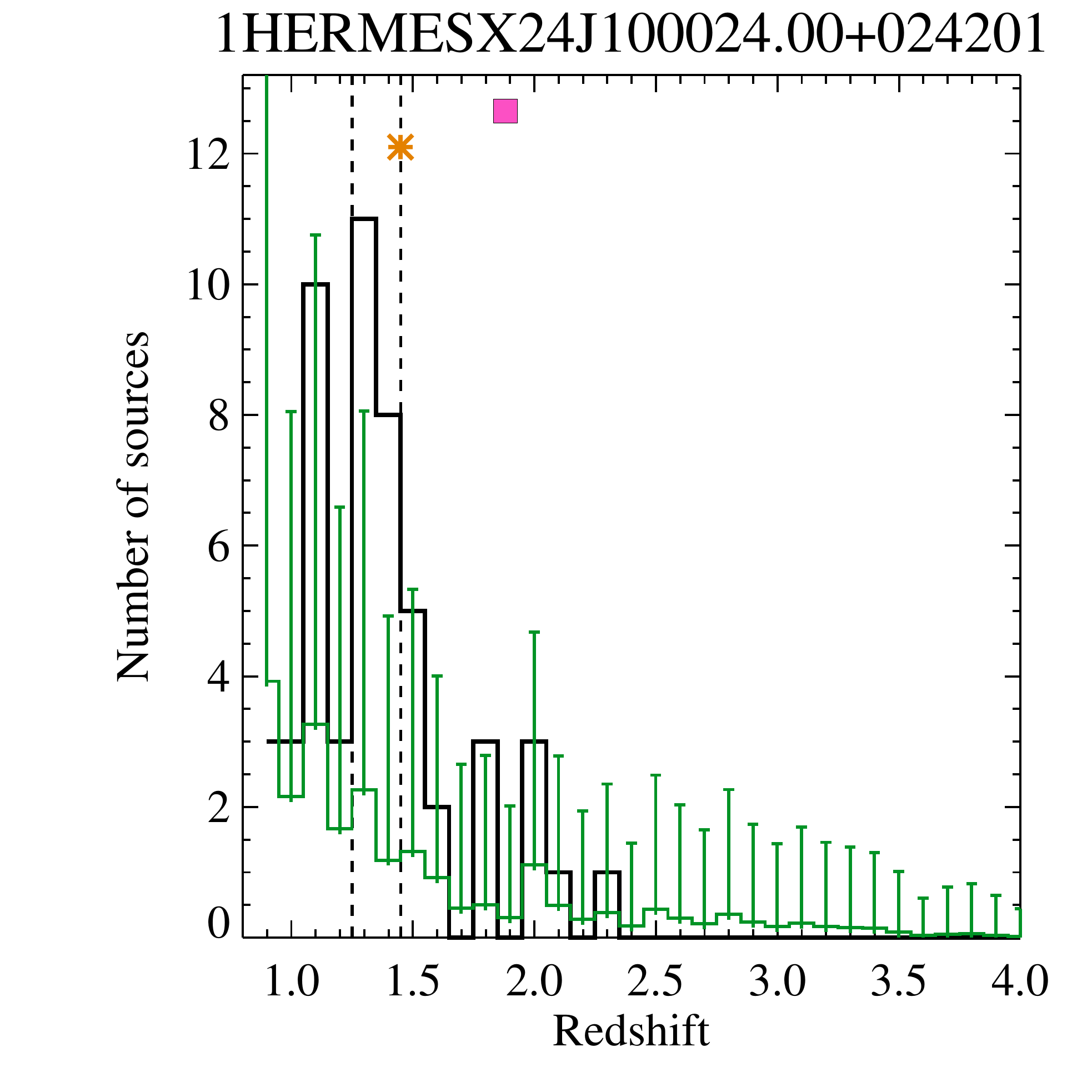}
\includegraphics[scale=0.25]{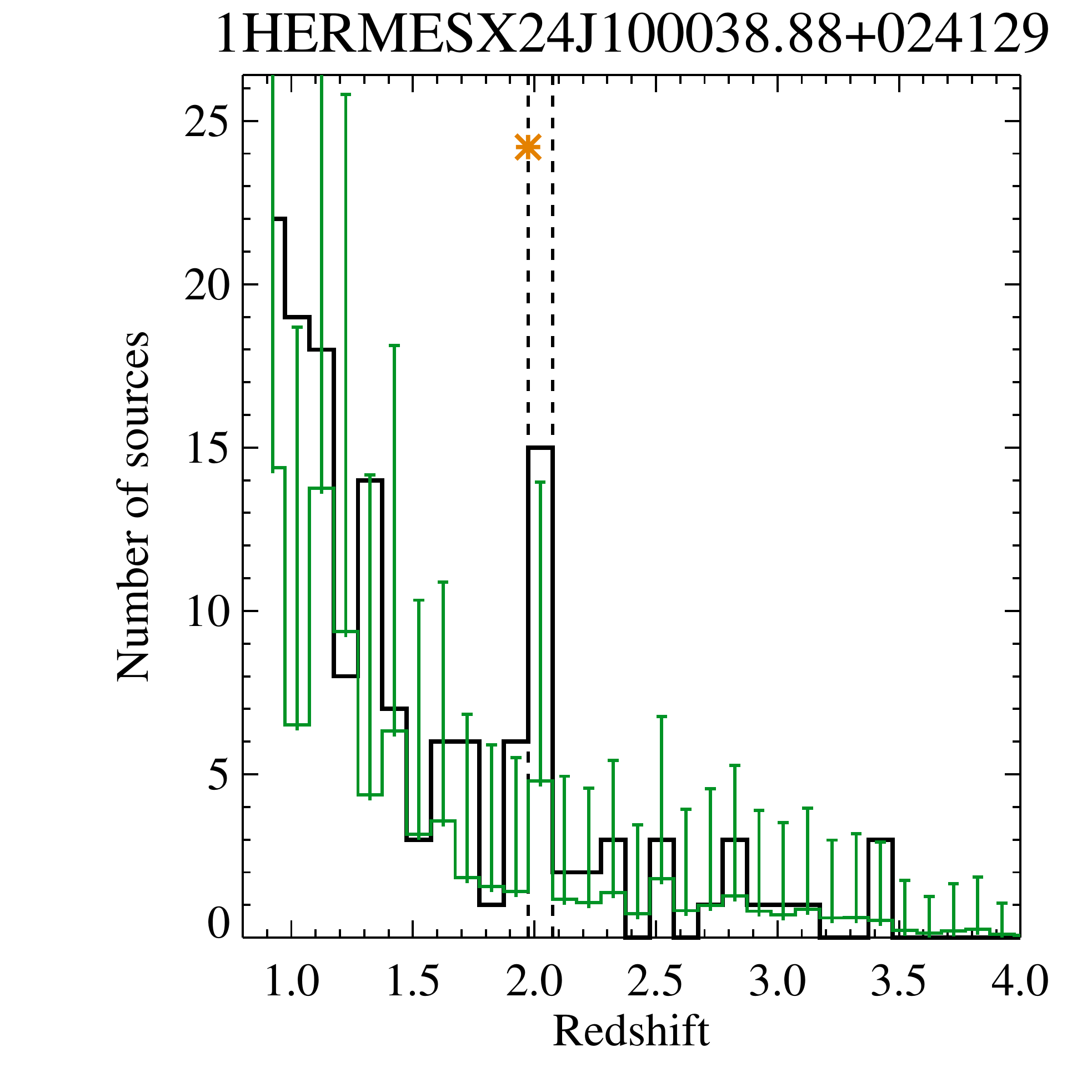}\\

\includegraphics[scale=0.25]{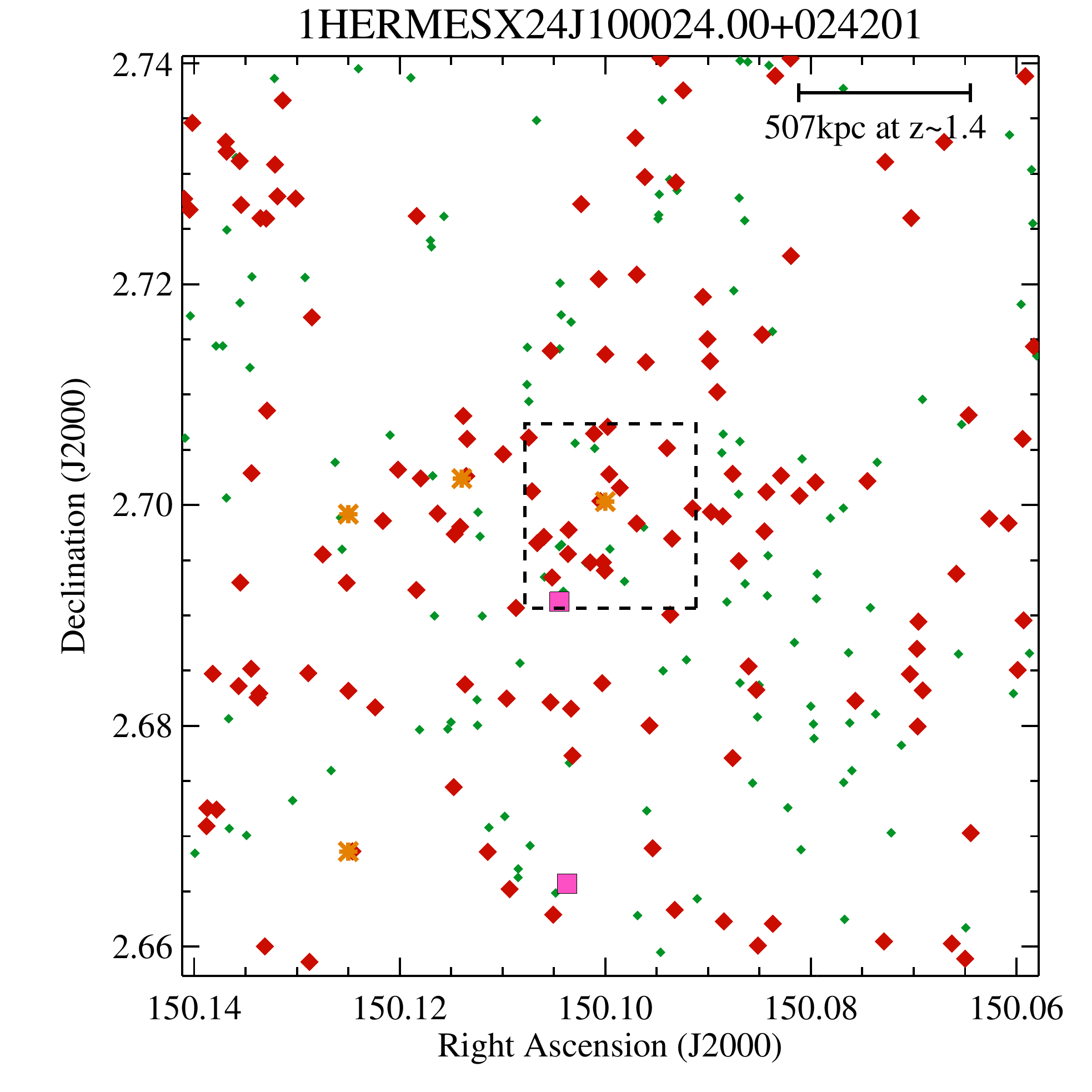}
\includegraphics[scale=0.25]{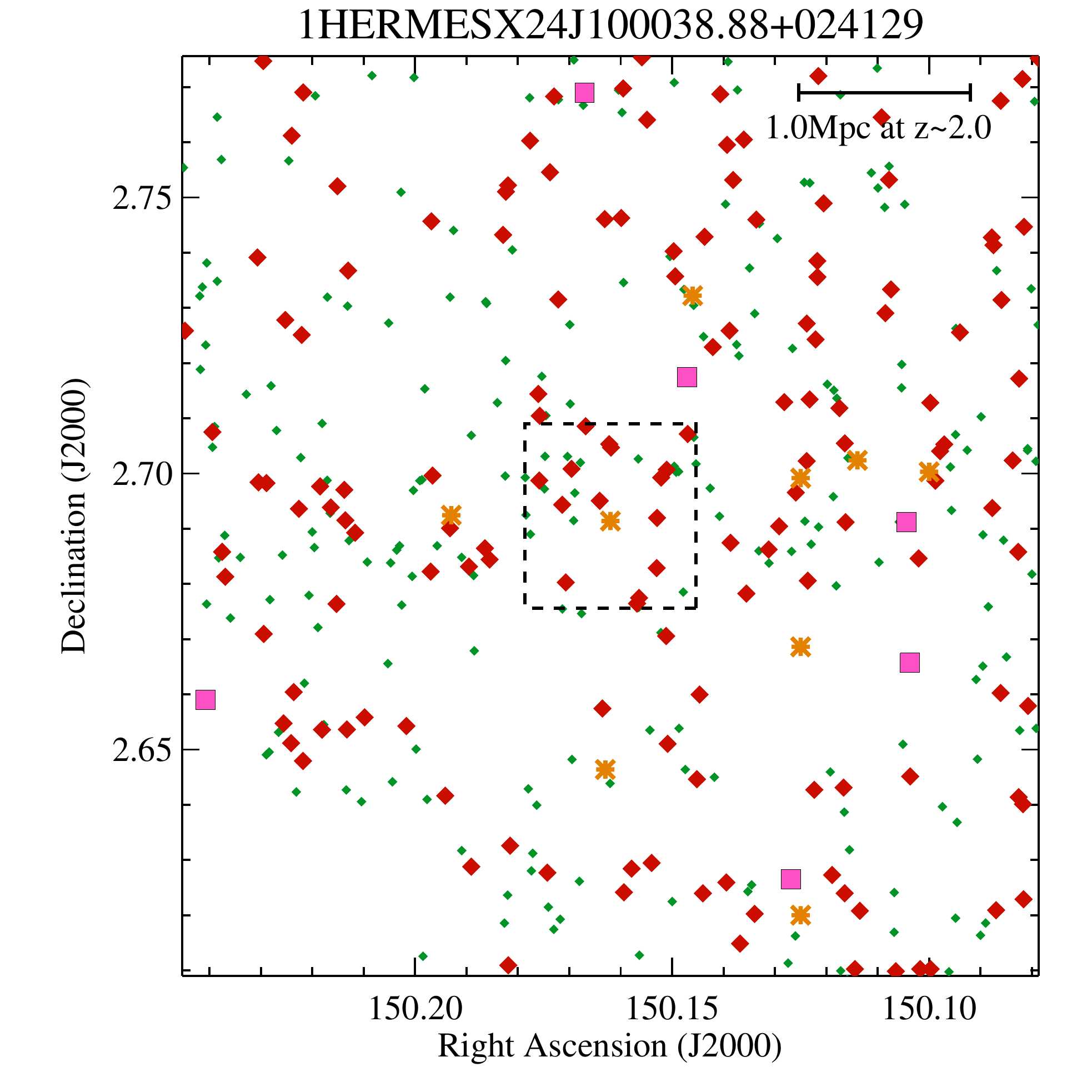}\\

\caption[]{Same as Figure \ref{fig:a_10_2d} but for regions surrounding SMGs with known redshifts but not associated with a 2D over-density. Only one source HERMESX24J100024.00+024201 has a QSO within the over-density region, but this falls at a much higher redshift than the SMG.}
\label{fig:spec_COS}

\end{center} 
\end{figure*}

\begin{figure*}
\begin{center}

\includegraphics[scale=0.26]{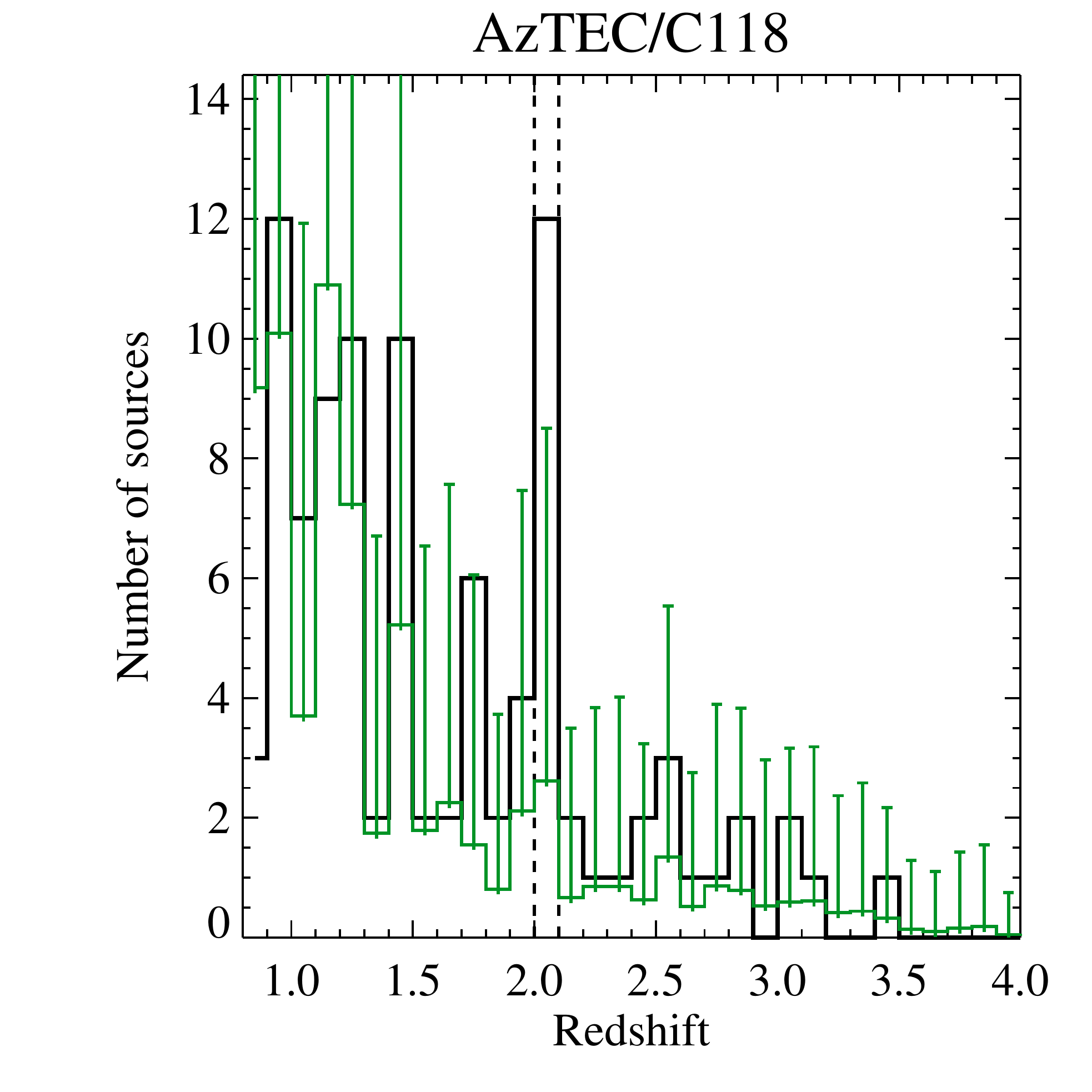}
\includegraphics[scale=0.26]{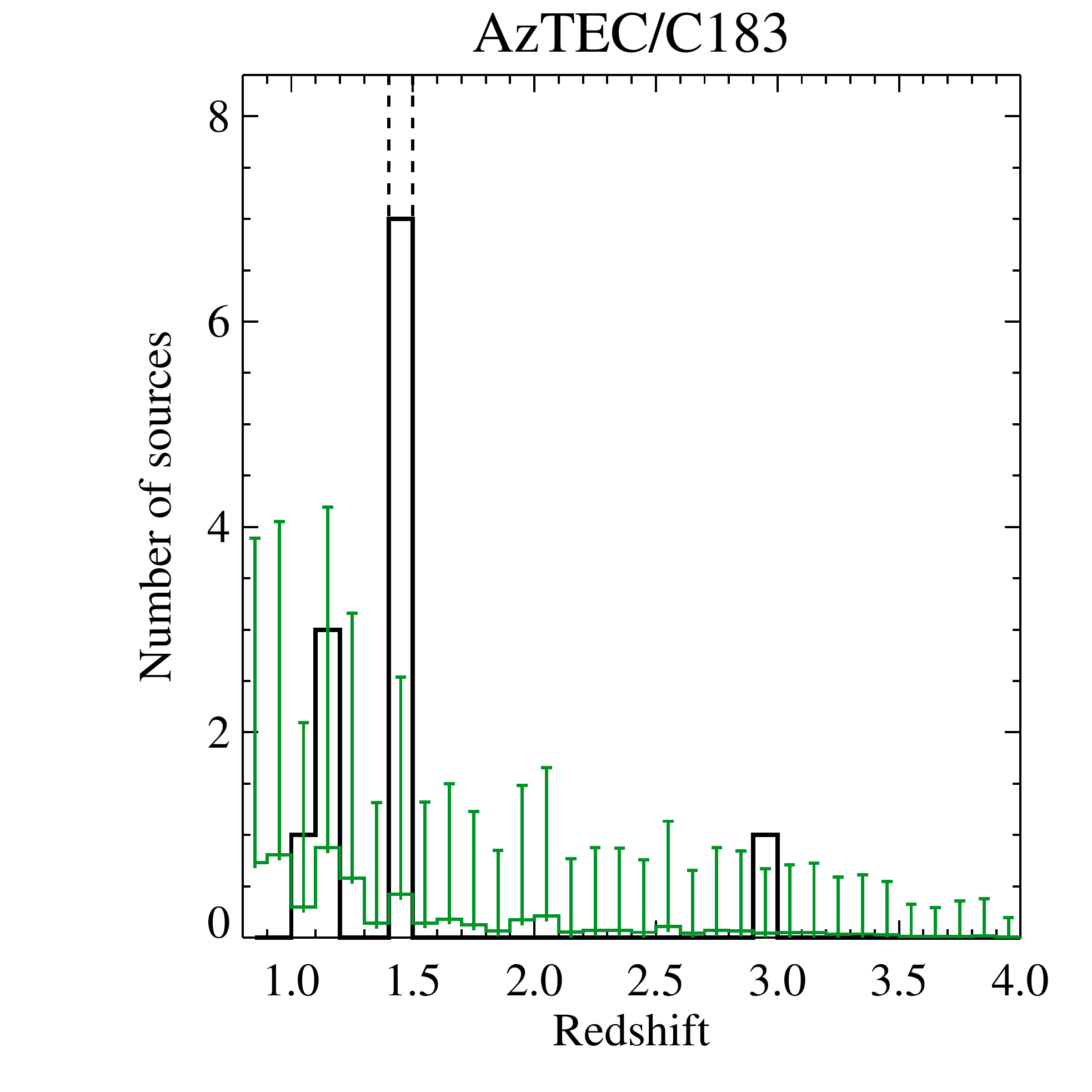}
\includegraphics[scale=0.26]{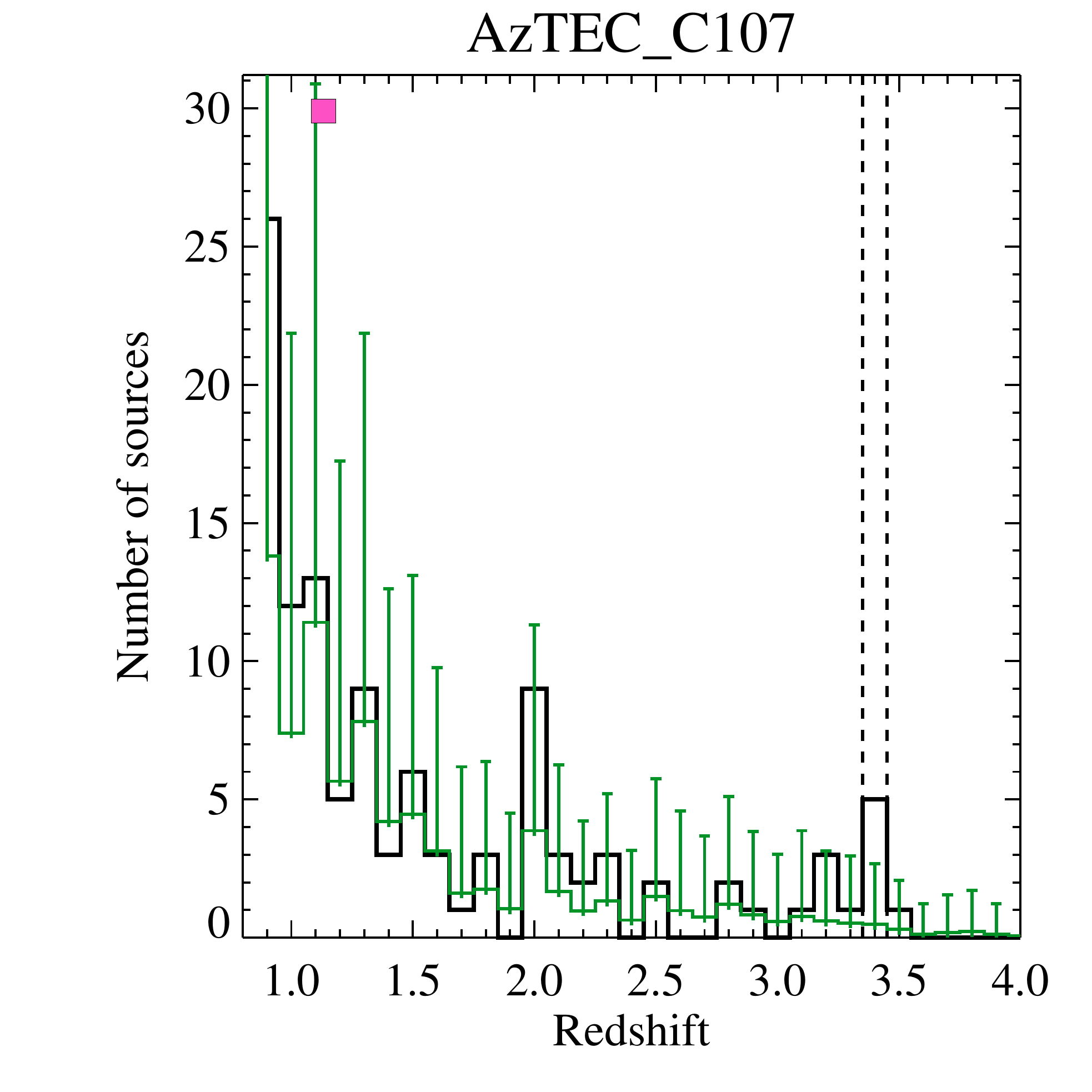}\\

\includegraphics[scale=0.26]{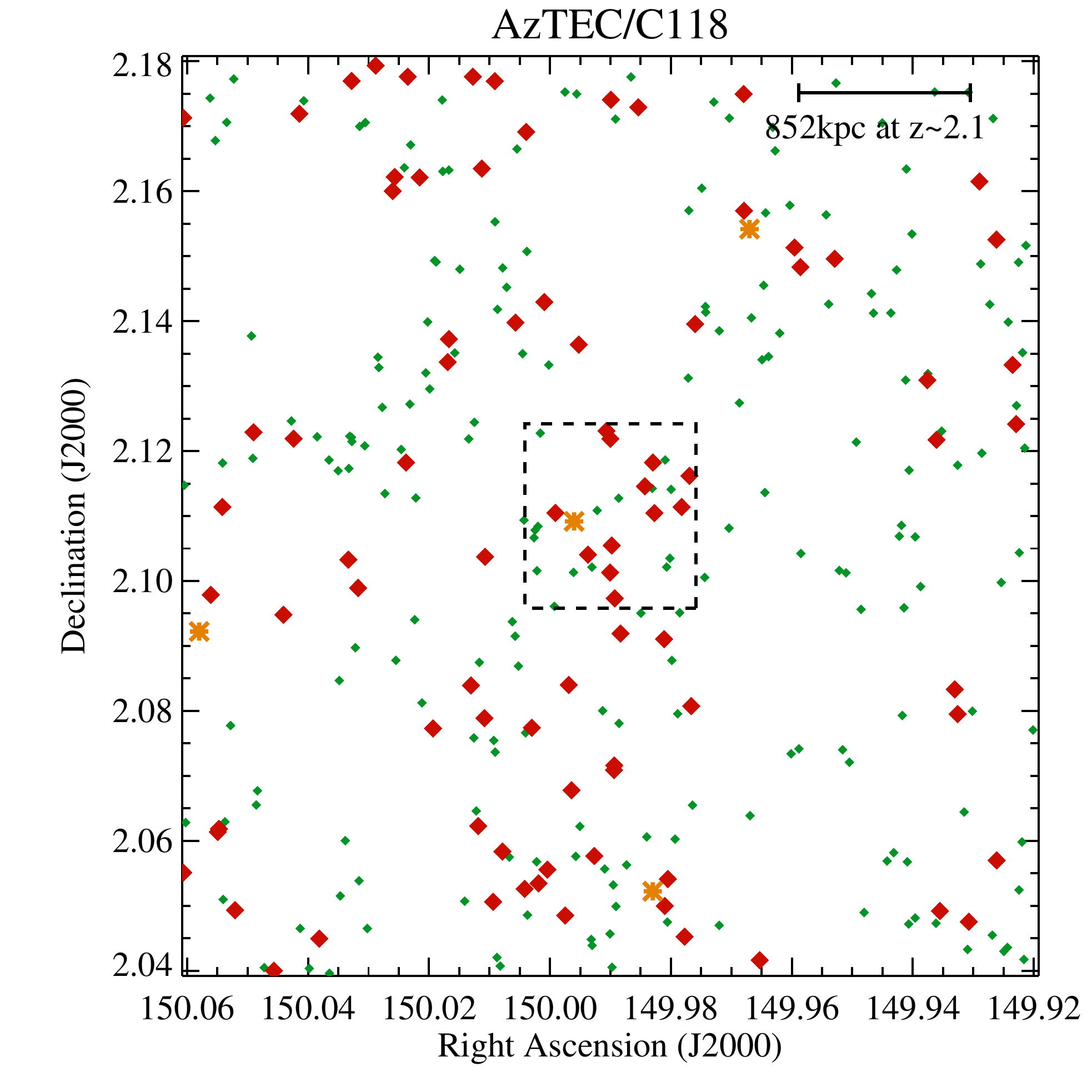}
\includegraphics[scale=0.26]{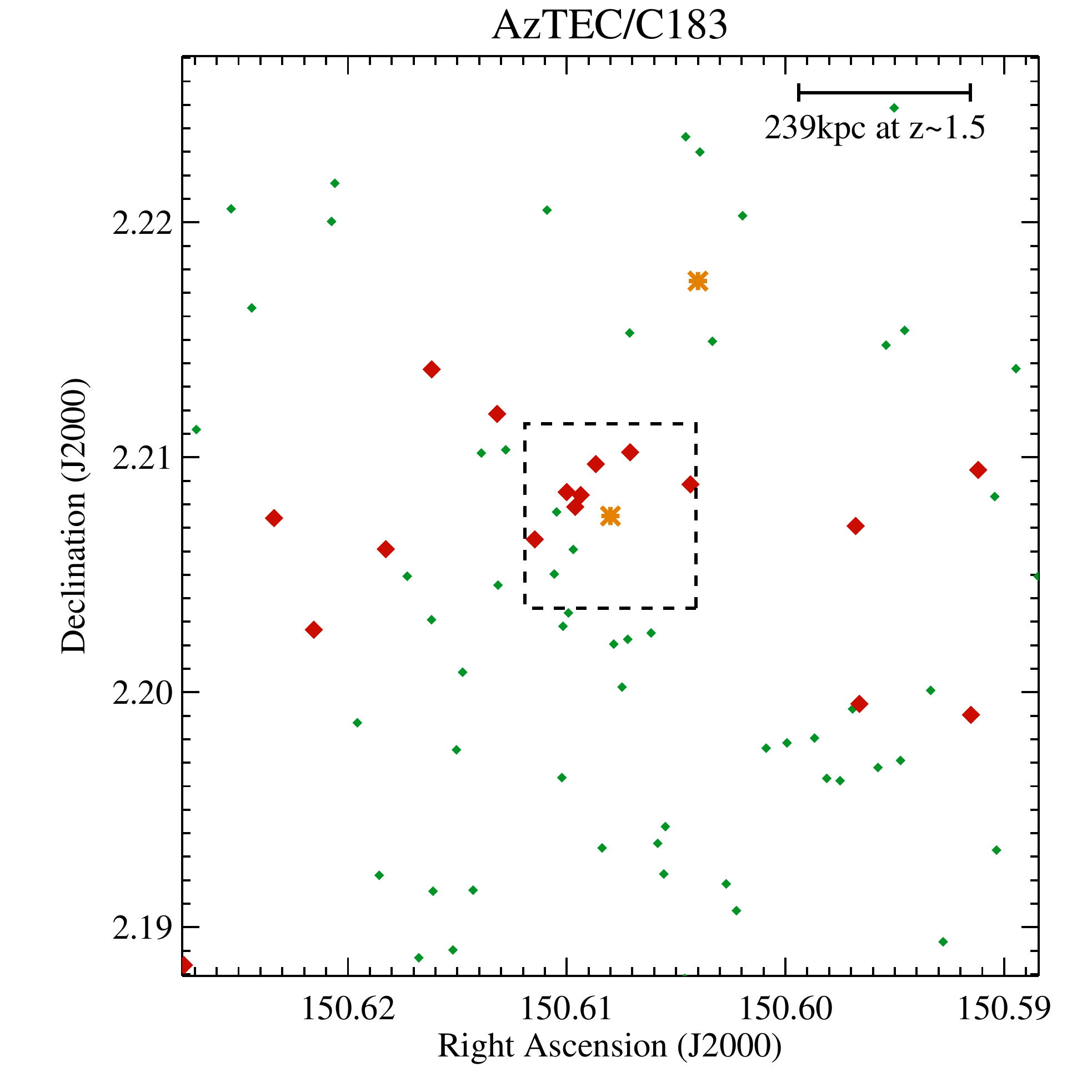}
\includegraphics[scale=0.26]{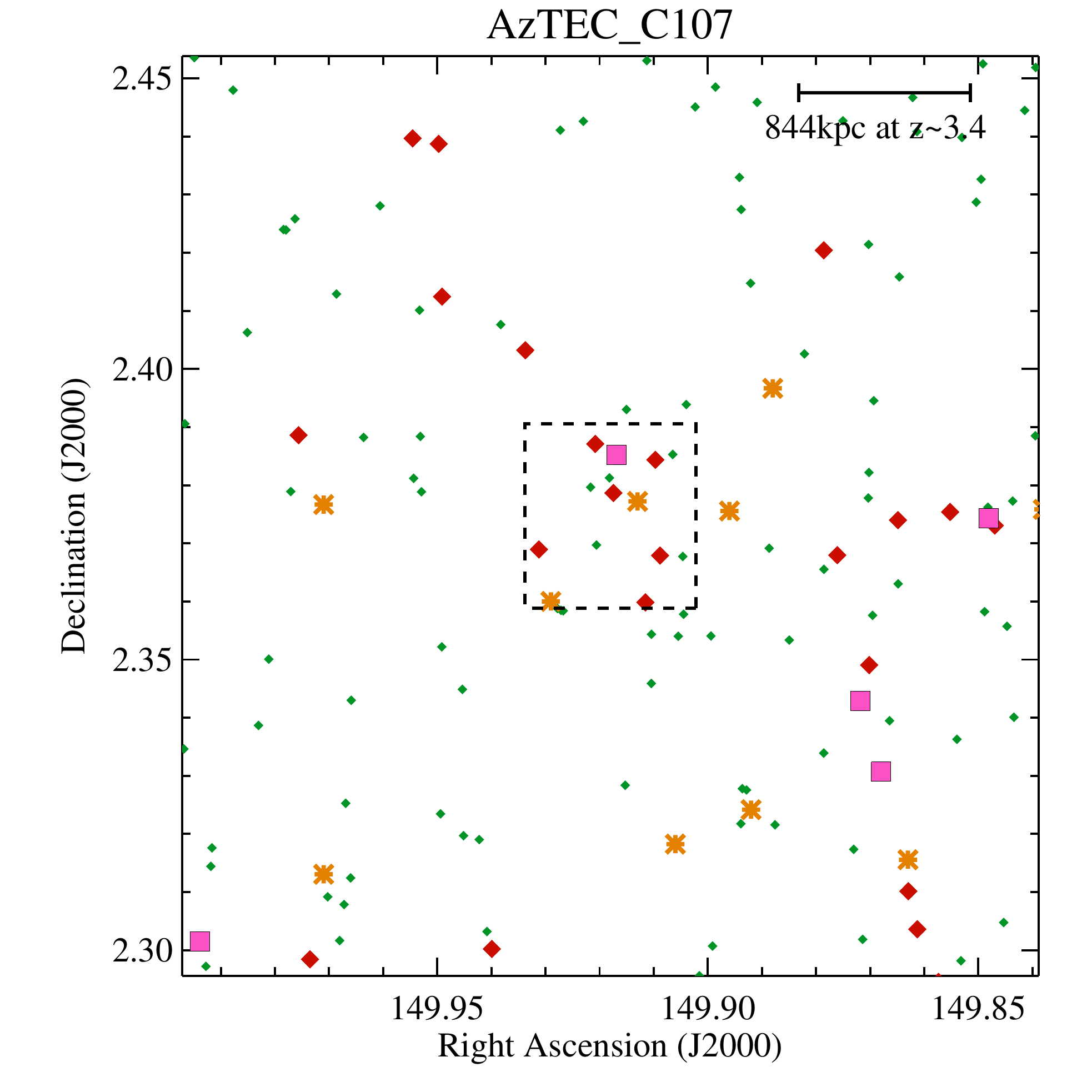}

\caption{Examples of over-densities found in the spatial region surrounding SMGs without known redshifts - full list of regions can be found in Table A1. Symbols are the same as Figure \ref{fig:a_10_2d}.}

\label{fig:pos_COS}
\end{center} 
\end{figure*}

\begin{figure*}
\begin{center}

\includegraphics[scale=0.27]{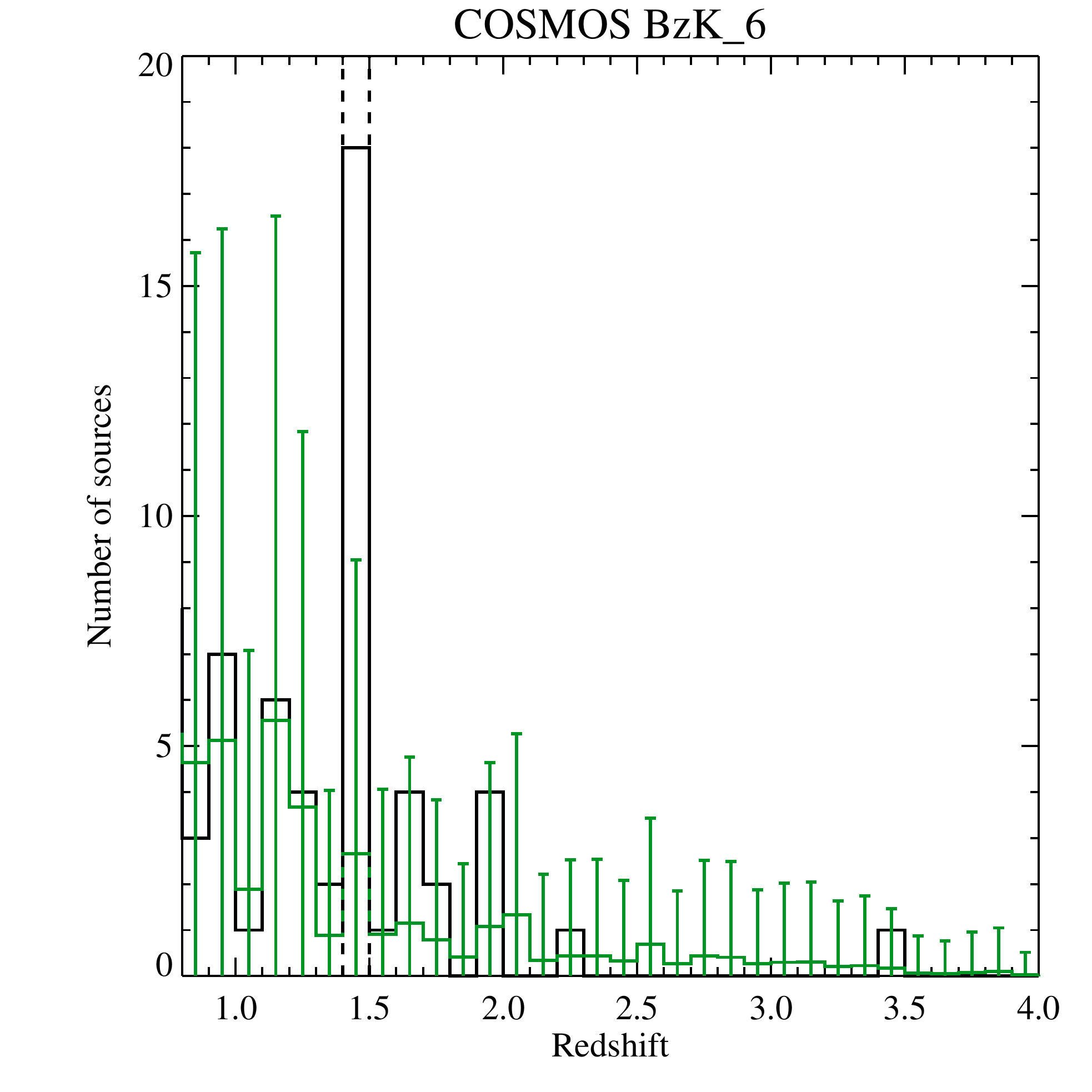}
\includegraphics[scale=0.27]{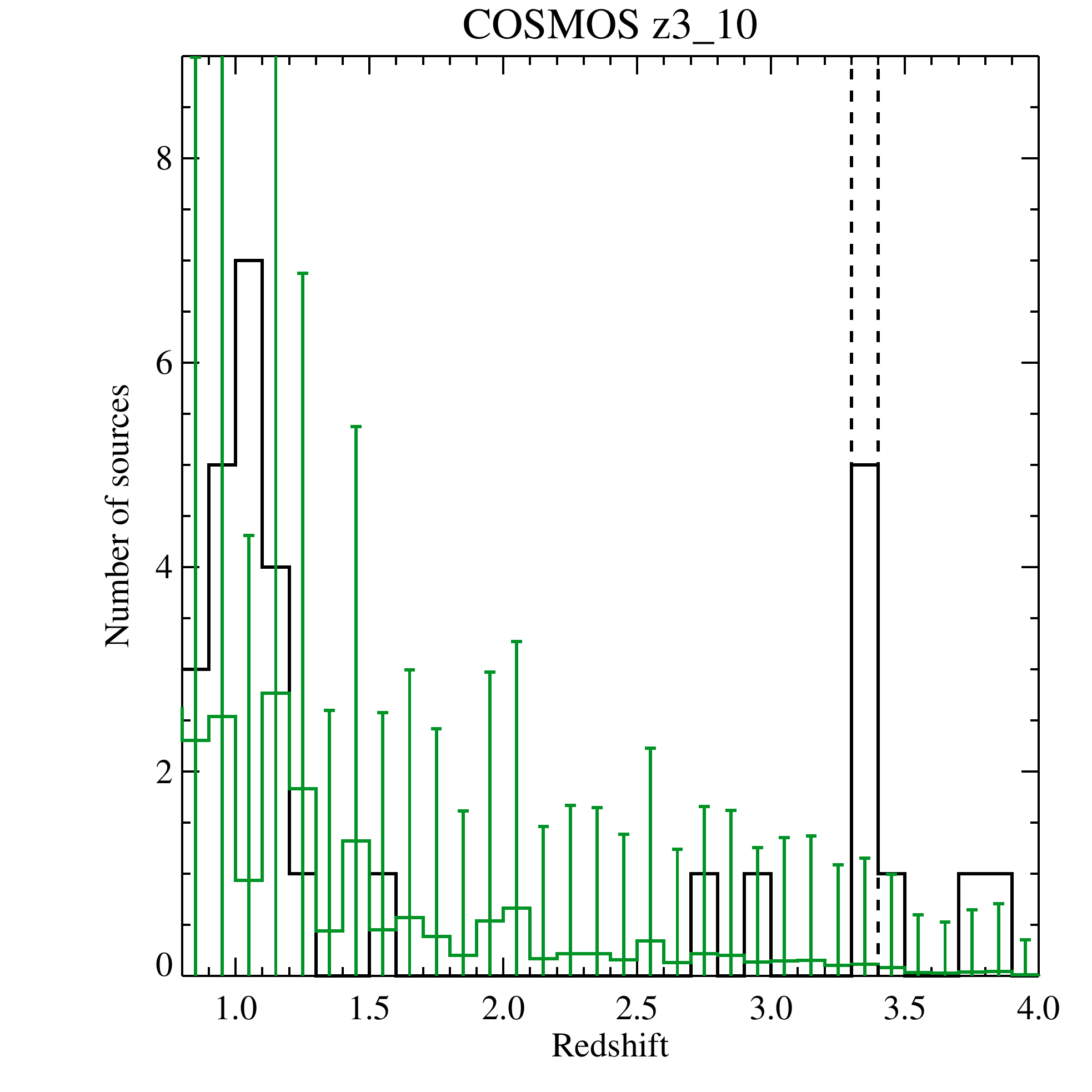}
\includegraphics[scale=0.27]{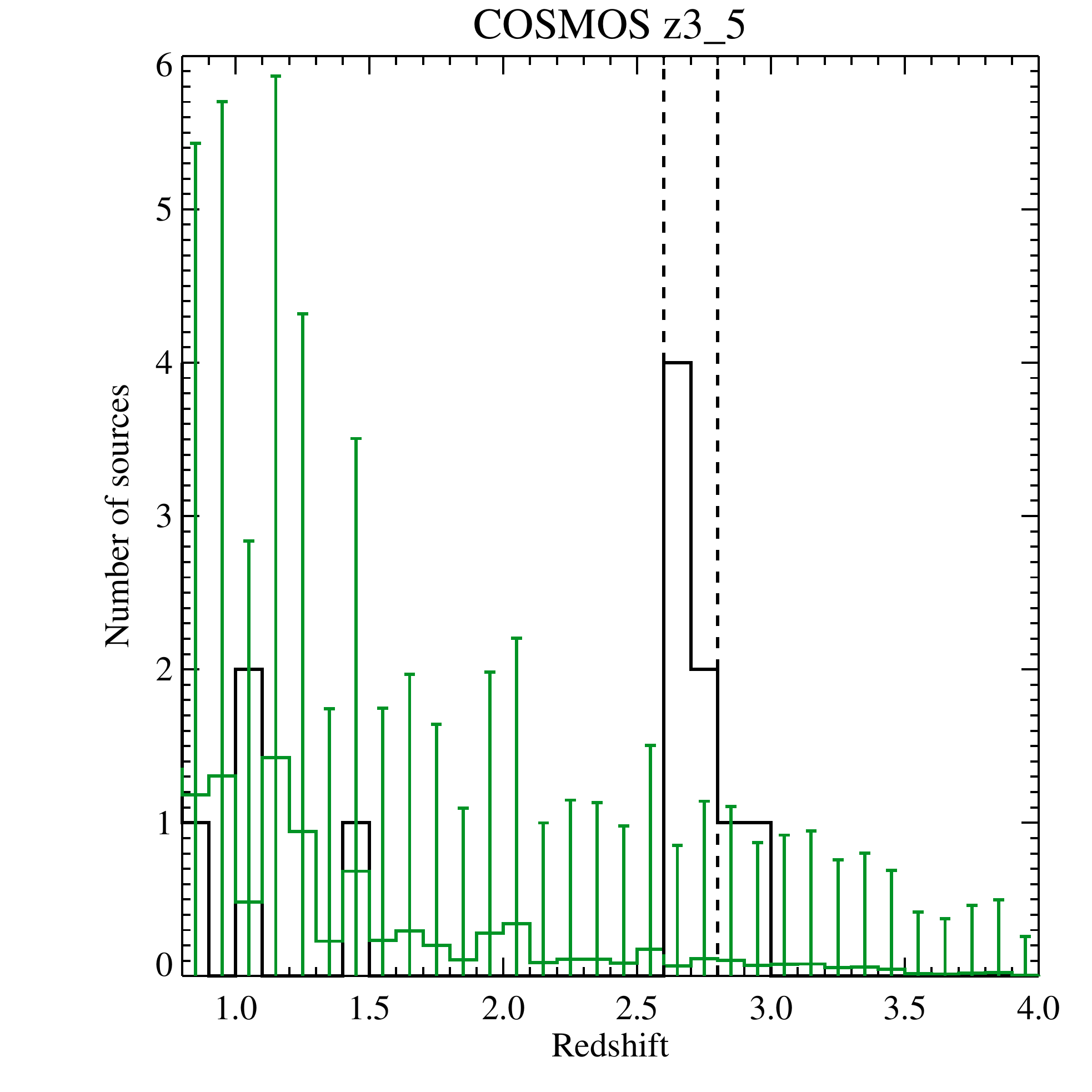}\\

\includegraphics[scale=0.27]{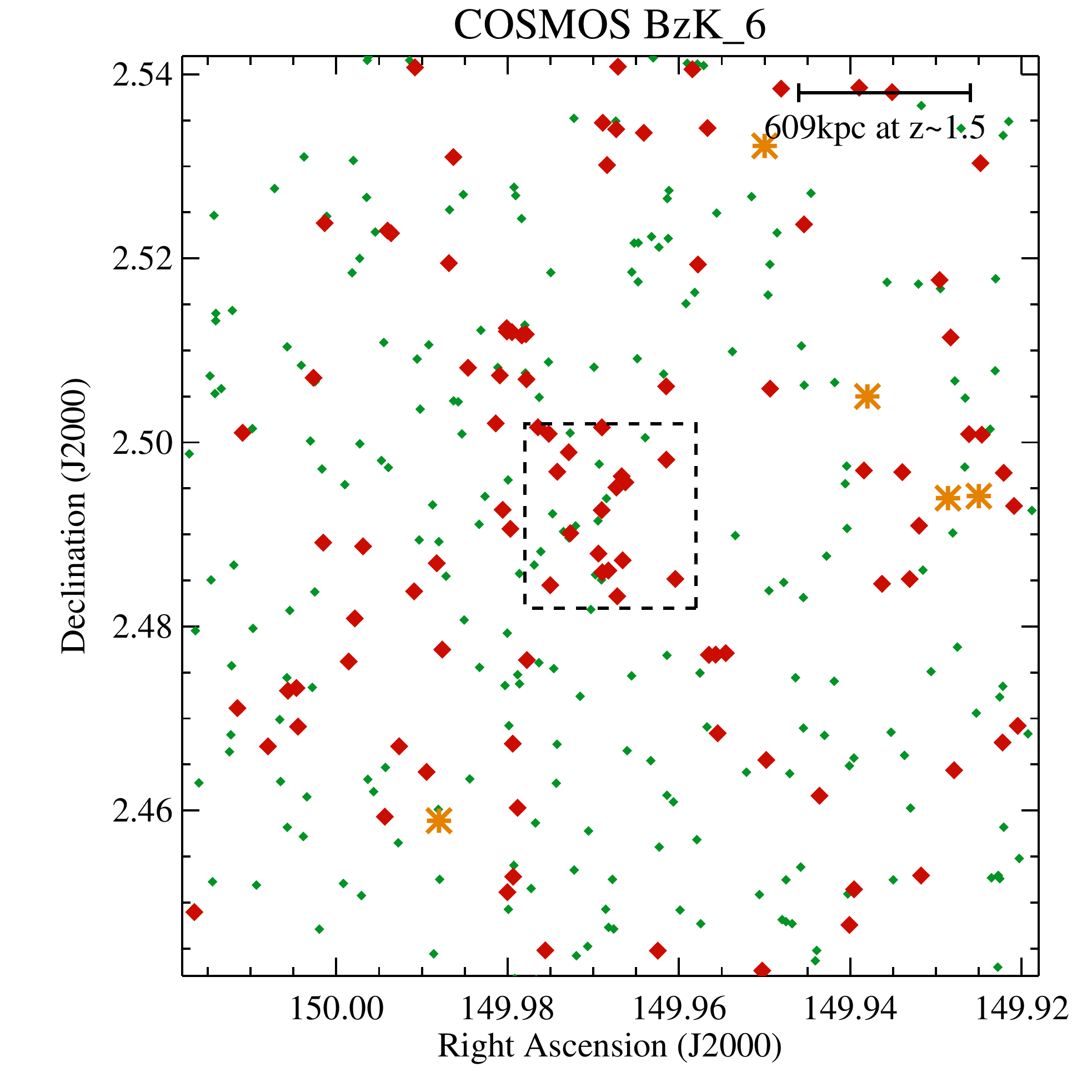}
\includegraphics[scale=0.27]{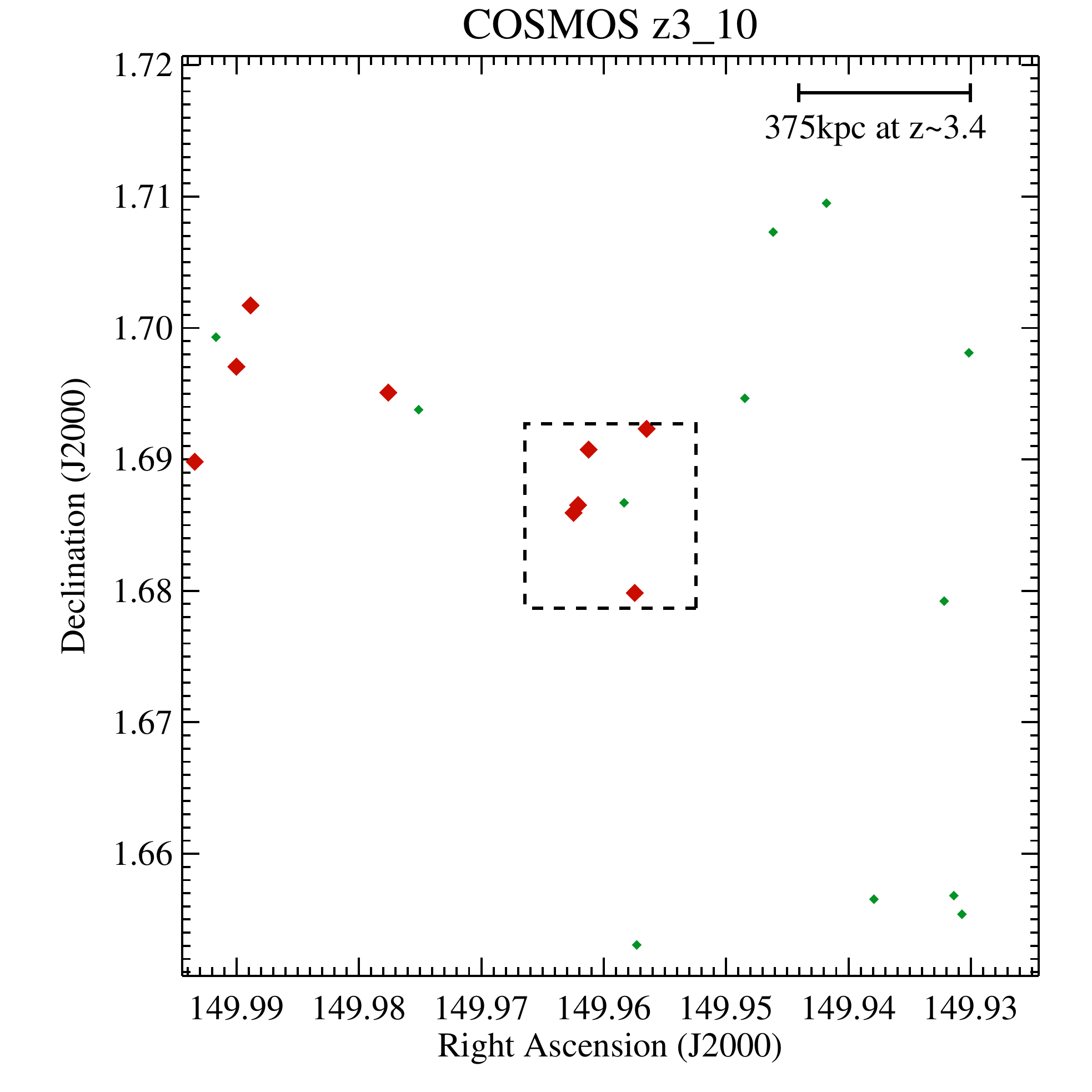}
\includegraphics[scale=0.27]{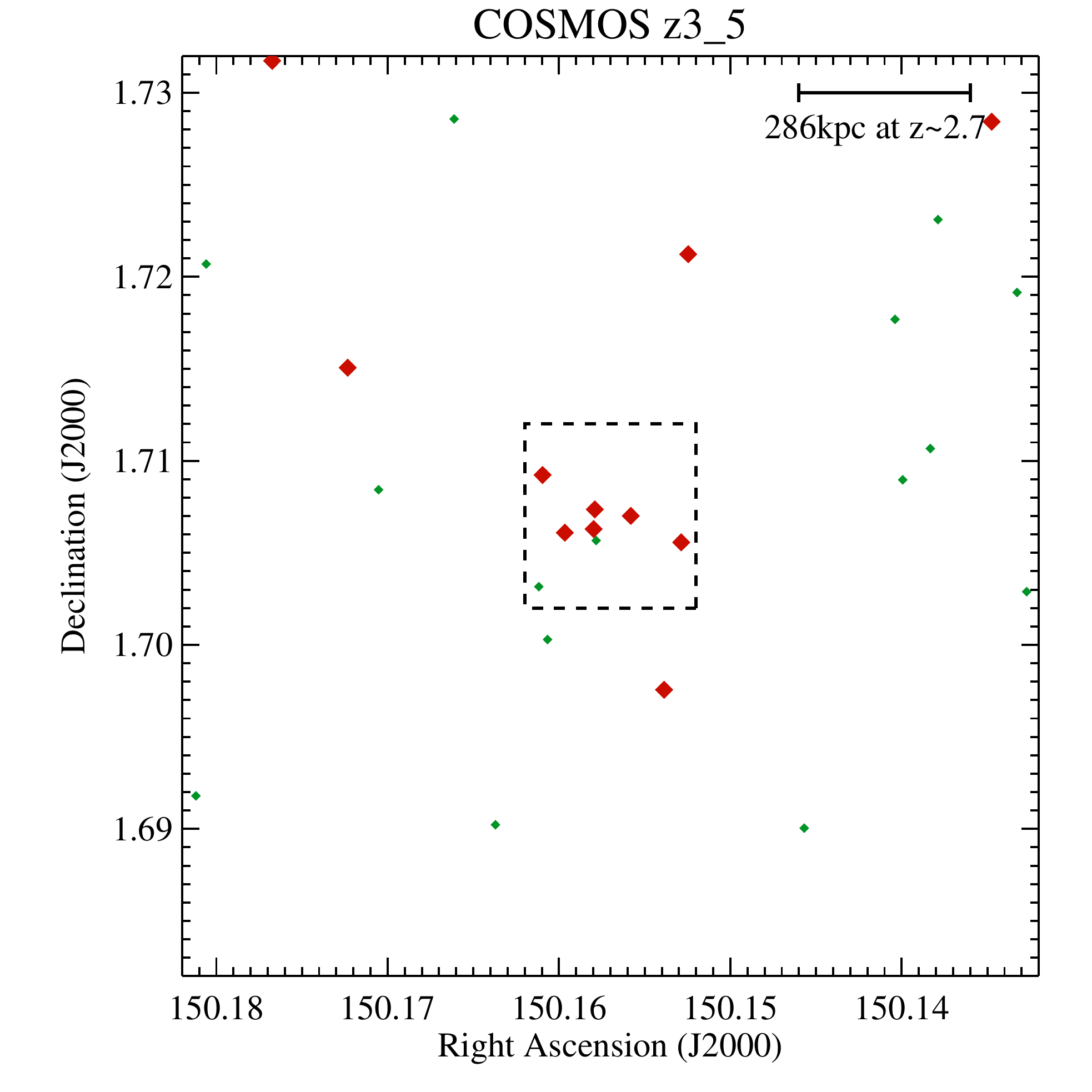}\\

\caption{Same as Figure \ref{fig:spec_COS} but for regions of high projected 2D number density and no SMG.}

\label{fig:pos_COS2}
\end{center} 
\end{figure*}

\begin{figure*}
\begin{center}

\includegraphics[scale=0.25]{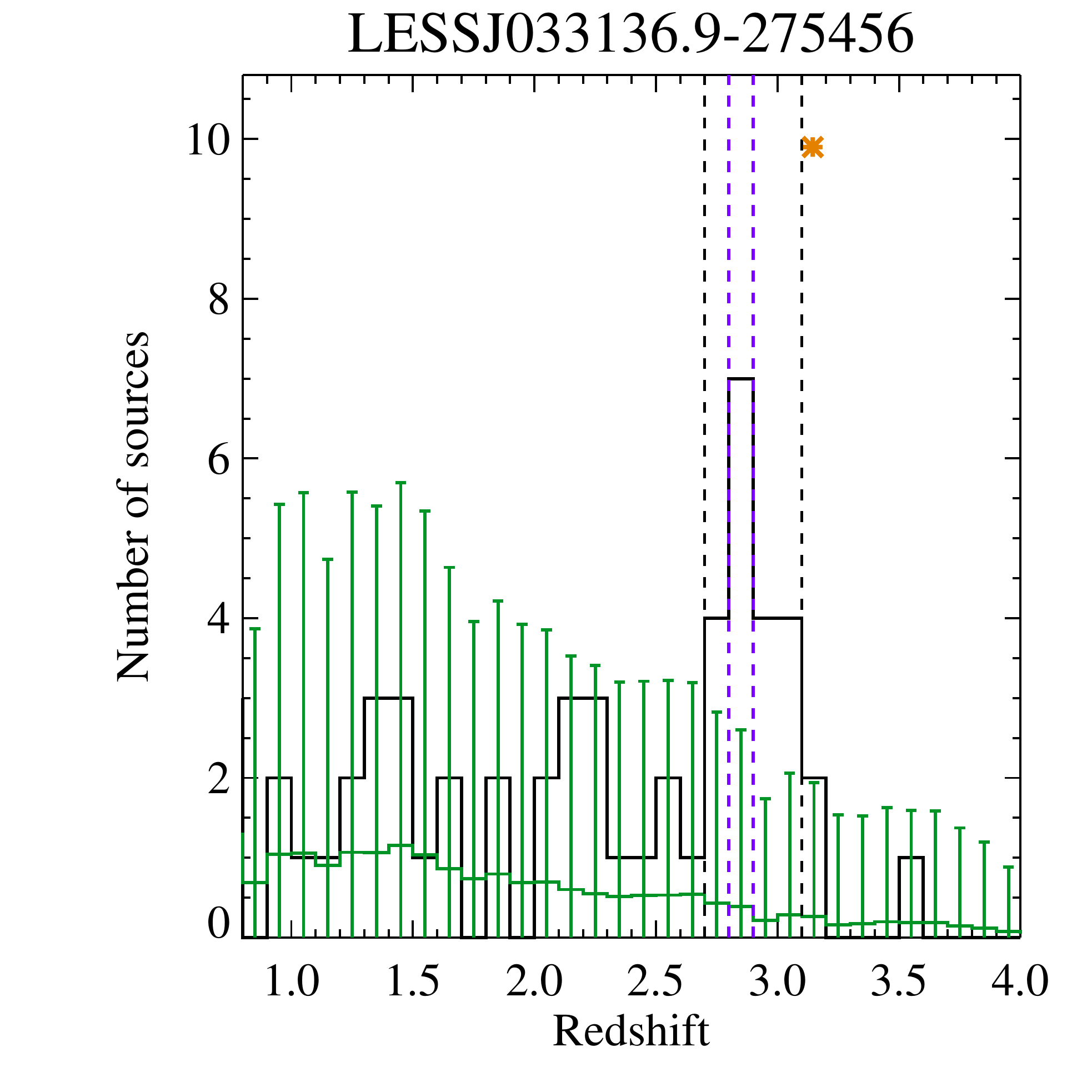}
\includegraphics[scale=0.25]{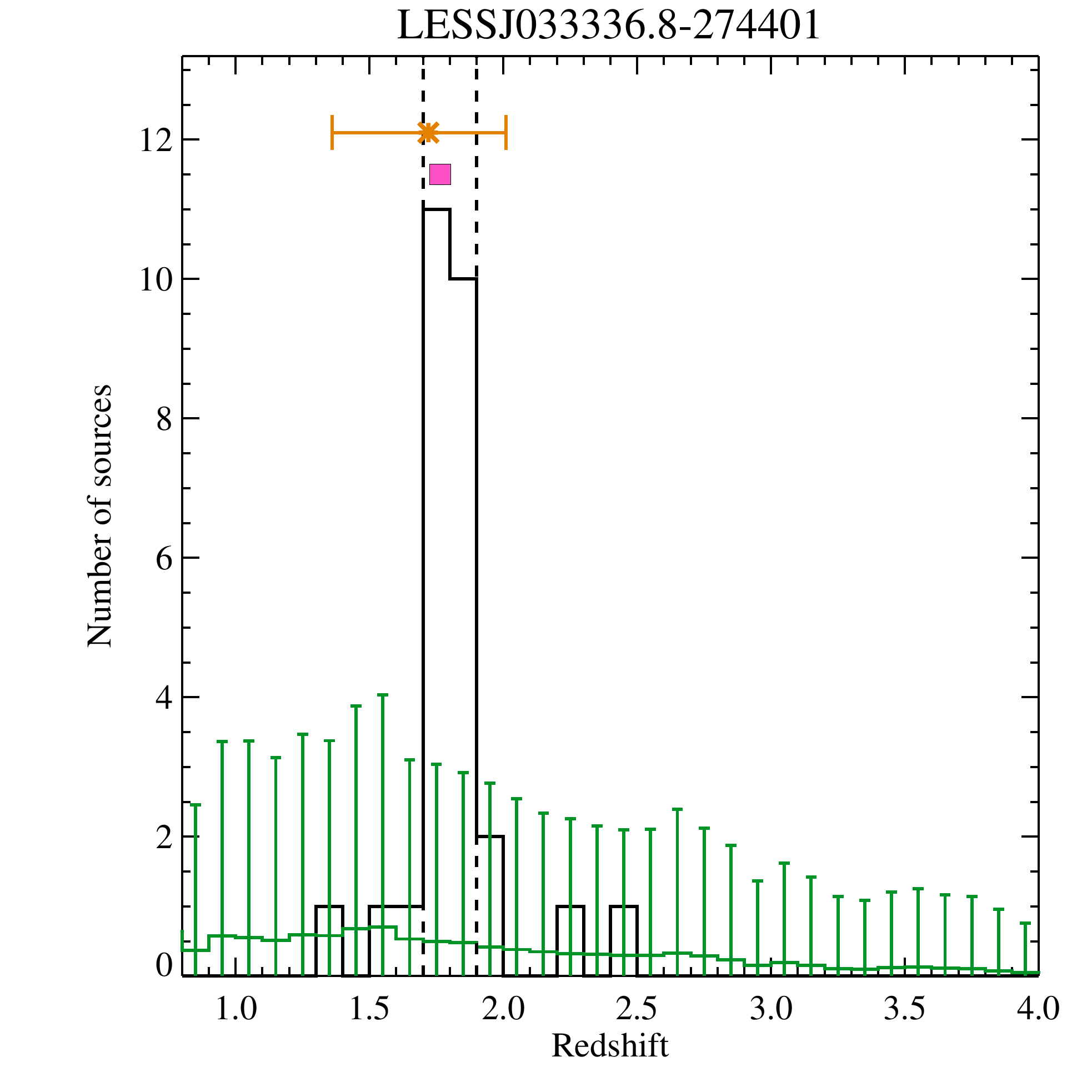}\\

\includegraphics[scale=0.25]{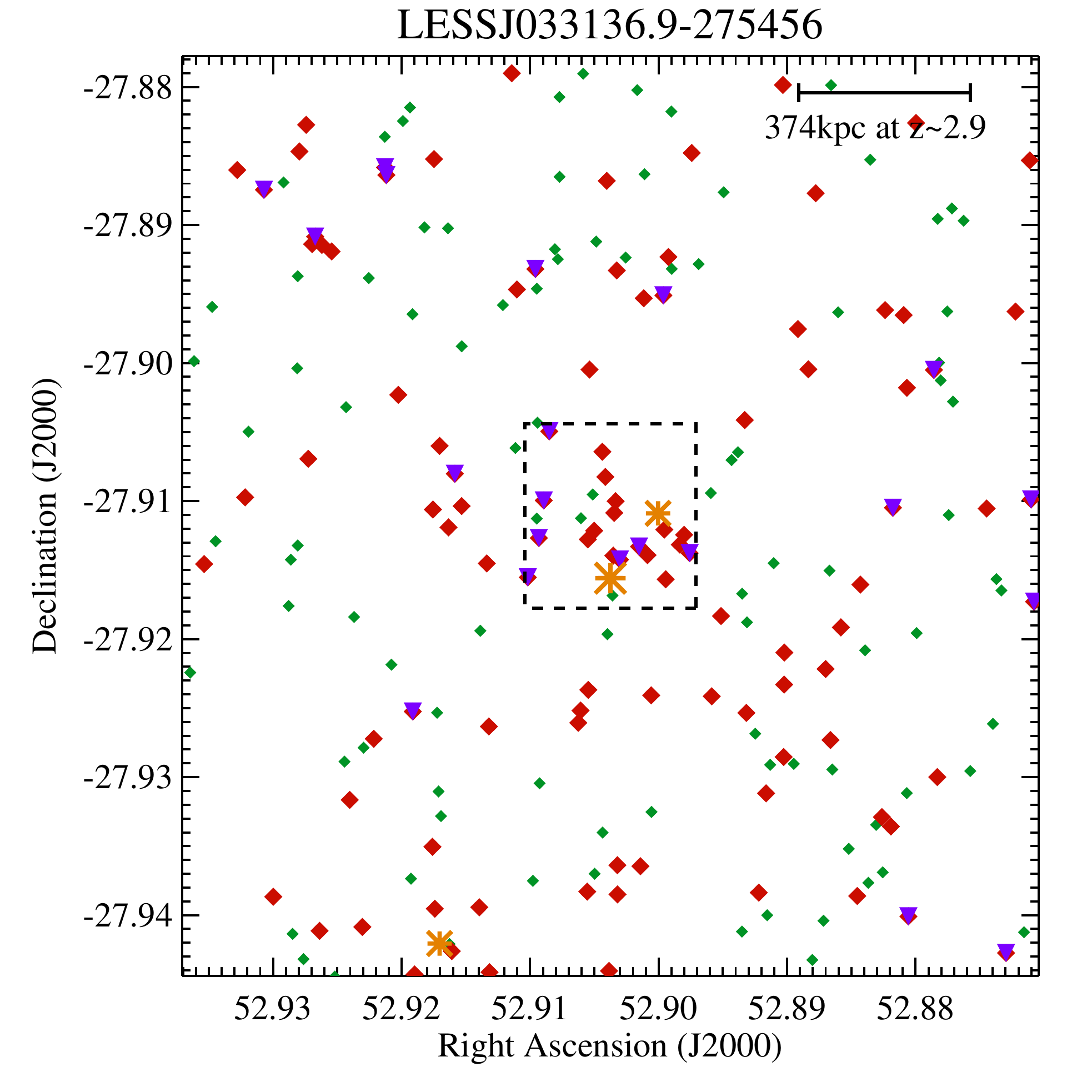}
\includegraphics[scale=0.25]{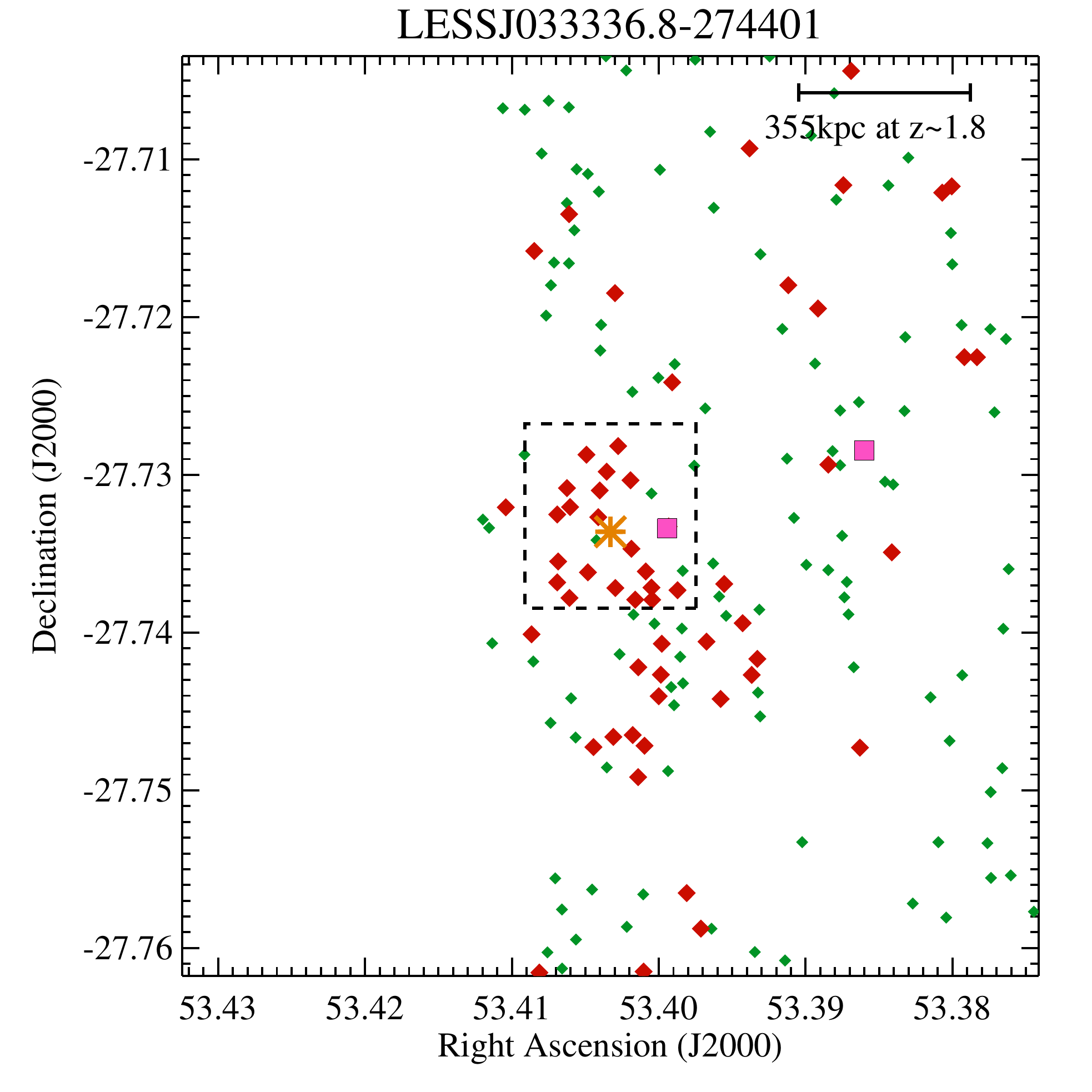}\\

\caption[]{The same as Figure \ref{fig:pos_COS} but for two interesting regions in the ECDF-S. The light purple triangles in the left panel display sources in the $\Delta z=0.1$ peak of the distribution (bounded by the purple line in the redshift distribution), while sources the full $\Delta z=0.4$ over-density is displayed as red diamonds. }
\label{fig:pos_ECDFS}
\end{center} 
\end{figure*}

\begin{figure*}
\begin{center}

\includegraphics[scale=0.25]{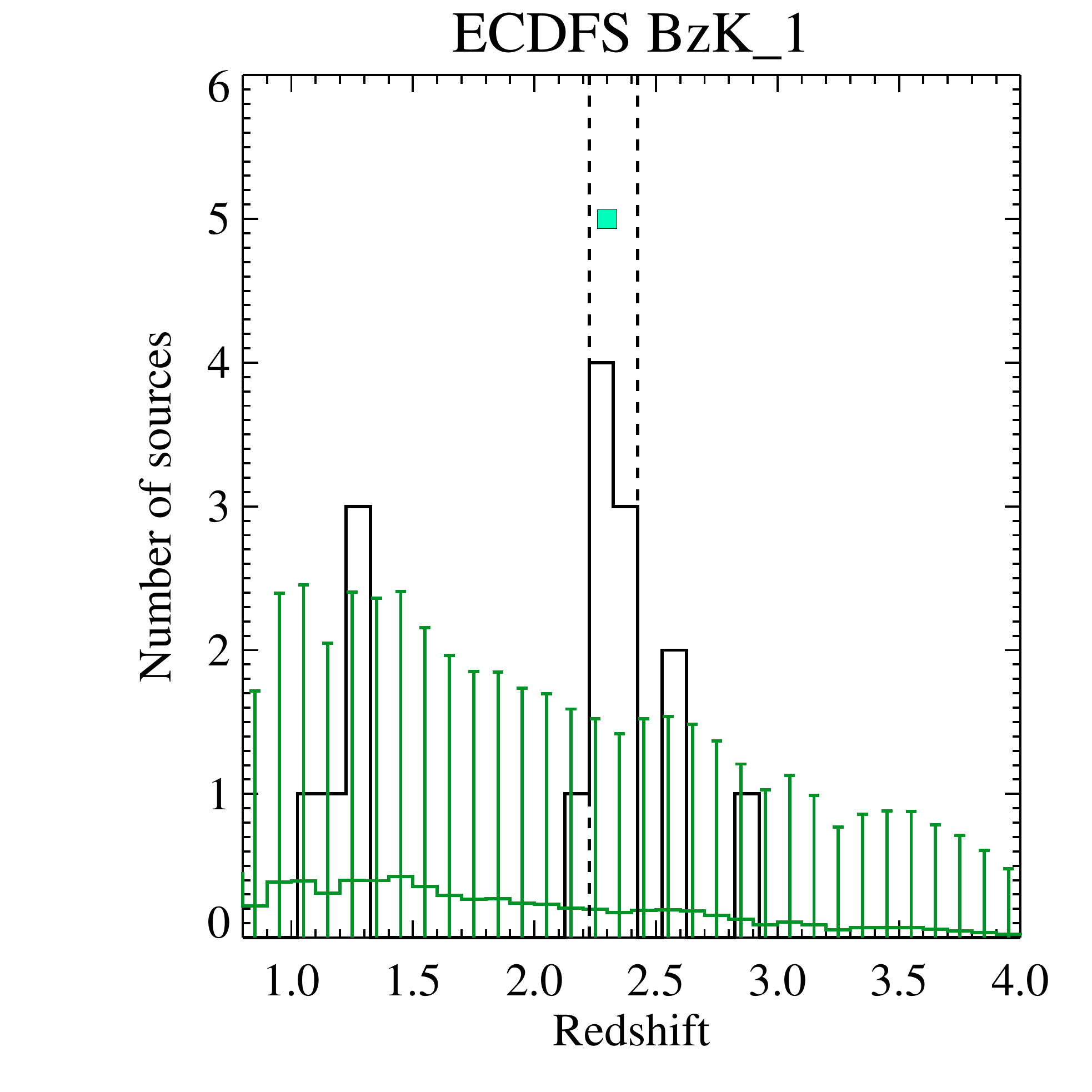}
\includegraphics[scale=0.25]{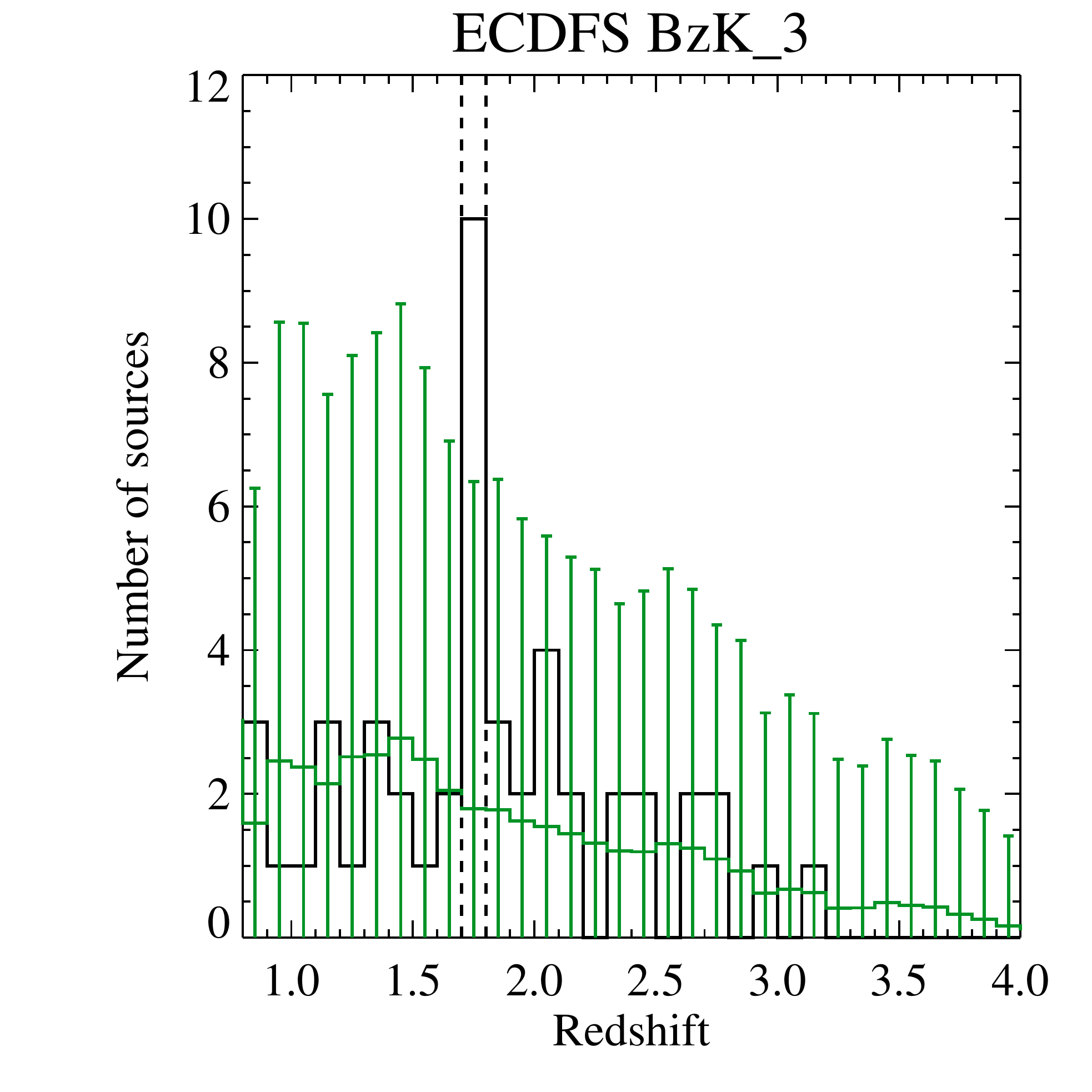}
\includegraphics[scale=0.25]{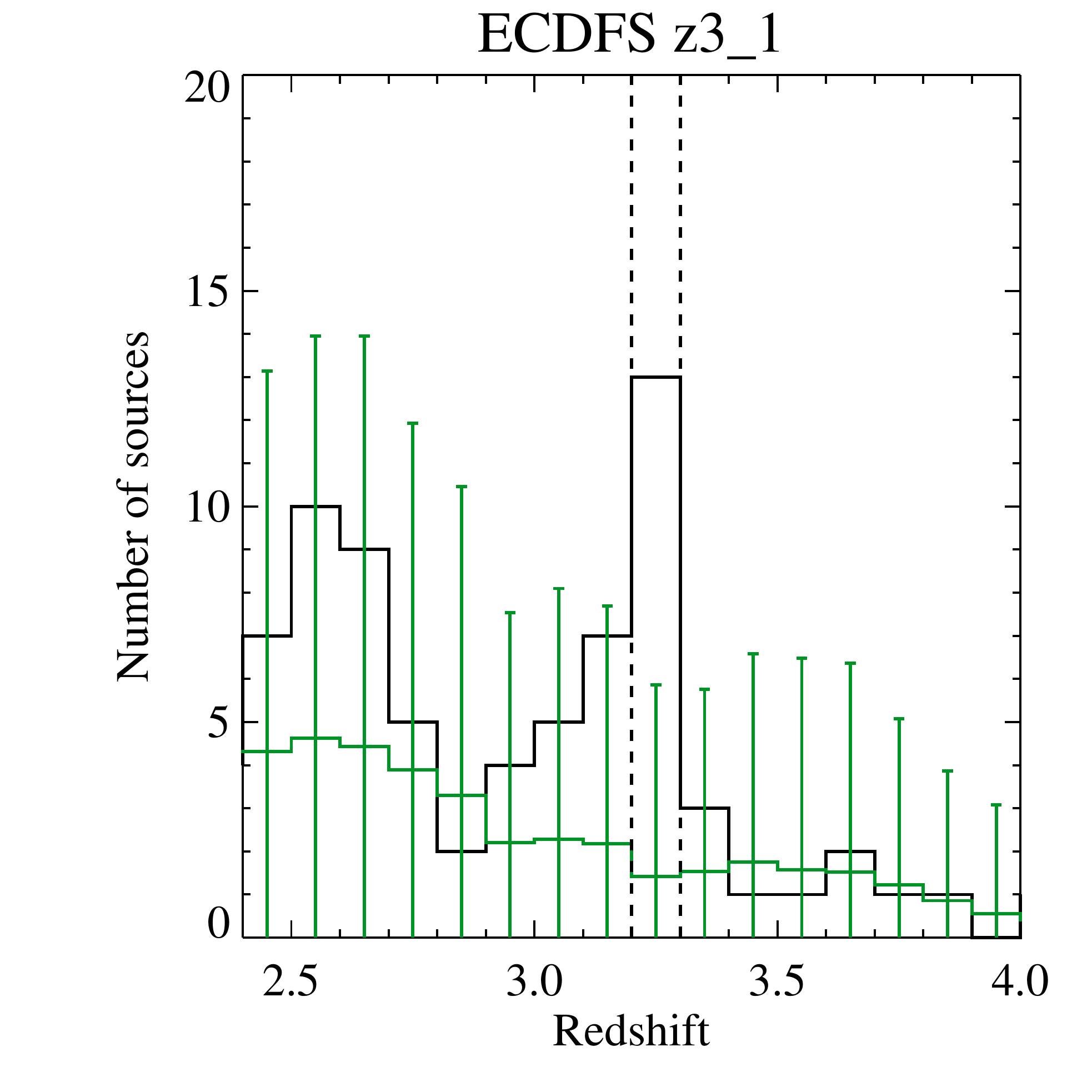}\\
\includegraphics[scale=0.25]{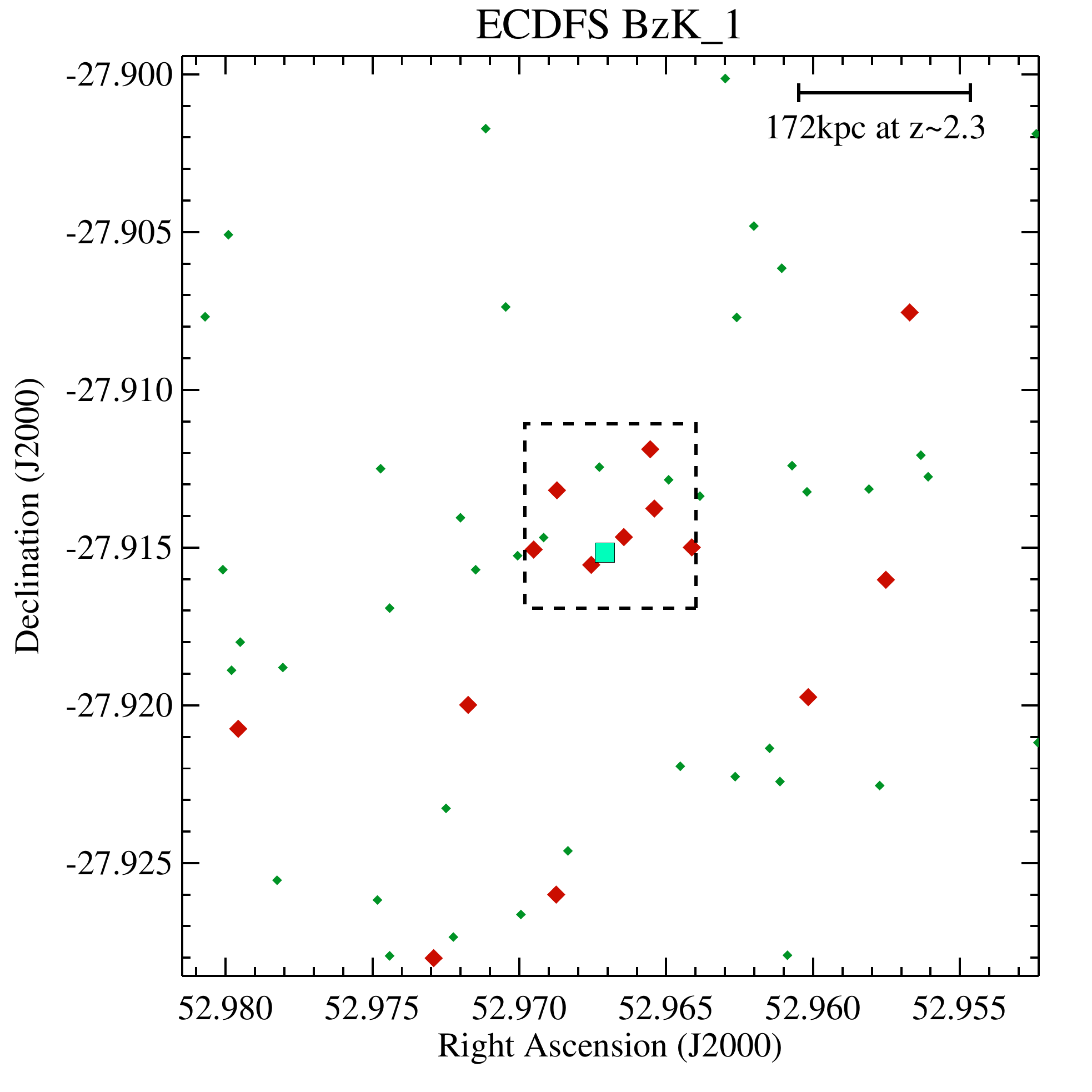}
\includegraphics[scale=0.25]{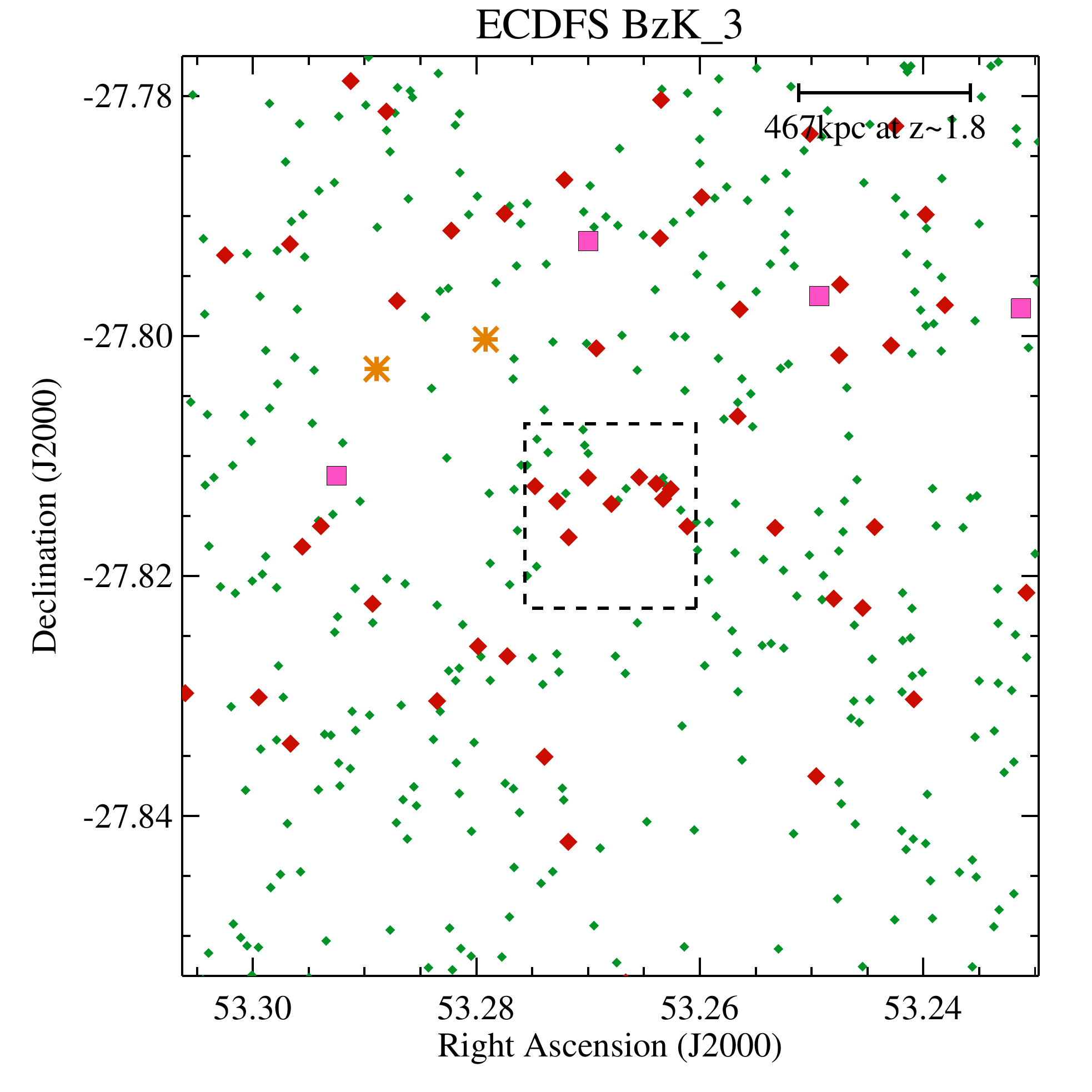}
\includegraphics[scale=0.25]{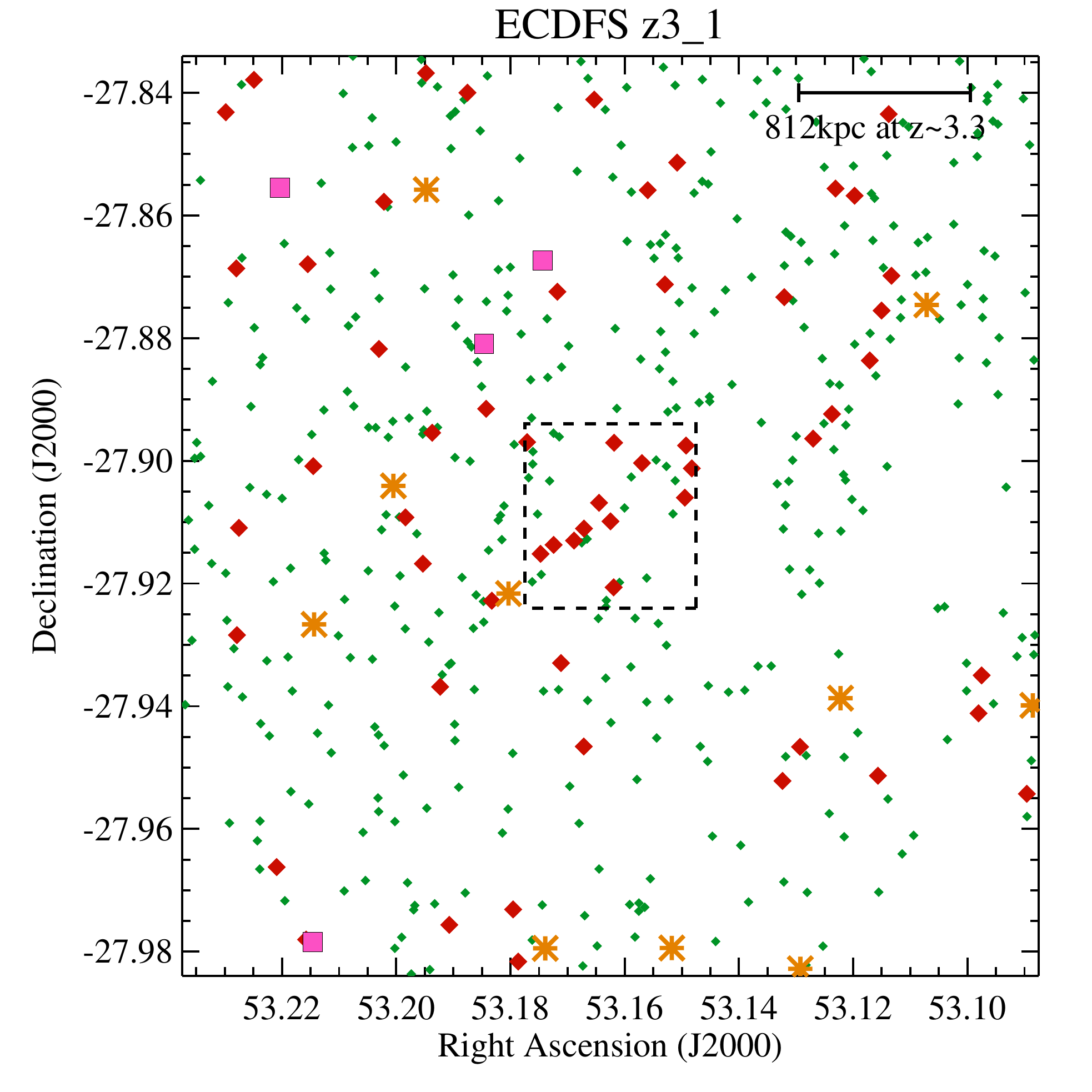}

\caption[]{The same as Figure \ref{fig:pos_COS2} but for regions of high 2D projected number density of sources in the ECDF-S.}
\label{fig:pos_ECDFS_no_sub}
\end{center} 
\end{figure*}

Below we discuss a number of the most interesting regions in our sample of over-densities in more detail. However, we note that the majority of the regions we have identified in this work are ideal candidates for further study. In the following discussion, we estimate the stellar masses of the member galaxies in our clustered regions at $1.4<z<2.5$ using their K-band magnitudes and the form outlined in \cite{Daddi04}:

\begin{equation}
log(M_{*}/10^{11}M_{\odot})=-0.4(K-K^{11})
\end{equation}

where $K^{11}$ is the K-band magnitude corresponding to an average stellar mass of $10^{11}$\,M$_{\odot}$ - K$^{11}$=20.14 in Vega (hence, we convert our AB magnitudes to Vega prior to estimating masses). For clustered regions at higher redshift we apply a similar scaling empirically derived from the SED estimated masses for $z\sim3$ LBGs of \cite{Shapley01}:

\begin{equation}
log(M_{*}/10^{11}M_{\odot})=-0.85(K-K^{11})
\end{equation} 

where $K^{11}$=20.28 in the AB system.

In addition, we also investigate the potential fate of the structures identified in this work in a $\Lambda$CDM Universe. In order to do this we select over-densities of comparable size, galaxy numbers and redshift in the Millennium Run Simulation (MR) semi analytic light cones of \cite{Guo11} and \cite{Henriques12}. For each region we discuss here, we take all sources in the simulation within $\Delta z=\pm0.2$ of the over-density redshift (only including sources which are likely to be detected in the COSMOS and MUSYC data). We then split the simulation into bins of the same size as the over-dense region, both spatially and in redshift. We identify any bin in the simulation which has at least the number of galaxies in our over-dense region. If the region in question contains an SMG, we also apply the additional constraint that there must be a M$_{*}>10^{10}$\,M$_{\odot}$, SFR$>100$\,M$_{\odot}$\,yr$^{-1}$ galaxy in the bin for it to be selected, a proxy for the presence of an SMG (hereafter SMG$_{MR}$ - this is in fact a very conservative estimate for an SMG which is likely to be much more massive and more intensely star-forming). We then trace these systems forward in time and investigate their host dark matter halos at $z=0$.

\subsubsection{1HERMESX24J100046.31+024132}

The region surrounding 1HERMESX24J100046.31+024132 contains ten sources in $\Delta z_{photo}=0.2$ range centered at $z\sim1.9$ (the SMG redshift). All of these sources fall in a compact $\sim0.6$\,Mpc diameter region around the SMG.  This represents a $\sim5\sigma$ over-density in two adjacent $\Delta z=0.1$ redshift bins on these scales and at this epoch. Therefore, the presence of this over-density and the SMG at an identical redshift suggests that this region is highly likely to be a forming massive structure at $z\sim1.9$. Assuming 1HERMESX24J100046.31+024132 has the typical stellar mass of an SMG at this redshift \citep[$\sim10^{11\pm0.5}$\,M$_{\odot}$, $e.g.$][]{Borys05,Hainline11} and estimating the total stellar mass of the K-band detected sources in the over-dense structure (as detailed above), we obtain a total mass of this system to be $\sim10^{11.6-11.9}$\,M$_{\odot}$.  While the total observed stellar mass is not sufficient to form a cluster core at $z=0$, this only represents the mass in those sources which have sufficiently high surface brightnesses to be observed in either the rest-frame FIR (the SMG) or the rest-frame UV (COSMOS sources require an I-Band detection for photometric redshifts). As such, we may be missing a significant fraction of the cluster population which are not detectable at these wavelengths. In addition, this region already contains a significant fraction of stellar mass required to form a cluster core at just $z\sim1.9$. Therefore, it is likely to grow considerably as it evolves to $z=0$. Considering similar regions in the MR simulation we identify just four regions with $\gtrsim$10 sources in a $\Delta z=0.1$ bin at $1.7<z<2.1$, which also contains an SMG$_{MR}$ source. Tracing these systems forward in time, we find that the mean halo mass that their constituent galaxies occupy at $z=0$ is $\sim10^{13}$\,M$_{\odot}$, but only $\sim20\%$ are hosted by a $>10^{13}$\,M$_{\odot}$ halo (one of the four structures evolves into a cluster). Hence, it is unclear whether or not regions like those surrounding 1HERMESX24J100046.31+024132 are likely to evolve into groups and clusters at low redshift.

\subsubsection{AzTEC/C183}

The AzTEC/C183 region contains a $\sim9\sigma$ over-density of seven sources in a $\Delta z_{photo}=0.1$ range centred at $z\sim1.45$. This region is intriguing as all seven sources fall in a compact $\sim250$\,kpc spatial region with a submm bright galaxy in the centre of the field of view. While the redshift of the SMG is unknown, it is potentially associated with these galaxies. In the following discussion we shall assume that this is the case. Once again taking the typical stellar mass of an SMG at this redshift we predict a total stellar mass of $\sim10^{12.2 -12.3}$\,M$_{\odot}$ in the region (already comparable to the most massive galaxies at $z=0$).  Assuming these systems are at an identical redshift, they could easily condense to form a single massive galaxy by just $z=1.35$ (assuming typical galaxy velocities of $\sim$500\,km/s) and later form the central region of a group or cluster, through accretion of other objects. Identifying similar structures in the MR simulation we find that no region of the simulation contains both seven galaxies and an SMG$_{MR}$ source in a $0.47^{\prime} \times 0.47^{\prime}   \times \Delta z=0.1$ volume at $1.25<z<1.65$. This is surprising and suggests that either 1) AzTEC/C183 is not truly associated with the over-density galaxies in this region, 2) that the galaxies in this region are not truly clustered in redshift as the $z_{photo}$ distribution suggests, 3) that we have not successfully identified comparable regions in the simulation, or 4) that the semi-analytics do not accurately produce the most extreme over-densities in the galaxy population at this redshift (see Section \ref{sec:LESS24}). If we relax this criteria and do not require the presence of a SMG$_{MR}$ source, we find 7 AzTEC/C183 like volumes in the simulation. Tracing these forward, we find that the mean mass halo hosting the galaxies in these structures at $z=0$ is $\sim10^{13.1}$\,M$_{\odot}$. However, once again we find that only $14\%$ of these galaxies are contained in a $>10^{13}$\,M$_{\odot}$ halo.

\subsubsection{COSMOS z3\_10}

While lower in number counts than the previously discussed clustered regions and not associated with an SMG, the COSMOS z3\_10 region displays a highly significant over-density of sources spanning $\Delta z=0.1$ in redshift at a much earlier epoch ($z\sim3.35$). The system contains 5 sources all within a compact 50$^{\prime\prime}$ ($\sim380$\,kpc at $z\sim3.35$) region. This represents a $>17\sigma$ over-density for its redshift and on these scales. The close proximity of such a large number of sources at this epoch suggests that we are once again witnessing a massive galaxy or group in formation.        
The total observed K-band estimated mass of this system is  $\sim10^{11.2}$\,M$_{\odot}$ - not yet large enough to form a massive galaxy. However, if we identify similar regions in the MR simulation and trace these structures forward, we find that $\sim45\%$ of the galaxies in these systems will reside in a $>10^{13}$\,M$_{\odot}$ halo at $z=0$ - a much higher fraction than the SMG over-densities discussed previously.

\subsubsection{COSMOS z3\_13/COSMOS z3\_14}

These two regions both contain three sources in an incredibly small projected volume at $z\sim3.25$ and $z\sim2.95$ respectively. While the number counts in these regions are small, the close proximity of the sources ($\sim$6$^{\prime\prime}$ or 46\,kpc at $z\sim3.25$ for COSMOS z3\_14) make them ideal candidates for a forming galaxy at high redshift. They contain $>20 \times$ the predicted number of sources in a field of this size at their redshift and hence are highly significant over-densities. We use the MR simulation to identify similar regions and find 23 such systems. Tracing these forward in time we find that $\sim50\%$ of the galaxies in such structures will reside in a  $>10^{13}$\,M$_{\odot}$ halo at $z=0$ -  once again a much higher fraction than the SMG over-densities. This is surprising and, coupled with the predictions from the previous regions, suggests that potentially, the most efficient method of targeting the sites of early group and cluster formation is to search for highly significant over-densities of a small number of sources on small projected angular scales at high redshift. In a $\Lambda$CDM frame work such structures are more likely to reside in groups and clusters at $z=0$ than more rich systems at later epochs - including those which contain an SMG.

\subsubsection{LESSJ033136.9-275456}

The LESSJ033136.9-275456 region contains a significant over-density of sources spanning $\Delta z=0.4$ in redshift and peaking at $z\sim2.9$. All four $\Delta z=0.1$ redshift bins in this region are $>4 \times $ the mean predicted at their redshift, with the central peak displaying a $\sim9 \times$over-density (7 sources). The structure constrains a total of 19 galaxies at $2.7<z<3.1$ all within a 42$^{\prime\prime}$ ($\sim330$\,kpc at $z\sim2.9$) region. This region also contains two submm bright galaxies, one with a redshift consistent with the very top end of the over-density ($z=3.14$) and the second with a unknown redshift. In order to roughly estimate the redshift of the second SMG, we consider its FIR SED derived from the $Herschel$ SPIRE \citep[taken from the $Herschel$ Multitiered Extragalactic survey, HerMES,][]{Oliver12}  and LESS maps of the field. For reasonable assumptions of dust temperature (38K$<T_{dust} <$42K) and power law emissivity index ($\beta =2.0$), the dust SED is best fit by a model at $2.3 < z < 3.0$ - consistent with the over-density.   

Assuming both SMGs are associated with this over-density, and the typical stellar mass of an SMG at this redshift, we observe a total K-band stellar mass of $\sim10^{12.3}$\,M$_{\odot}$. Therefore, the stellar mass in this region is already comparable to the most massive galaxies $z=0$. We also find no structures in the MR simulation with the same volume density of sources as the region surrounding  LESSJ033136.9-275456, even without including the constraint of the presence of an SMG$_{MR}$ source ($i.e.$ there are no $42^{\prime\prime} \times 42^{\prime\prime} \times \Delta z=0.4$ volumes in the simulation at $z=2.9\pm0.2$ which contain $\gtrsim19$ sources). This is interesting given that we no longer have any potential error introduced in selection of the MR simulation regions via the presence of an SMG$_{MR}$ source (as  there may have been for AzTEC/C183). This suggests that in the case of LESSJ033136.9-275456 either the galaxies are not truly clustered as the photometric redshift distribution would imply, or the most over-dense structures at this redshift are inconsistent with those predicted by the MR simulation semi-analytic models.

\begin{figure*}
\begin{center}

\includegraphics[scale=0.9]{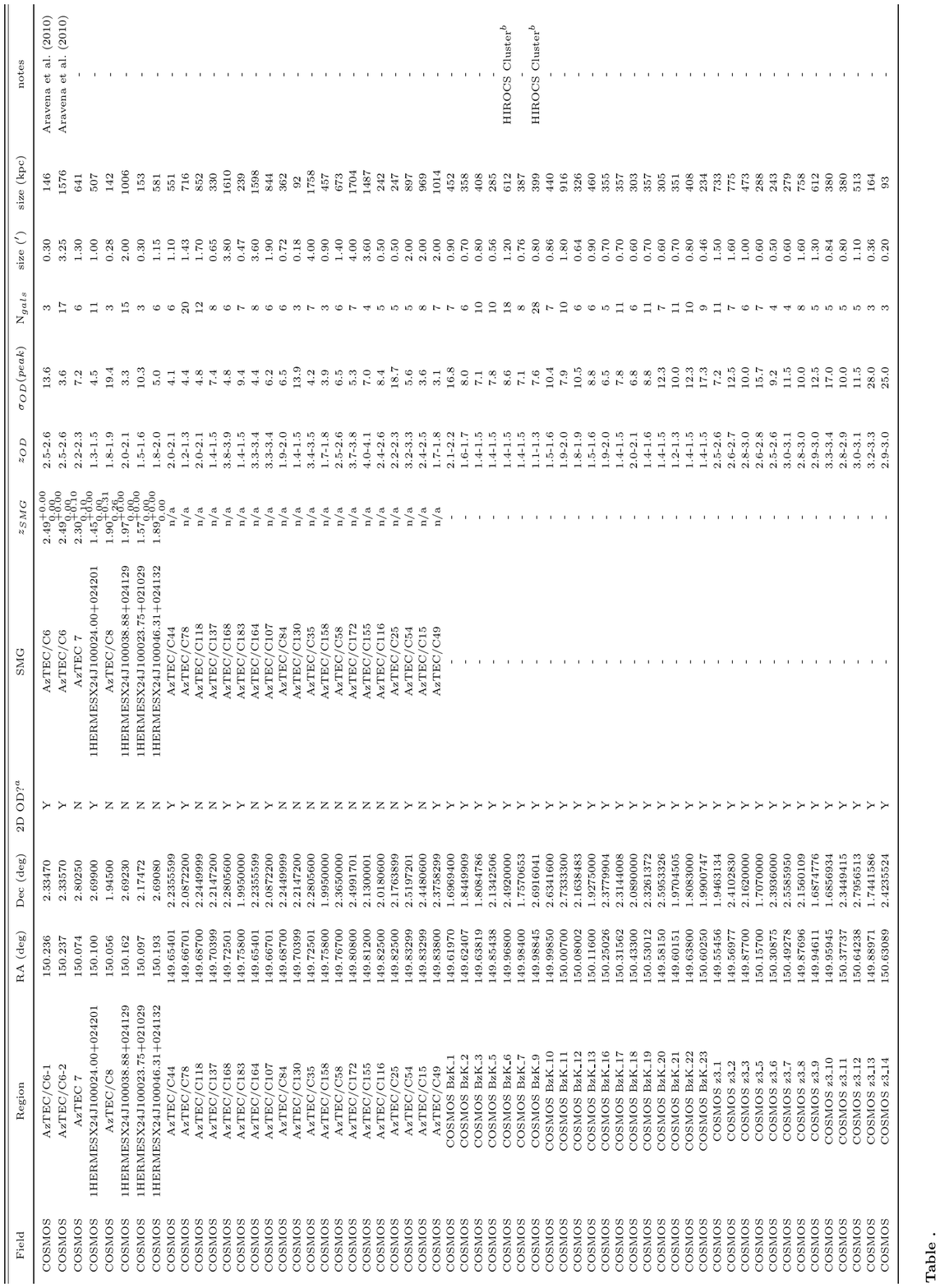}

\caption*{}
\label{tab:props}
\end{center} 
\end{figure*}

\begin{figure*}
\begin{center}

\includegraphics[scale=0.9]{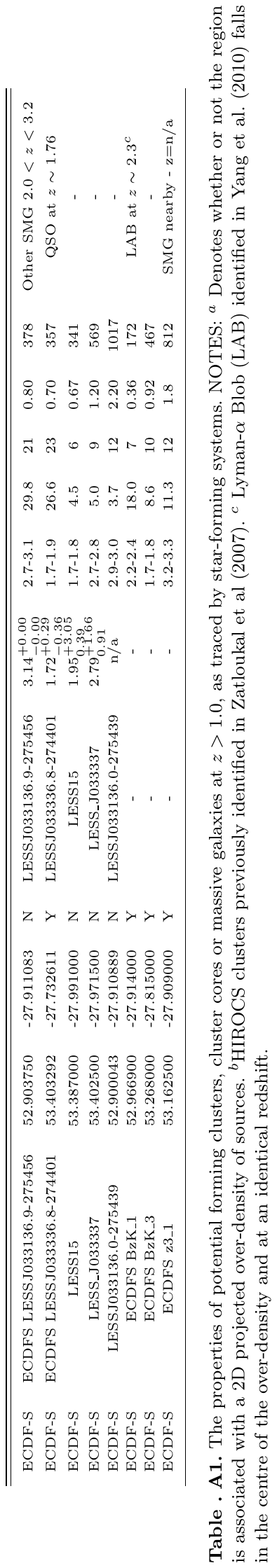}

\caption*{}
\label{tab:props}
\end{center} 
\end{figure*}


\begin{thebibliography}{}



\bibitem[Allen et al.(2011)]{Allen11} Allen, S.~W., Evrard, A.~E., \& Mantz, A.~B.\ 2011, \araa, 49, 409 


\bibitem[\protect\citeauthoryear{Angulo et al.}{2012}]{Angulo12} 
Angulo R.~E., Springel V., White S.~D.~M., Jenkins A., Baugh C.~M., Frenk 
C.~S., 2012, MNRAS, 426, 2046 

\bibitem[Aretxaga et al.(2011)]{Aretxaga11} Aretxaga, I., Wilson, 
G.~W., Aguilar, E., et al.\ 2011, \mnras, 415, 3831 

\bibitem[Adelberger et al.(2004)]{Adelberger04} Adelberger, K.~L., 
Steidel, C.~C., Shapley, A.~E., et al.\ 2004, \apj, 607, 226 


\bibitem[Aravena et al.(2010)]{Aravena10} Aravena, M., Bertoldi, 
F., Carilli, C., et al.\ 2010, \apjl, 708, L36


\bibitem[Bassett et al.(2013)]{Bassett13} Bassett, R., Papovich, 
C., Lotz, J.~M., et al.\ 2013, \apj, 770, 58 

\bibitem[Bauer et al.(2011)]{Bauer11} Bauer, A.~E., 
Gr{\"u}tzbauch, R., J{\o}rgensen, I., Varela, J., 
\& Bergmann, M.\ 2011, \mnras, 411, 2009 

\bibitem[Bertoldi et al.(2007)]{Bertoldi07} Bertoldi, F., Carilli, 
C., Aravena, M., et al.\ 2007, \apjs, 172, 132  

\bibitem[Biggs et al.(2011)]{Biggs10} Biggs, A.~D., Ivison, 
R.~J., Ibar, E., et al.\ 2011, \mnras, 413, 2314 


\bibitem[Borys et al.(2005)]{Borys05} Borys, C., Smail, I., 
Chapman, S.~C., et al.\ 2005, \apj, 635, 853 

\bibitem[\protect\citeauthoryear{Bower, Kodama, 
\& Terlevich}{1998}]{Bower98} Bower R.~G., Kodama T., Terlevich A., 1998, MNRAS, 299, 1193


\bibitem[Capak et al.(2008)]{Capak08} Capak, P., Aussel, H., 
Ajiki, M., et al.\ 2008, VizieR Online Data Catalog, 2284, 0

\bibitem[Capak et al.(2011)]{Capak11} Capak, P.~L., Riechers, 
D., Scoville, N.~Z., et al.\ 2011, \nat, 470, 233 

\bibitem[Cardamone et al.(2010)]{Cardamone10} Cardamone, C.~N., van 
Dokkum, P.~G., Urry, C.~M., et al.\ 2010, \apjs, 189, 270 

\bibitem[Casey et al.(2011)]{Casey2011} Casey, C.~M., Chapman, 
S.~C., Smail, I., et al.\ 2011, \mnras, 411, 2739 

\bibitem[\protect\citeauthoryear{Casey et al.}{2012a}]{Casey12a} 
Casey C.~M., et al., 2012a, \apj, 761, 139 

\bibitem[Casey et al.(2012b)]{Casey12b} Casey, C.~M., Berta, S., 
B{\'e}thermin, M., et al.\ 2012b, \apj, 761, 140 

\bibitem[\protect\citeauthoryear{Casey et al.}{2013}]{Casey13} 
Casey C.~M., et al., 2013, MNRAS, 2487


\bibitem[Castellano et al.(2007)]{Casellano07} Castellano, M., 
Salimbeni, S., Trevese, D., et al.\ 2007, \apj, 671, 1497 


\bibitem[Chapman et al.(2005)]{Chapman05} Chapman, S.~C., Blain, 
A.~W., Smail, I., \& Ivison, R.~J.\ 2005, \apj, 622, 772 

\bibitem[Cimatti et al.(2008)]{Cimatti08} Cimatti, A., Cassata, P., Pozzetti, L., et al.\ 2008, \aap, 482, 21

\bibitem[Daddi et al.(2004)]{Daddi04} Daddi, E., Cimatti, A., 
Renzini, A., et al.\ 2004, \apj, 617, 746 

\bibitem[\protect\citeauthoryear{Davies et al.}{2013}]{Davies13} 
Davies L.~J.~M., Bremer M.~N., Stanway E.~R., Lehnert M.~D., 2013, MNRAS, 
1544 

\bibitem[Davies et al.(2013b)]{Davies13b} Davies, L.~J.~M., Maraston, C., Thomas, D., et al.\ 2013, \mnras, 434, 296

\bibitem[D{\'{\i}}az-S{\'a}nchez et al.(2007)]{Diaz07} 
D{\'{\i}}az-S{\'a}nchez, A., Villo-P{\'e}rez, I., P{\'e}rez-Garrido, A., 
\& Rebolo, R.\ 2007, \mnras, 377, 516 


\bibitem[\protect\citeauthoryear{Delaye et al.}{2013}]{Delaye13} 
Delaye L., et al., 2013, arXiv, arXiv:1307.0003 

\bibitem[Douglas et al.(2009)]{Douglas09} Douglas, L.~S., Bremer, 
M.~N., Stanway, E.~R., Lehnert, M.~D., \& Clowe, D.\ 2009, \mnras, 400, 561 

\bibitem[Dressler(1980)]{Dressler80} Dressler, A.\ 1980, \apj, 
236, 351 

\bibitem[Haines et al.(2007)]{Haines07} Haines, C.~P., Gargiulo, A., Mercurio, A., et al.\ 2007, Cosmic Frontiers, 379, 206 

\bibitem[Ivison et al.(2011)]{2011MNRAS.412.1913I} Ivison, R.~J., et al.\ 2011, \mnras, 412, 1913 

\bibitem[Fynbo et al.(1999)]{Fynbo99} Fynbo, J.~U., M{\o}ller, 
P., \& Warren, S.~J.\ 1999, \mnras, 305, 849 


\bibitem[Gawiser et al.(2006a)]{Gawiser06} Gawiser, E., van 
Dokkum, P.~G., Herrera, D., et al.\ 2006a, \apjs, 162, 1 

\bibitem[Gawiser et al.(2006b)]{Gawiser06b} Gawiser, E., van 
Dokkum, P.~G., Gronwall, C., et al.\ 2006b, \apjl, 642, L13 

\bibitem[Guillaume et al.(2006)]{Guillaume06} Guillaume, M., 
Llebaria, A., Aymeric, D., Arnouts, S., 
\& Milliard, B.\ 2006, \procspie, 6064, 332 

\bibitem[Gladders \& Yee(2000)]{Gladders00} Gladders, M.~D., \& Yee, H.~K.~C.\ 2000, \aj, 120, 2148 



\bibitem[Gonz{\'a}lez et al.(2011)]{Gonzalez11} Gonz{\'a}lez, 
J.~E., Lacey, C.~G., Baugh, C.~M., \& Frenk, C.~S.\ 2011, \mnras, 413, 749 

\bibitem[Greve et al.(2010)]{Greve10} Greve, T.~R., 
Wei{$\beta$}, A., Walter, F., et al.\ 2010, \apj, 719, 483 

\bibitem[Gr{\"u}tzbauch et al.(2012)]{Grutzbauch12} Gr{\"u}tzbauch, 
R., Bauer, A.~E., J{\o}rgensen, I., \& Varela, J.\ 2012, \mnras, 423, 3652 

\bibitem[Guo et al.(2011)]{Guo11} Guo, Q., White, S., 
Boylan-Kolchin, M., et al.\ 2011, \mnras, 413, 101



\bibitem[Haberzettl et al.(2012)]{Haberzettl12} Haberzettl, L., 
Williger, G., Lehnert, M.~D., Nesvadba, N., 
\& Davies, L.\ 2012, \apj, 745, 96 

\bibitem[Hainline et al.(2011)]{Hainline11} Hainline, L.~J., 
Blain, A.~W., Smail, I., et al.\ 2011, \apj, 740, 96 


\bibitem[Hatch et al.(2011)]{Hatch11} Hatch, N.~A., De Breuck, 
C., Galametz, A., et al.\ 2011, \mnras, 410, 1537 


\bibitem[\protect\citeauthoryear{Hayward et 
al.}{2011}]{Hayward11} Hayward C.~C., Kere{\v s} D., Jonsson P., 
Narayanan D., Cox T.~J., Hernquist L., 2011, ApJ, 743, 159 


\bibitem[Henriques et al.(2012)]{Henriques12} Henriques, B.~M.~B., 
White, S.~D.~M., Lemson, G., et al.\ 2012, \mnras, 421, 2904 

\bibitem[Hilton et al.(2010)]{Hilton10} Hilton, M., 
Lloyd-Davies, E., Stanford, S.~A., et al.\ 2010, \apj, 718, 133

\bibitem[\protect\citeauthoryear{Huang et al.}{2013}]{Huang13} 
Huang S., Ho L.~C., Peng C.~Y., Li Z.-Y., Barth A.~J., 2013, ApJ, 768, L28

\bibitem[\protect\citeauthoryear{Husband et 
al.}{2013}]{Husband13} Husband K., Bremer M.~N., Stanway E.~R., 
Davies L.~J.~M., Lehnert M.~D., Douglas L.~S., 2013, MNRAS, 432, 2869

\bibitem[Ilbert et al.(2008)]{Ilbert08} Ilbert, O., Salvato, M., 
Capak, P., et al.\ 2008, Panoramic Views of Galaxy Formation and Evolution, 
399, 169 

\bibitem[Kurk et al.(2009)]{Kirk09} Kurk, J., Cimatti, A., Zamorani, G., et al.\ 2009, \aap, 504, 331

\bibitem[Lehnert \& Bremer(2003)]{Lehnert03} Lehnert, M.~D., \& Bremer, M.\ 2003, \apj, 593, 630 

\bibitem[Maraston(2005)]{Maraston05} Maraston, C.\ 2005, \mnras, 
362, 799 

\bibitem[McCracken et al.(2010)]{McCracken10} McCracken, H.~J., 
Capak, P., Salvato, M., et al.\ 2010, \apj, 708, 202 

\bibitem[McCracken et al.(2012)]{McCracken2012} McCracken, H.~J., 
Milvang-Jensen, B., Dunlop, J., et al.\ 2012, arXiv:1204.6586 



\bibitem[Mo \& White(2002)]{Mo02} Mo, H.~J., \& White, S.~D.~M.\ 2002, \mnras, 336, 112 

\bibitem[Newman et al.(2012)]{Newman12} Newman, A.~B., Ellis, 
R.~S., Bundy, K., \& Treu, T.\ 2012, \apj, 746, 162 


\bibitem[Nipoti et al.(2009)]{Nipoti09} Nipoti, C., Treu, T., Auger, M.~W., \& Bolton, A.~S.\ 2009, \apjl, 706, L86 

\bibitem[Nipoti et al.(2012)]{Nipoti12} Nipoti, C., Treu, T., Leauthaud, A., et al.\ 2012, \mnras, 422, 1714 

\bibitem[{{Oke} \& {Gunn}(1983)}]{Oke83}
{Oke}, J.~B. \& {Gunn}, J.~E. 1983, \apj, 266, 713

\bibitem[Oliver et al.(2012)]{Oliver12} Oliver, S.~J., Bock, J., 
Altieri, B., et al.\ 2012, \mnras, 424, 1614 

\bibitem[Papovich et al.(2010)]{Papovich10} Papovich, C., 
Momcheva, I., Willmer, C.~N.~A., et al.\ 2010, \apj, 716, 1503


\bibitem[Robin et al.(2007)]{Robin07} Robin, A.~C., Rich, 
R.~M., Aussel, H., et al.\ 2007, \apjs, 172, 545 

\bibitem[Santos et al.(2013)]{Santos13} Santos, J.~S., Altieri, 
B., Popesso, P., et al.\ 2013, \mnras, 433, 1287


\bibitem[Scott et al.(2008)]{Scott08} Scott, K.~S., Austermann, 
J.~E., Perera, T.~A., et al.\ 2008, \mnras, 385, 2225 


\bibitem[\protect\citeauthoryear{Shapley et 
al.}{2001}]{Shapley01} Shapley A.~E., Steidel C.~C., Adelberger 
K.~L., Dickinson M., Giavalisco M., Pettini M., 2001, ApJ, 562, 95 

\bibitem[Siringo et al.(2009)]{Siringo09} Siringo, G., Kreysa, E., Kov{\'a}cs, A., et al.\ 2009, \aap, 497, 945 


\bibitem[Smolcic et al.(2012)]{Smolcic12} Smolcic, V., Aravena, 
M., Navarrete, F., et al.\ 2012, arXiv:1205.6470 


\bibitem[Springel et al.(2006)]{Springel06} Springel, V., Frenk, 
C.~S., \& White, S.~D.~M.\ 2006, \nat, 440, 1137 

\bibitem[Stanford et al.(2012)]{Stanford12} Stanford, S.~A., 
Brodwin, M., Gonzalez, A.~H., et al.\ 2012, \apj, 753, 164 

\bibitem[Stanway et al.(2003)]{Stanway03} Stanway, E.~R., Bunker, 
A.~J., \& McMahon, R.~G.\ 2003, \mnras, 342, 439 

\bibitem[Stanway et al.(2008)]{Stanway08} Stanway, E.~R., Bremer, 
M.~N., \& Lehnert, M.~D.\ 2008, \mnras, 385, 493 

\bibitem[{{Steidel} {et~al.}(1995)}]{Steidel95}
{Steidel} C.~C., {Pettini} M., {Hamilton} D. 1995, \aj, 110, 2519

\bibitem[Steidel et al.(2004)]{Steidel04} Steidel, C.~C., 
Shapley, A.~E., Pettini, M., et al.\ 2004, \apj, 604, 534 


\bibitem[Strazzullo et al.(2013)]{Strazzullo13} Strazzullo, V., Gobat, R., Daddi, E., et al.\ 2013, \apj, 772, 118 

\bibitem[\protect\citeauthoryear{Tanaka et al.}{2013}]{Tanaka13} 
Tanaka M., et al., 2013, ApJ, 772, 113


\bibitem[\protect\citeauthoryear{Thomas et al.}{2010}]{Thomas10} 
Thomas D., Maraston C., Schawinski K., Sarzi M., Silk J., 2010, MNRAS, 404, 
1775 

\bibitem[Tran et al.(2010)]{Tran10} Tran, K.-V.~H., Papovich, 
C., Saintonge, A., et al.\ 2010, \apjl, 719, L126

\bibitem[Trevese et  al.(2007)]{Trevese07} Trevese, D., Castellano, M., Fontana, A., \& Giallongo, E.\ 2007, \aap, 463, 853 

\bibitem[Treister et al.(2009)]{Treister09} Treister, E., Virani, 
S., Gawiser, E., et al.\ 2009, \apj, 693, 1713

\bibitem[Treu et al.(2003)]{Treu03} Treu, T., Ellis, R.~S., Kneib, J.-P., et al.\ 2003, \apj, 591, 53

\bibitem[\protect\citeauthoryear{van de Sande et 
al.}{2013}]{van de Sande13} van de Sande J., et al., 2013, ApJ, 771, 
85 

\bibitem[van Dokkum et al.(2008)]{VanDokkum08} van Dokkum, P.~G., 
Franx, M., Kriek, M., et al.\ 2008, \apjl, 677, L5 

\bibitem[Verma et al.(2007)]{Verma07} Verma, A., Lehnert, 
M.~D., F{\"o}rster Schreiber, N.~M., Bremer, M.~N., 
\& Douglas, L.\ 2007, \mnras, 377, 1024 


\bibitem[Wardlow et al.(2011)]{Wardlow11} Wardlow, J.~L., Smail, 
I., Coppin, K.~E.~K., et al.\ 2011, \mnras, 415, 1479 

\bibitem[Wei{\ss} et al.(2009)]{Weiss09} Wei{\ss}, A., Kov{\'a}cs, A., Coppin, K., et al.\ 2009, \apj, 707, 1201 

\bibitem[Wen et al.(2010)]{Wen10} Wen, Z.~L., Han, J.~L., 
\& Liu, F.~S.\ 2010, \mnras, 407, 533

\bibitem[Wilson et al.(2006)]{Wilson06} Wilson, G., Muzzin, A., 
Lacy, M., et al.\ 2006, arXiv:astro-ph/0604289 

\bibitem[Yang et al.(2010)]{Yang10} Yang, Y., Zabludoff, A., 
Eisenstein, D., \& Dav{\'e}, R.\ 2010, \apj, 719, 1654 

\bibitem[Yun et al.(2012)]{Yun12} Yun, M.~S., Scott, K.~S., 
Guo, Y., et al.\ 2012, \mnras, 420, 957

\bibitem[Zatloukal et al.(2007)]{Zatloukal07} Zatloukal, M., R{\"o}ser, H.-J., Wolf, C., Hippelein, H., \& Falter, S.\ 2007, \aap, 474, L5

\bibitem[Ziparo et al.(2013)]{Ziparo13} Ziparo, F., Popesso, P., 
Biviano, A., et al.\ 2013, \mnras, 434, 3089 

\end{thebibliography}
\end{document}